\documentclass[aps,prd,reprint,onecolumn,superscriptaddress,12pt]{revtex4-2}

\usepackage{bm}
\usepackage{amsfonts}
\usepackage{latexsym}
\usepackage[latin1]{inputenc}
\usepackage{graphicx}
\usepackage{amsmath}
\usepackage[style=default]{caption}

\usepackage{subcaption}
\usepackage[mathscr]{eucal}
\usepackage{enumerate}
\usepackage{palatino}
\usepackage{mathpazo}
\usepackage{textcomp}
\usepackage{booktabs}
\usepackage{dcolumn}
\usepackage{ragged2e}
\usepackage{xcolor}
\usepackage{orcidlink}
\usepackage{commath}
\usepackage{comment}
\usepackage[italicdiff]{physics}
\usepackage[rightcaption]{sidecap}
\usepackage{array, makecell, rotating}
\usepackage{multirow}
\usepackage{eqparbox}
\usepackage{bigstrut} 
\usepackage{hyperref}
\hypersetup{colorlinks,citecolor=blue}
\hypersetup{colorlinks=true,linkcolor=red,filecolor=magenta, urlcolor=blue}

\newcommand{\fixme}[1]{{\color{red}{\bf{#1}}}}

\newcommand{\ti}[1]{{\Tilde{#1}}}

\begin{document}
\title{Optical Signatures of q-deformed solution in Einstein-Maxwell-dilaton Gravity}

\author{Chawit Sakkawattana}
\email{sakkawattana\_c@silpakorn.edu}
\affiliation{Strong Gravity Group, Department of Physics, Faculty of Science, Silpakorn University, Nakhon Pathom 73000, Thailand}

\author{Chatchai Promsiri} \email{chatchaipromsiri@gmail.com} 
\affiliation{Quantum Computing and Information Research Centre (QX), Faculty of Science, King Mongkut's University of Technology Thonburi,  Bangkok 10140, Thailand}

\author{Supakchai Ponglertsakul}
\email{supakchai.p@gmail.com}
\affiliation{Strong Gravity Group, Department of Physics, Faculty of Science, Silpakorn University, Nakhon Pathom 73000, Thailand}

\date{\today}

\begin{abstract}

We consider null geodesics in the background of spherically symmetric object in Einstein-Maxwell-Dilaton (EMD) theory with coupling function $f(\Phi)=e^{-2\lambda \Phi}$. The spherical solution is characteristically described by dilaton coupling $\lambda$, integrated dilaton flux $D$ and magnetic charge $P$. Then, we derive geodesic equations by using the Hamilton-Jacobi approach. The radial photon orbital equation on equatorial plane and effective potential are analyzed. The total deflection angle and trajectories of photon as a function of impact parameter $b$ are plotted with the variation of $\lambda,D$ and $P$. Furthermore, the relation between photon ring's width and the Lyapunov exponent is also explored.  In addition, we use the Gralla-Lupsasca-Marrone (GLM) model to model intensity profile of optically thin accretion disk around the object. Hence, we construct optical images of the object surrounded by three distinct emission profiles. Lastly, we investigate the radius of innermost stable circular orbit (ISCO) for timelike geodesics. 

\end{abstract}

\maketitle

\section{Introduction}

The achievement in capturing the first direct image of a black hole by the event horizon telescope (EHT) \cite{EventHorizonTelescope:2019dse,EventHorizonTelescope:2019uob,EventHorizonTelescope:2019jan,EventHorizonTelescope:2019ths,EventHorizonTelescope:2019pgp,EventHorizonTelescope:2019ggy}, has triggered a vast number of phenomenologically studies of black hole since 2019.
The trajectories of photons radiated by luminous sources in the vicinity of black hole are deflected by its highly curved spacetime. The deflected photon trajectories results in the characteristic image observed by a distant observer, which black hole appears as a central non-luminous region known as the black hole shadow, designated by one or more photon rings \cite{10.1098/rspa.1959.0015,10.1093/mnras/131.3.463,10.1119/1.15126,Zakharov:2005ek,Zakharov:2014lqa,Bardeen1970,Bardeen:1973tla}.

Many studies of non-vacuum solutions in Einstein's general theory of relativity as well as in alternative theories of gravity suggest that the final state of gravitational collapse is not necessarily a black hole. 
The resulting objects can exhibit a shadow region and a photon ring.
As a result, they could be black hole mimickers.
Black hole mimickers can be broadly categorized into two classes, the exotic compact objects (ECOs), such as wormhole \cite{Morris:1988cz,Bambi:2021qfo,PhysRevD.104.024071}, boson stars \cite{Schunck:2003kk,Liebling:2012fv,Yoshida:1997qf}, gravastars \cite{Mazur:2001fv,Ray:2020yyk} and fuzzball \cite{Mathur:2005zp,Cipriani:2025ini}. 
The others are the naked singularity geometries, where the singularity is visible to distant observers due to the absence of an event horizon.
For example, the Janis-Newman-Winicour (JNW) spacetime \cite{PhysRevLett.20.878,Virbhadra:1997ie}, the Joshi-Malafarina-Narayan (JMN) \cite{Joshi:2011zm} spacetime and the $q$-metric spacetimes \cite{Quevedo:2010mn}.
The existence of naked singularities in nature is highly debated because it violates the cosmic censorship hypothesis \cite{Penrose:1969pc}.
However, some theoretical work has shown that it might be possible for a collapsing object to form a singularity that is not hidden by an event horizon \cite{Shapiro:1991zza}.
Thus, in this EHT era, it is important to study the optical appearance of these spacetime geometries and compare them with the EHT results. 
This could help us constrain the parameter of these spacetimes and provides a pathway to identifying observational signatures that can distinguish them from black holes.

In addition to ECOs and naked singularities, another class of black hole mimickers arises from the low-energy limit of string theory. In this context, Garfinkle, Horowitz and Strominger (GHS) discover a magnetically charged black hole solution with dilaton scalar hair \cite{Garfinkle:1990qj}. This solution becomes known later as the GHS black hole. This solution belonging to the Einstein-Maxwell-dilaton (EMD) theory, admits an arbitrary dilaton coupling constant $\lambda$ and represents one of the earliest examples of dilatonic black holes. Later developments that extended the GHS solution to electrically charged are studied in \cite{Gurses:1995fw,Nozawa:2020wet}. More recently, Porfyriadis and Remmen \cite{Porfyriadis:2023qqo} construct a three-parameter family of static, spherically symmetric, asymptotically flat black hole solutions in the string frame, generalizing the original GHS black hole. The three global charges characterizing this family are the Komar mass $M$, the magnetic charge $P$ and the integrated dilaton flux $D$. 
A deformation parameter $q$ is also introduced, which quantifies the deviation from the GHS configuration: $q=1$ recovers the original GHS solution, while $q \neq 1$ leads to a family of dilatonic geometries with distinct near-horizon properties. 
In the string frame, the horizon area can vanish, remain finite, or diverge depending on $q$, whereas in the Einstein frame the corresponding solutions reduce to \textit{pointlike objects} for $q>0$. 
This sensitivity to the choice of frame emphasizes the importance of frame dependence when analyzing the near-horizon geometry and associated observational signatures. 
The parameter $q$ itself is constrained by $q^2 \leq 1 + \lambda^2$ to ensure a Lorentzian metric signature. 
In what follows, we shall refer to this generalized three-parameter family as the \textit{$q$-deformed solution}.

In this work, based on the work done in \cite{Promsiri:2023rez}, we discuss an optical appearance of static spherically symmetric q-deformed spacetime in the EMD gravity \cite{Porfyriadis:2023qqo}. We probe into how corresponding parameters $\lambda,D,P$ affect photon trajectories and black hole shadow. Therefore, this paper is organized as follow. In section \ref{sec:setup}, we briefly explain static spherically symmetric solution in the EMD gravity. The general framework describing null geodesics is given in section \ref{sec:null}. More specifically, we discuss the effective potential of circular photon orbit and visualizing photon trajectories in this section. In section \ref{sec:shadow}, we display shadow radius of the EMD solution using celestial coordinates. An optical image of the EMD theory is constructed in section \ref{sec:images} where luminous source is modeled by optically thin accretion disk with GLM emission profile. We summarize our findings in the last section \ref{sec:conclude}.

\section{q-deformed solution in EMD gravity}\label{sec:setup}
The action of Einstein-Maxwell-dilaton gravity (EMD) in Einstein frame is given by 
\begin{align}
S_\text{EMD} &=\int d^{4}x\sqrt{-g_E}\left( \mathcal{R} - 2g_E^{\mu \nu} \nabla_\mu \Phi \nabla_\nu \Phi - f(\Phi)F_{\mu \nu}F^{\mu \nu}  \right), \label{action}
\end{align}
where the coupling function between Maxwell and dilaton fields is $f(\Phi)=e^{-2\lambda \Phi}$ and a constant $\lambda$ denotes the dilaton coupling constant. The static spherically symmetric solution of such a system is explored in \cite{Porfyriadis:2023qqo}. In the Einstein frame, it reads
\begin{align}
    ds_E^2 &= -f(r)dt^2+g(r)dr^2+h(r)\left(d\theta^2 + \sin^2\theta d\phi^2\right), \label{bh}
\end{align}
where 
\begin{align}
    f(r) &= \left(\frac{r-r_+}{r-r_-}\right)^{\pm \Delta}\left(\frac{(r-r_-)(r-r_+)}{r^2}\right)^{\frac{1}{1+\lambda^2}}, \\
    g(r) &= \left(\frac{r-r_+}{r-r_-}\right)^{\frac{2}{q}\mp \Delta }\left[\frac{(r-r_-)(r-r_+)}{r^2}\right]^{\frac{\lambda^2}{1+\lambda^2}}\left(\frac{r_+-r_-}{q}\right)^4\frac{r^2\Theta(r)^{-4}}{(r-r_+)^3(r-r_-)^3}, \\
    h(r) &= \left(\frac{r-r_+}{r-r_-}\right)^{-1+\frac{1}{q}\mp \Delta }\left[\frac{r(r_+-r_-)}{q(r-r_-)\Theta(r)}\right]^2.
\end{align}
We introduce short-hand notations $\Delta\equiv\frac{\lambda\sqrt{1-q^2+\lambda^2}}{q(1+\lambda^2)}$ and $\Theta (r)\equiv  
\left(\frac{r-r_+}{r-r_-}\right)^{1/q} - 1$. As $r\to\infty$, the solution approaches Minkowski spacetime i.e., the solution is asymptotically flat. Other than $\lambda$, the solution depends on three parameters $r_\pm$ and $q$. Without loss of generality, we will choose $r_-<r_+$ where $r_+>0$ \cite{Porfyriadis:2023qqo}. And in this work, we will specifically focus in the region $r>r_+$. To ensure that the solution is non-complex, one requires $q^2 \leq 1+\lambda^2$. 

In addition, the above solution reduces to Garfinkel, Horowitz and Strominger (GHS) \cite{Garfinkle:1990qj} black hole when $q=1$ and extremal GHS as two horizons are degenerate i.e., $r_\pm \to r_0$. It should be emphasized that in the Einstein frame, this solution becomes pointlike object when $q>0$ \cite{Porfyriadis:2023qqo}.

\begin{figure}[htbp]
    \centering
    \includegraphics[width = 9 cm]{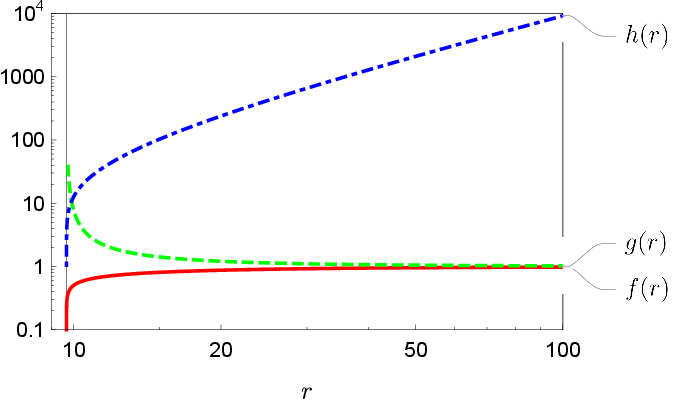}
    \caption{The metric functions $f(r),g(r),h(r)$ (red,green,blue) plotted against coordinate $r$. In this case, we choose $M=1,\lambda=2, \ti{D}=-2$ and $\ti{P}=0.8$. The vertical line marks the location of reduced horizon  
    $\ti{r}_+=9.67.$ \justifying} 
  \label{fig1}
\end{figure}

The Maxwell tensor and the dilaton field are given by
\begin{align}
 F &= P\sin \theta d\theta \wedge d\phi, \\
\Phi (r) &= -\frac{\lambda}{2(1+\lambda^2)}\log\left[\frac{(r-r_+)(r-r_-)}{r^2}\right]\pm \frac{\lambda\Delta}{2}\log\left[\frac{r-r_+}{r-r_-}\right] + \phi_0,
\end{align}
where $\phi_0$ is integration constant and will be set to zero throughout this paper.
Parameters $q$ and $\lambda$ are related to Komar mass (M), magnetic charged (P) and integrated dilaton flux (D) \cite{Porfyriadis:2023qqo} as follow 
\begin{eqnarray}
    M&=&\frac{r_++r_-}{2(1+\lambda^2)}+\frac{\lambda (r_+-r_-)}{2(1+\lambda^2)}\frac{\sqrt{1-q^2+\lambda^2}}{q}, \\
    P&=&\sqrt{\frac{2r_+r_-}{1+\lambda^2}}, \\
    D&=&-\frac{\lambda (r_++r_-)}{2(1+\lambda^2)}+\frac{r_+-r_-}{2(1+\lambda^2)}\frac{\sqrt{1-q^2+\lambda^2}}{q}.
\end{eqnarray}
It is convenient to define reduced magnetic charge and reduced dilaton flux as
\begin{eqnarray}
    \Tilde{P}\equiv\frac{P}{\sqrt{2}M} \ \ \ \text{and} \ \ \ \Tilde{D}\equiv\frac{D}{M}.
\end{eqnarray}
Thus, parameter $q$ can be written in term of $\ti{P}$ and $\ti{D}$ as follow
\begin{eqnarray}
    q^2=\frac{(1-\lambda \Tilde{D})^2-(1+\lambda^2)\Tilde{P}^2}{1+\Tilde{D}^2-\Tilde{P}^2}. \label{cond}
\end{eqnarray}
Moreover, the reduced horizon radii are given by
\begin{eqnarray}
    \Tilde{r}_+&\equiv&\frac{r_+}{M}=1-\lambda \Tilde{D}\pm \sqrt{(1-\lambda \Tilde{D})^2-(1+\lambda^2)\Tilde{P}^2}, \\
    \Tilde{r}_-&\equiv&\frac{r_-}{M}=\frac{(1+\lambda^2)\Tilde{P}^2}{1-\lambda \Tilde{D}\pm \sqrt{(1-\lambda \Tilde{D})^2-(1+\lambda^2)\Tilde{P}^2}}.
\end{eqnarray}
We also demand that $(1-\lambda \Tilde{D})^2 \geq (1+\lambda^2)\Tilde{P}^2$ to ensure that the $\ti{r}_+$ and $\ti{r}_-$ are real number. Thus, with \eqref{cond} and $q^2\leq1+\lambda^2$, the parameters $\Tilde{P},\Tilde{D}$ and $\lambda$ cannot be chosen arbitrarily. Moreover, the pointlike object \eqref{bh} can be reduced to the Schwarzschild black hole by setting $\ti{r}_+=2$ and $\ti{r}_-=\ti{P}=\ti{D}=\lambda=0$. Therefore, we choose only the ``plus" root of the $\ti{r}_+$ and $\ti{r}_-$. Remark that, throughout this paper, the Komar mass will be set to unity. As a demonstration, we show the behaviour of the metric functions $f,g,h$ in Fig.~\ref{fig1}. Clearly, the metric is asymptotically flat i.e., $f,g\to 1$ and $h\sim r^2$ as $r\to\infty$.

In the extremal limit, i.e., $\Tilde{r}_-\to \Tilde{r}_+$, it is possible to define the maximum reduced magnetic charge for a given dilaton flux at the boundary 
\begin{eqnarray}
    \Tilde{P}_\text{ext}^2=\frac{(1-\lambda \Tilde{D})^2}{1+\lambda^2}. \label{extformula}
\end{eqnarray}

\begin{figure}
    \centering
    \begin{subfigure}[t]{.3\linewidth}
        \includegraphics[width=5.8cm]{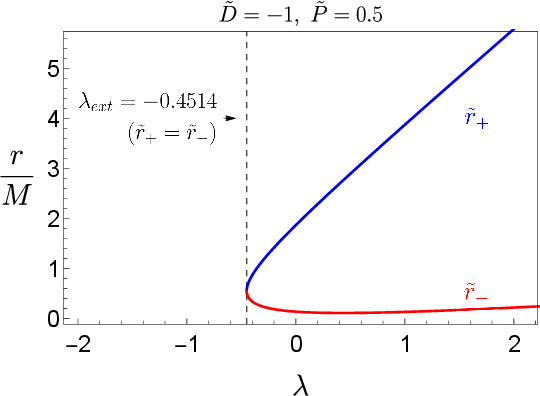}
    \end{subfigure}
    \hspace{6mm}
    \begin{subfigure}[t]{.3\linewidth}
        \includegraphics[width=5.8cm]{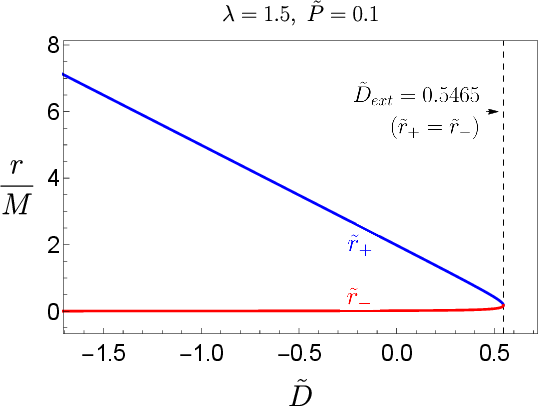}
    \end{subfigure}
    \hspace{6mm}
    \begin{subfigure}[t]{.3\linewidth}
        \includegraphics[width=5.8cm]{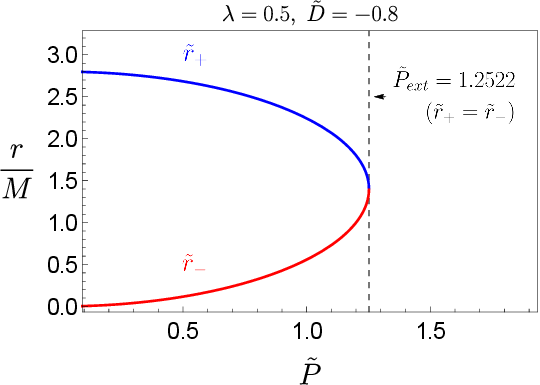}
    \end{subfigure}
        \caption{The radii $\tilde{r}_\pm$ plotted as function of $\lambda, \tilde{D}$ and $\tilde{P}$. The vertical dashed lines mark the location at which $\tilde{r}_-$ coincides with $\tilde{r}_+$.       
        \justifying} 
     \label{fig:extremalLines}
\end{figure}

For $\Tilde{P}> \Tilde{P}_{ext}$, we find that $\Tilde{r}_\pm$ becomes complex number. Thus, we will consider only the case where $\Tilde{P} \leq \Tilde{P}_{ext}$. In Fig.~\ref{fig:extremalLines}, we display how horizon radii are affected by reduced magnetic charged $\Tilde{P}$, dilaton coupling $\lambda$ and reduced dilaton flux $\Tilde{D}$. From the right figure, we notice that $\Tilde{r}_- (\Tilde{r}_+)$ increases (decreases) with magnetic charged. Both radii meet at the extremal point $\Tilde{P}=1.2522$ which can be calculated from \eqref{extformula}. For varying $\lambda$, $\Tilde{r}_+$ monotonically increases with $\lambda$. On the other hand, $\Tilde{r}_-$ initially decreases at small $\lambda$ and increases at large $\lambda$. Finally, $\Tilde{r}_+$ decreases with $\Tilde
D$ while $\Tilde{r}_-$ increases slightly as $\Tilde{D}$ gets larger.

\begin{figure}
    \centering
    \begin{subfigure}[t]{.3\linewidth}
        \includegraphics[width=5.5cm]{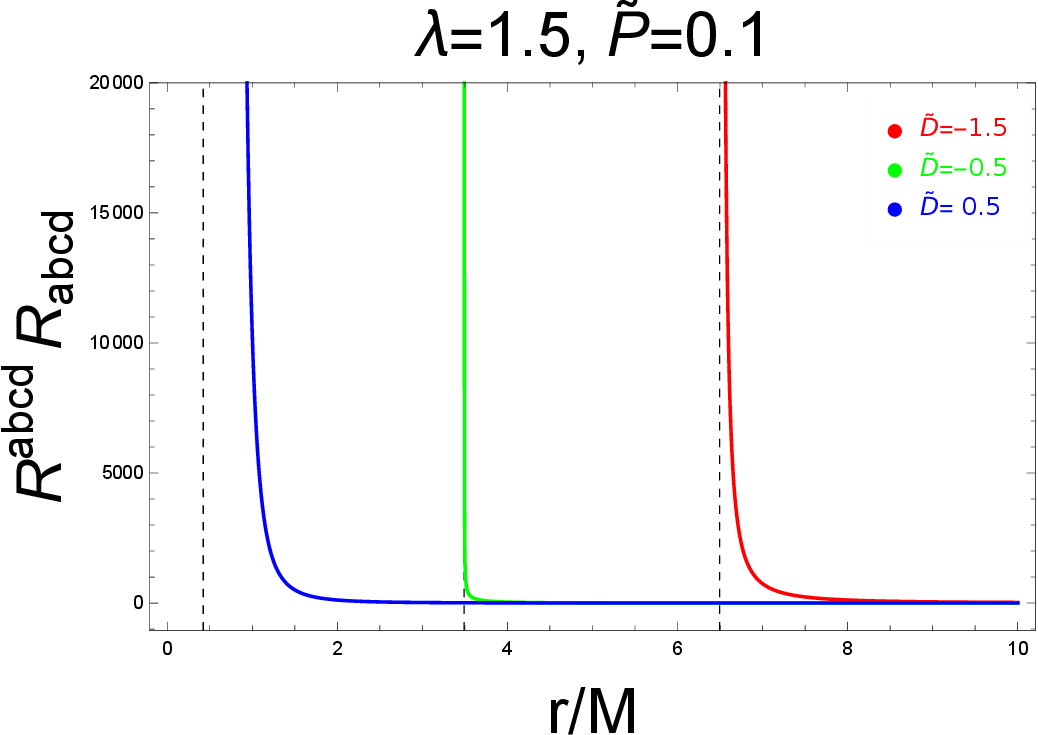}
    \end{subfigure}
    \hspace{6mm}
    \begin{subfigure}[t]{.3\linewidth}
        \includegraphics[width=5.5cm]{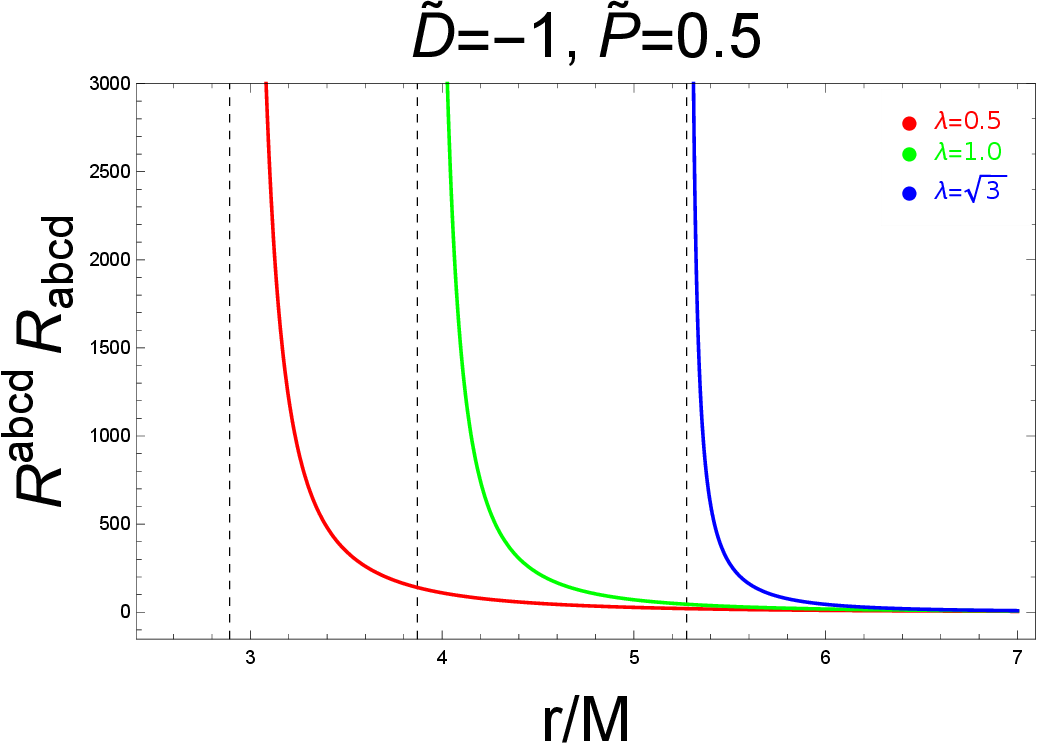}
    \end{subfigure}
    \hspace{6mm}
    \begin{subfigure}[t]{.3\linewidth}
        \includegraphics[width=5.5cm]{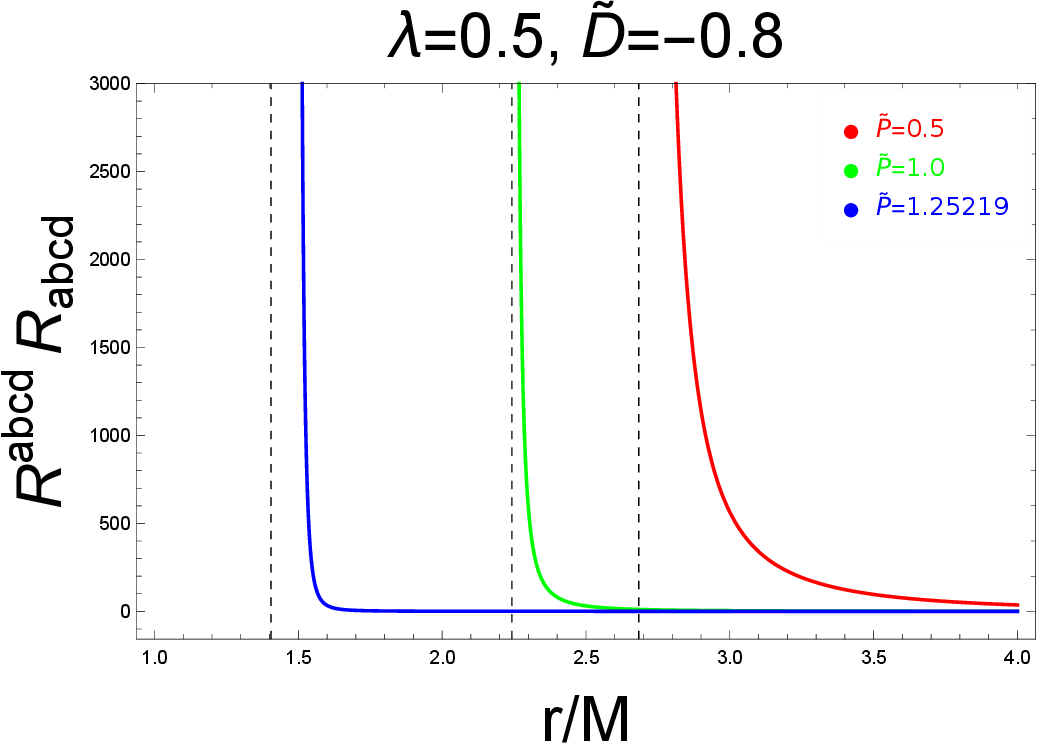}
    \end{subfigure}
        \caption{The Kretschmann scalar $R_{abcd}R^{abcd}$ as a function of $r/M$. The dashed vertical lines denote the horizon radii $r_+/M$ of each respective cases.   
        \justifying} 
     \label{fig:KM}
\end{figure}

We illustrate the behaviour of Kretschmann scalar as a function of $\tilde{r}$ in Fig.~\ref{fig:KM}. The vertical lines indicate the location of horizon radii of each respective cases. Notably, we observe that the curvature scalar diverges at $\tilde{r}_+ > 0$. This is in contrast with conventional essential singularity which usually lies hidden behind a black hole's event horizon.

\section{Null Geodesics}\label{sec:null}
In this section, we analyze null geodesics on the black hole spacetime. Here, we follow the Lagrangian and Hamilton-Jacobi approach discussed in \cite{Chandra}. From the line element \eqref{bh}, the equation of motion can be derived from the following Lagrangian
\begin{align}
    \mathcal{L}&=\frac{1}{2}g_{\mu\nu}\Dot{x}^{\mu}\Dot{x}^{\nu} \nonumber \\
    &=\frac{1}{2}\left[-f(r)\Dot{t}^2+g(r)\Dot{r}^2+h(r)\left(\Dot{\theta}^2 + \sin^2{\theta} \Dot{\phi}^2\right)\right]\label{eq5},
\end{align}
where $\dot{x}^{\mu}=dx^{\mu}/d\sigma$ is a tangent vector along geodesic curve $x^{\mu}(\sigma)$ with affine parameter $\sigma$.  The canonical momenta of particles can be calculated by  $p_{\mu}=\frac{\partial\mathcal{L}}{\partial\Dot{x}^\mu}=g_{\mu\nu}\Dot{x^{\nu}}$
then 
\begin{align}
    p_{t}&=-f(r)\Dot{t} = -E\label{eq6},\\
    p_{r}&=g(r)\Dot{r}\label{eq7},\\
    p_{\theta}&=h(r)\Dot{\theta}\label{eq8},\\
    p_{\phi}&=h(r)\sin^2{\theta}\Dot{\phi}=L\label{eq9}.
\end{align}
Two conserved quantities, the energy $E$ and the angular momentum $L$ are introduced as a result of two spacetime symmetries i.e., time translation and angular rotation. To analyze the geodesics in $r$ and $\theta$ coordinate, we implement the Hamilton-Jacobi approach. Let us start by defining the Jacobi action as
\begin{align}
    S &= \frac{1}{2}\delta\sigma -Et+L\phi+S_{r}(r)+S_{\theta}(\theta),
\end{align}
where $\delta=0$ and $1$ for null and timelike geodesics, respectively. The Hamilton-Jacobi equation can be computed via the differentiation of the action $S$ with respect to affine parameter $\sigma$
\begin{align}
    \frac{\partial S}{\partial \sigma}&=-\frac{1}{2}g^{\mu\nu}\frac{\partial S}{\partial x^{\mu}}\frac{\partial S}{\partial x^{\nu}}=-\frac{1}{2}g^{\mu\nu}p_{\mu}p_{\nu}=\frac{1}{2}\delta.\label{eq10}
\end{align}
With \eqref{eq5} and \eqref{eq6}--\eqref{eq9}, we obtain the following equation
\begin{align}
    h(r)\delta-\frac{h(r)}{f(r)}E^2+\frac{h(r)}{g(r)}\left(\frac{\partial S_r}{\partial r}\right)^2+L^2=-\left(\frac{\partial S_{\theta}}{\partial \theta}\right)^2-L^2\cot^2\theta. \label{eq22}
\end{align}
We notice that the LHS of the above equation is $r-$dependence while the RHS is $\theta-$dependence. Therefore, we equate each side of \eqref{eq22} to the Carter constant $\mathcal{Q}$ \cite{PhysRev.174.1559}. These give us two following equations
\begin{align}
\frac{h^2(r)f(r)}{g(r)}\left(\frac{\partial S_r}{\partial r}\right)^2 &=\mathcal{R}(r) \label{eq13}, \\
\left(\frac{\partial S_{\theta}}{\partial\theta}\right)^2 &=\Theta(\theta), \label{eq14}
\end{align}
where
\begin{align}
    \mathcal{R}(r)&=h^2(r)E^2-f(r)h(r)\left[h(r)\delta+L^2+\mathcal{Q}\right],\label{eq26}\\
    \Theta(\theta)&=\mathcal{Q}-L^2\cot^2\theta. \label{eq27}
\end{align}
Recall that, $\frac{S_r}{\partial r}=p_r$ and $\frac{S_\theta}{\partial \theta}=p_\theta$, thus \eqref{eq13} and $\eqref{eq14}$ are 
\begin{align}
    h(r)\left(\dv{r}{\sigma}\right)&=\pm\sqrt{\frac{\mathcal{R}(r)}{f(r)g(r)}},\label{eq32}\\
    h(r)\left(\dv{\theta}{\sigma}\right)&=\pm\sqrt{\Theta(\theta)}.\label{eq33}
\end{align}
Similar relations are obtained in \cite{Promsiri:2023rez} for hairy black hole in EMD theory. The $\pm$ sign can be interpreted as the positive and negative radial and angular directions. These equations together with \eqref{eq6} and \eqref{eq9} fully described particle motion (timelike and null-like) in static spherically symmetric spacetime.


In spherically symmetric spacetime, without loss of generality, one can consider a photon moving on the equatorial plane $\theta=\frac{\pi}{2}$ i.e., $\dot{\theta}=0$ and $\delta=0$. The RHS of \eqref{eq22} implies that $\mathcal{Q}=0$. We redefine a new affine parameter $\Tilde{\sigma} = \sigma L$ and a useful parameter, namely the impact parameter $b=\frac{L}{E}$. Hence, the radial equation \eqref{eq13} is now
\begin{align}
    \left(\frac{dr}{d\tilde{\sigma}}\right)^2 &= \frac{1}{f(r)g(r)b^2}\left(1-b^2V_{eff}\right), \label{radialEq}
\end{align}
where the effective potential is 
\begin{align}
    V_{eff} &\equiv \frac{f(r)}{h(r)}, \nonumber \\
    &=  \left(\frac{q}{r_--r_+}\right)^2\left(\left(\frac{r-r_+}{r-r_-}\right)^\frac{1}{q}-1\right)^2\left(\frac{r-r_+}{r-r_-}\right)^{\frac{-(1+\lambda^2)+2\lambda\sqrt{1-q^2+\lambda^2}}{q(1+\lambda^2)}}\left(\frac{(r-r_-)(r-r_+)}{r^2}\right)^{\frac{2}{1+\lambda^2}}. \label{veff}
\end{align}
The radius of photon sphere $(r_{ph})$ can be found by solving the following equations 
\begin{align}
    V_\text{eff}'(r_{ph}) = 0,~~~V_\text{eff}(r_{ph}) = \frac{1}{b_{ph}^2}, \label{effcondition}
\end{align}
where prime denotes derivative with respect to $r$ with $\Tilde{r}_-<\Tilde{r}_+ < r_{ph}$ and $b_{ph}$ refers to critical impact parameter. Since, the expression of the effective potential is complicated, thus the photon radius will be only determined numerically. 

\begin{figure}[htbp]
\centering
       \includegraphics[scale=.47]{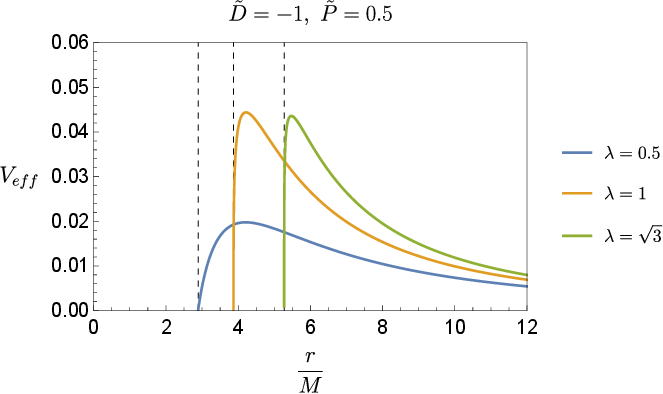}
       \includegraphics[scale=.47]{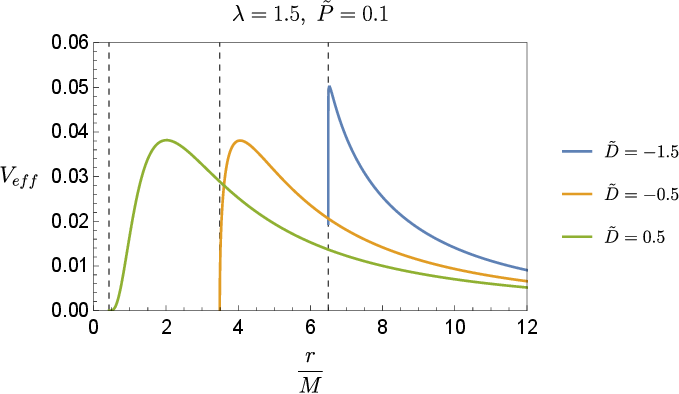}
       \includegraphics[scale=.47]{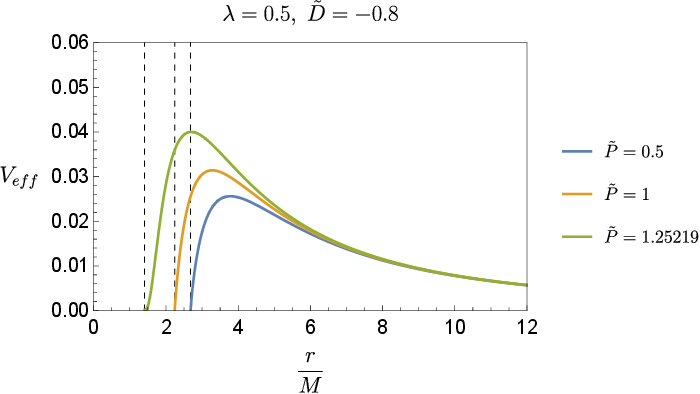}
\caption{The effective potential $V_\text{eff}$ as a function of distance $r/M$ for different values of parameters $\lambda,~\Tilde{D},$ and $\Tilde{P}$. Dashed lines represent by the outer horizon for each case of parameters.\justifying}
 \label{fig:veff}
\end{figure}

We illustrate the behavior of the effective potential \eqref{veff} in Fig.~\ref{fig:veff} where $\Tilde{D},\Tilde{P}$ and $\lambda$ are fixed in their respective figures. The photon radius can be evaluated from the radius which $V_{eff}$ attains its peak. It is observed that, as $\lambda$ becomes larger, the photon radii $r_{ph}$ are closer to the outer horizons (vertical dashed lines). The opposite trend is found in the middle panel. As $\Tilde{D}$ becomes more positive, a distance between the horizon radius and photon radius becomes grower. On the other hand, the correlation of $\Tilde{P}$ and separation of $\Tilde{r}_+$ and $\Tilde{r}_{ph}$ is not straightforward. A closer investigation reveals that, the separation distant is decreasing as the magnetic charge increases. The minimum distance occurs at $\Tilde{P}=1.09247$. After that, the distance between two surfaces grows apart and reaches the maximum at the extremal value $\Tilde{P}_{ext}=1.2522$. For parameter explored in this figure, we have listed corresponding parameters in Table~\ref{Tab:I}. The detail about $r_{ISCO}$ can be found in Appendix~\ref{app:isco}. 

\begin{table}[htbp]
\caption{Corresponding parameters for each individual case used in Fig.~\ref{fig:veff}}
\centering
\begin{tabular}{|*{4}{c|}}
  \cline{2-4}
\multicolumn{1}{c|}{} & \multicolumn{3}{c|}{$\Tilde{D}=-1 ,\Tilde{P}=0.5$} \\
  \cline{2-4}
 \multicolumn{1}{c|}{} & \eqmakebox[H]{$\lambda=0.5$} & \eqmakebox[H]{$\lambda=1$} & \eqmakebox[H]{$\lambda=\sqrt{3}$} \\ \hline
  $\Tilde{r}_+$ & 2.8920 & 3.8708 & 5.2745 \\ \hline
  $r_{ph}$ & 4.2057 & 4.2164 & 5.4729 \\ \hline
  $b_{ph}$ & 7.1126 & 4.7465 & 4.7897 \\ \hline
  $r_{ISCO}$ & 8.2343 & 6.5720 & 7.5886 \\ \hline
\end{tabular}

\vspace{5mm}

\begin{tabular}{|*{4}{c|}}
  \cline{2-4}
\multicolumn{1}{c|}{} & \multicolumn{3}{c|}{$\lambda=1.5 ,\Tilde{P}=0.1$} \\
  \cline{2-4}
 \multicolumn{1}{c|}{} & \eqmakebox[H]{$\Tilde{D}=-1.5$} & \eqmakebox[H]{$\Tilde{D}=-0.5$} & \eqmakebox[H]{$\Tilde{D}=0.5$} \\ \hline
  $\Tilde{r}_+$ & 6.4950 & 3.4907 & 0.4232 \\ \hline
  $r_{ph}$ & 6.5264 & 4.0548 & 2.0171 \\ \hline
  $b_{ph}$ & 4.4662 & 5.1242 & 5.1186 \\ \hline
  $r_{ISCO}$ & 8.1333 & 6.7965 & 5.0834 \\ \hline
\end{tabular}

\vspace{5mm}

\begin{tabular}{|*{4}{c|}}
  \cline{2-4}
\multicolumn{1}{c|}{} & \multicolumn{3}{c|}{$\lambda=0.5 ,\Tilde{D}=-0.8$} \\
  \cline{2-4}
 \multicolumn{1}{c|}{} & \eqmakebox[H]{$\Tilde{P}=0.5$} & \eqmakebox[H]{$\Tilde{P}=1$} & \eqmakebox[H]{$\Tilde{P}=\Tilde{P}_{ext}(1.25219)$} \\ \hline
  $\Tilde{r}_+$ & 2.6836 & 2.2426 & 1.4050 \\ \hline
  $r_{ph}$ & 3.7864 & 3.2839 & 2.7002 \\ \hline
  $b_{ph}$ & 6.2492 & 5.6420 & 4.9994 \\ \hline
  $r_{ISCO}$ & 7.2948 & 6.2487 & 5.1792 \\ \hline
\end{tabular} \label{Tab:I}
\end{table}


To fully characterize dynamic of photon near the spherically symmetric solution of the EMD gravity, we are interested in the motion of particle in $r-\phi$ plane rather than $r(\Tilde{\sigma})$. One starts by realizing the following 
\begin{align}
\left(\frac{dr}{d\tilde{\sigma}}\right)^2 &= \left(\frac{dr}{d\phi}\frac{d\phi}{d\tilde{\sigma}}\right)^2
\left(\frac{dr}{d\phi}\right)^2 = \left(\frac{dr}{d\phi}\right)^2\left(\frac{1}{h(r)^2}\right),  \label{rdot}
\end{align}
Therefore, the radial equation \eqref{radialEq} can be recast into the form
\begin{align}
    \left(\frac{dr}{d\phi}\right)^2 &= \left(\frac{1}{f(r)g(r)b^2} - \frac{1}{h(r)g(r)}\right)h(r)^2.
\end{align}
It is useful to define variable $u=1/r$ such that infinity is now at $u=0$ and horizon is at $u_+=1/r_+$. After changing variable, we obtain 
\begin{align}
    \left(\frac{du}{d\phi}\right)^2 &= \frac{u^4}{g(u)}\left(\frac{1}{f(u)b^2} - \frac{1}{h(u)}\right)h(u)^2, \nonumber \\
    &\equiv u^4 P(u). \label{Pu}
\end{align}
We differentiate \eqref{Pu} with respect to $\phi$. Thus, we obtain second order differential equation
\begin{align}
    \frac{d^2u}{d\phi^2} &= 2u^3P(u) + \frac{u^4}{2}P'(u), \label{eqU}
\end{align}
where $P'(u)=dP/du$. We omit to express the full expression of the right hand side of \eqref{eqU} due to its intricacy. This differential equation serves as equation of motion of photon in spherically symmetric spacetime \eqref{bh}. As will be seen later, its solution, $u(\phi)$, describes how particle moves in the $r-\phi$ plane.

To help visualize an optical appearance of spherically symmetric solution in the EMD theory which will be presented later, it is useful to define a number of photon orbit $n$ in term of total deflection angle $\Delta\phi$ as $n=\Delta\phi/2\pi.$ One can classify range of $n$ into three categories by counting how many time photon has intersect with an equatorial plane \cite{Gralla:2019xty}. These are
\begin{itemize}
    \item $n>3/4$: the light rays intersect the equatorial plane at most one time or \textit{direct emission} \justifying.
    \item $3/4 < n < 5/4$: the light rays intersect the equatorial at least two times or \textit{lensing ring} \justifying.
    \item $5/4 < n$: the light rays intersect the equatorial plane at least three times or \textit{photon ring} \justifying.
\end{itemize}
For each fixed $b$, we can numerically solve \eqref{eqU} with initial conditions $u(0)=0$ and $\frac{du}{d\phi}=\frac{1}{b}$. The light ray will be traced back from infinity $u=0$ to either before entering the horizon $u_+=1/r_+$ or back to infinity again. As a result, we obtain a total deflection angle of light ray as a function of impact parameter $b$. Now let us first qualitatively describe a behavior of $n$ vs $b$. In case of zero impact parameter $b=0$, photon travels radially in a straight line toward the spherical object which corresponds with $n=0$. When the impact parameter is less than its critical value $b<b_{ph}$, photon trajectories are bent around the central object before entering the horizon. Thus, one expect that $n$ should increase with $b$. Interestingly, at $b=b_{ph}$, the incoming photon infinitely orbits around the gravitating object. As point out in \cite{Gralla:2019xty}, it is shown that the total deflection angle of the null-geodesics around the Schwarzschild black hole $\Delta\phi$ diverges logarithmically as $b\to b_{ph}$. Lastly, for the photon trajectory that is far away from the object $b>b_{ph}$, gravitational attraction is small, rendering a small bending of the incoming photon trajectory before it is scattered back to the distant region. Thus, $n$ should decrease as $b$ increases in this case.

\begin{figure}[htbp]
    \begin{subfigure}[t]{.3\linewidth}
        \includegraphics[width=5cm]{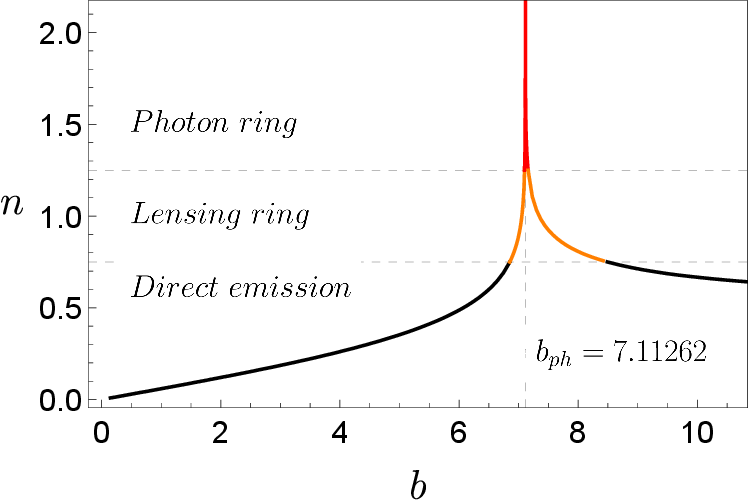}
    \end{subfigure}
    \begin{subfigure}[t]{.3\linewidth}
        \includegraphics[width=5cm]{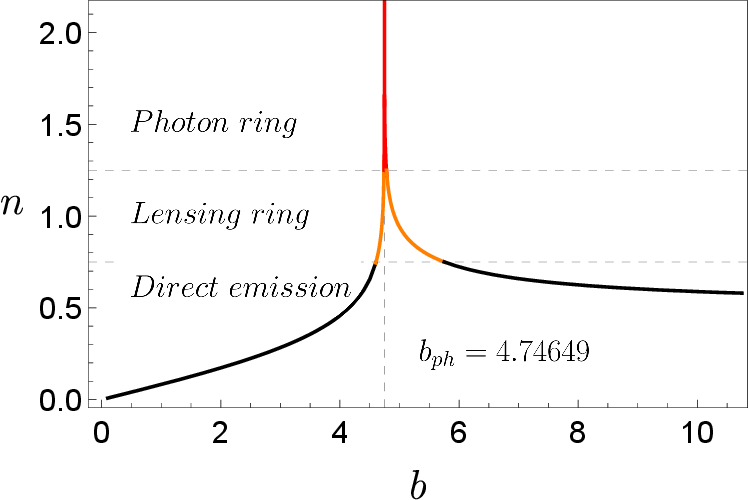}
    \end{subfigure}
    \begin{subfigure}[t]{.3\linewidth}
        \includegraphics[width=5cm]{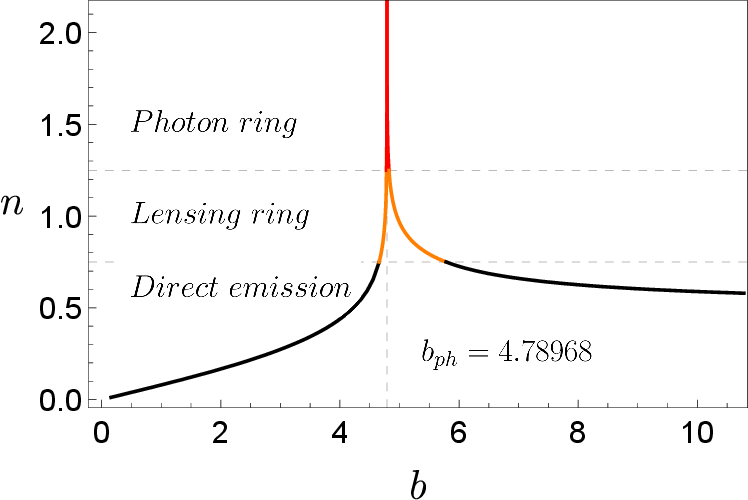}
    \end{subfigure}
    \begin{subfigure}[t]{.3\linewidth}
        \includegraphics[width=5cm]{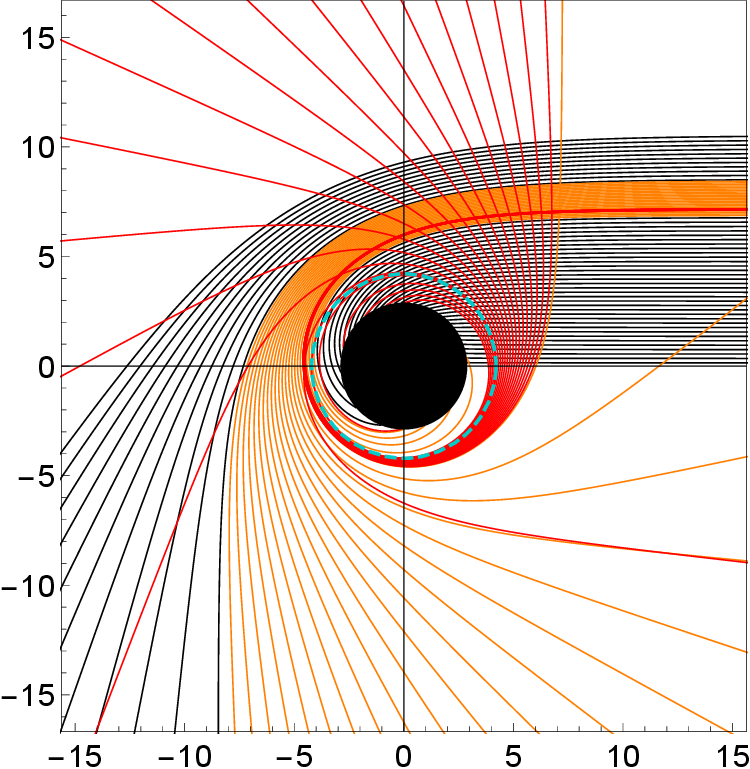}
        \caption{$\lambda=0.5$}
    \end{subfigure}
    \begin{subfigure}[t]{.3\linewidth}
        \includegraphics[width=5cm]{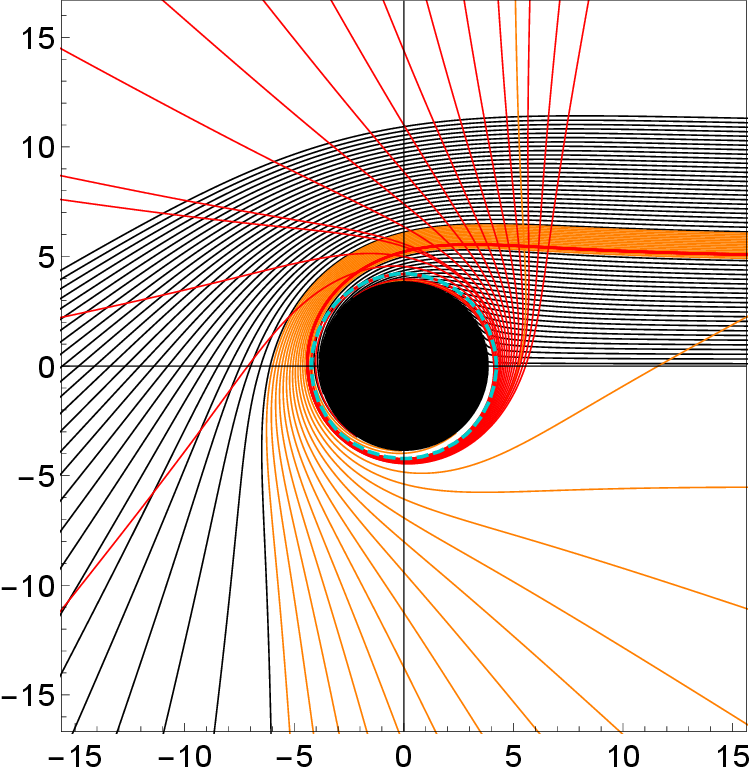}
        \caption{$\lambda=1$}
    \end{subfigure}
    \begin{subfigure}[t]{.3\linewidth}
        \includegraphics[width=5cm]{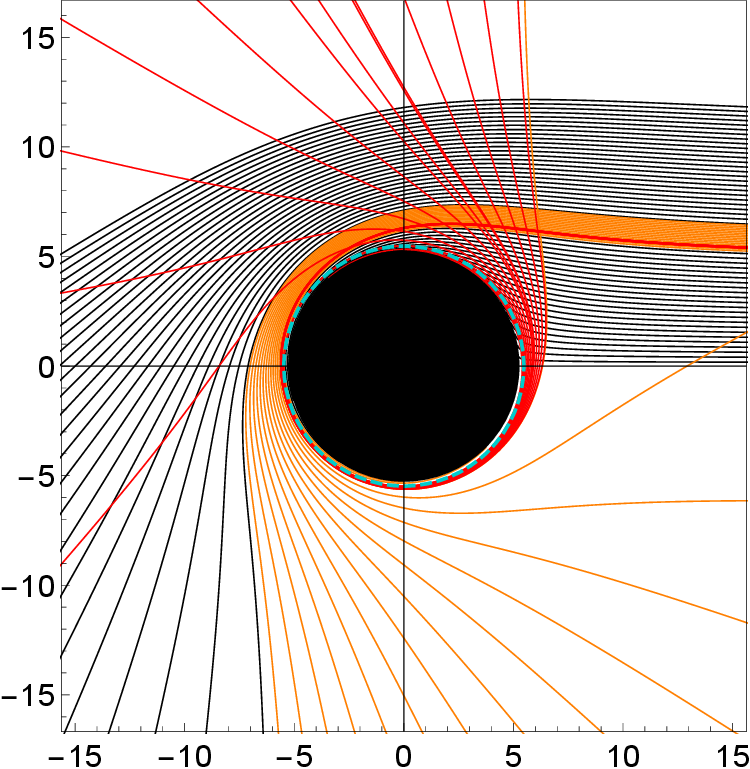}
        \caption{$\lambda=\sqrt{3}$}
    \end{subfigure}
    \caption{ The number of photon orbits $n$ as a function of impact parameter $b$ for fixed $\Tilde{D}=-1$ and $\Tilde{P}=0.5$ for three distinct values of $\lambda=0.5,1$ and $\sqrt{3}$ (top row, left to right). The trajectories of photon with direct emission (black), lensing ring (orange) and photon ring (red) for the same set of parameters (bottom row). The cyan dashed circle denotes the $r_{ph}$. \justifying    
   }
 \label{fig:raytrace1}
\end{figure}

\begin{figure}[htbp]
    \begin{subfigure}[t]{.3\linewidth}
         \includegraphics[width=5cm]{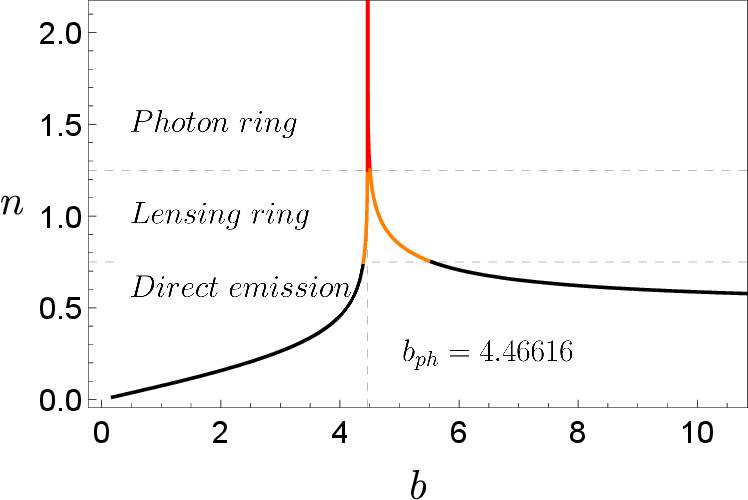}
    \end{subfigure}
    \begin{subfigure}[t]{.3\linewidth}
        \includegraphics[width=5cm]{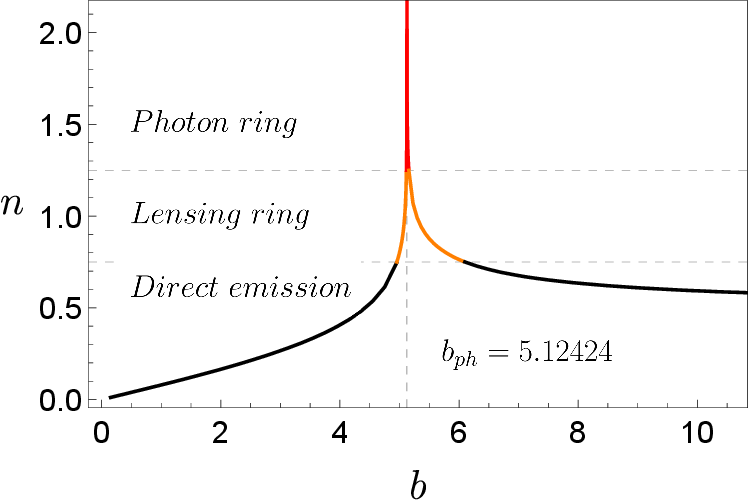}
    \end{subfigure}
    \begin{subfigure}[t]{.3\linewidth}
        \includegraphics[width=5cm]{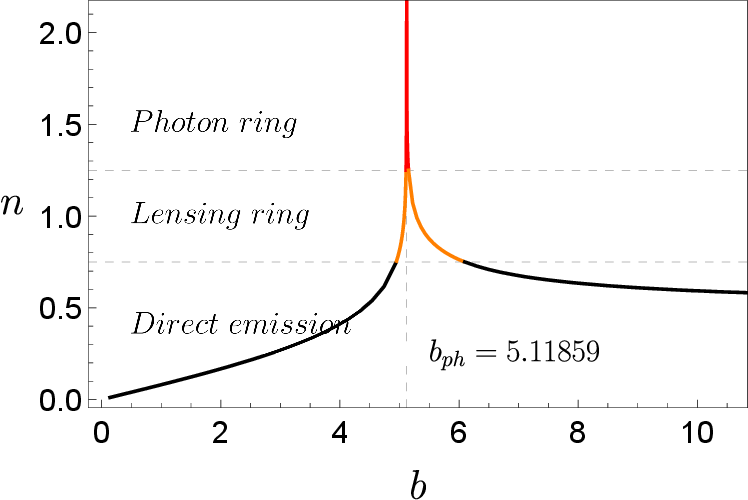}
    \end{subfigure}
    \begin{subfigure}[t]{.3\linewidth}
        \includegraphics[width=5cm]{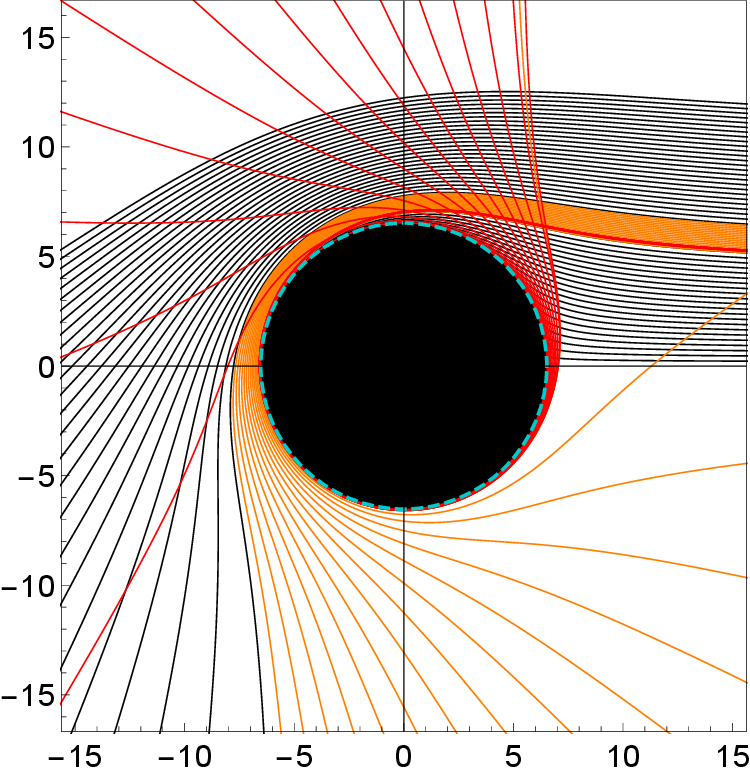}
        \caption{$\Tilde{D}=-1.5$}
    \end{subfigure}
    \begin{subfigure}[t]{.3\linewidth}
        \includegraphics[width=5cm]{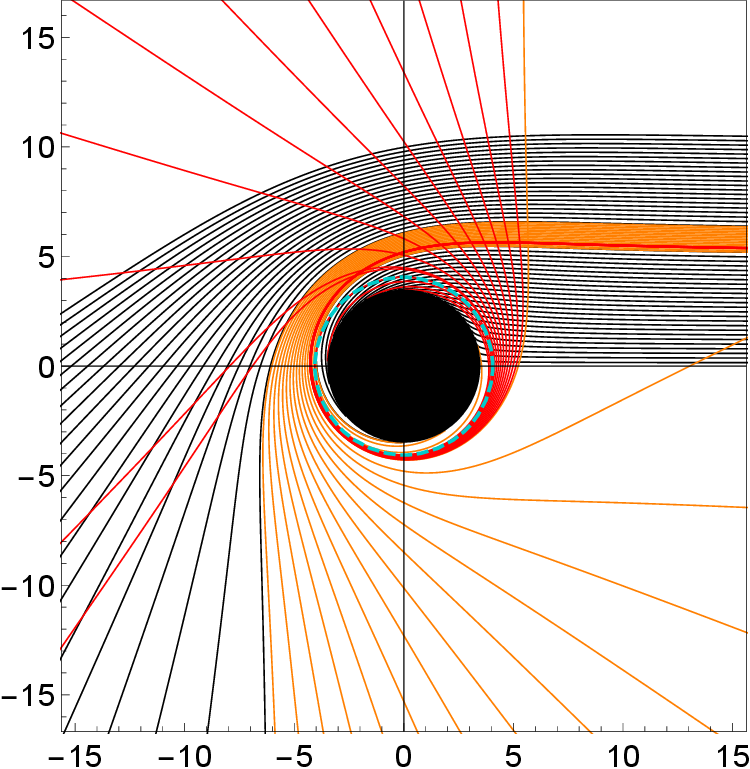}
        \caption{$\Tilde{D}=-0.5$}
    \end{subfigure}
    \begin{subfigure}[t]{.3\linewidth}
        \includegraphics[width=5cm]{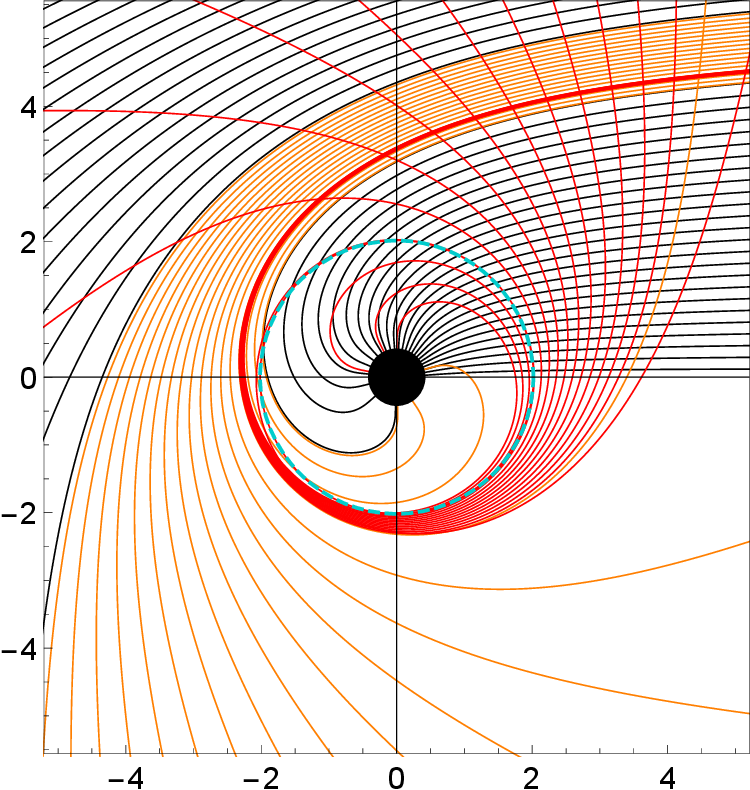}
        \caption{$\Tilde{D}=0.5$}
    \end{subfigure}
    \caption{ The number of photon orbits $n$ as a function of impact parameter $b$ for fixed $\lambda=1.5$ and $\Tilde{P}=0.1$ for three distinct values of $\Tilde{D}=-1.5,-0.5$ and $0.5$ (top row, left to right). The trajectories of photon with direct emission (black), lensing ring (orange) and photon ring (red) for the same set of parameters (bottom row). The cyan dashed circle denotes the $r_{ph}$. \justifying}
 \label{fig:raytrace2}
\end{figure}

\begin{figure}[htbp]
    \begin{subfigure}[t]{.3\linewidth}
        \includegraphics[width=5cm]{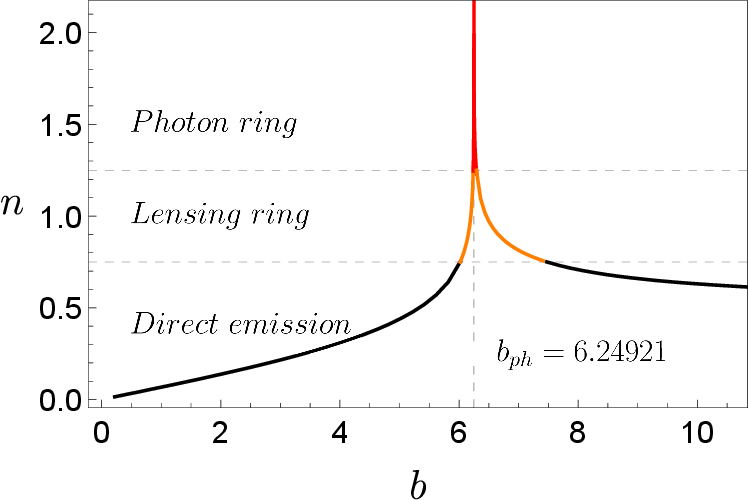}
    \end{subfigure}
    \begin{subfigure}[t]{.3\linewidth}
        \includegraphics[width=5cm]{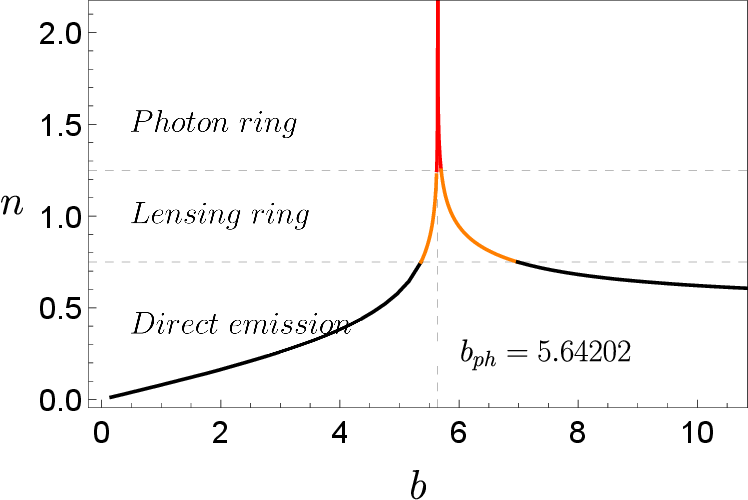}
    \end{subfigure}
    \begin{subfigure}[t]{.3\linewidth}
        \includegraphics[width=5cm]{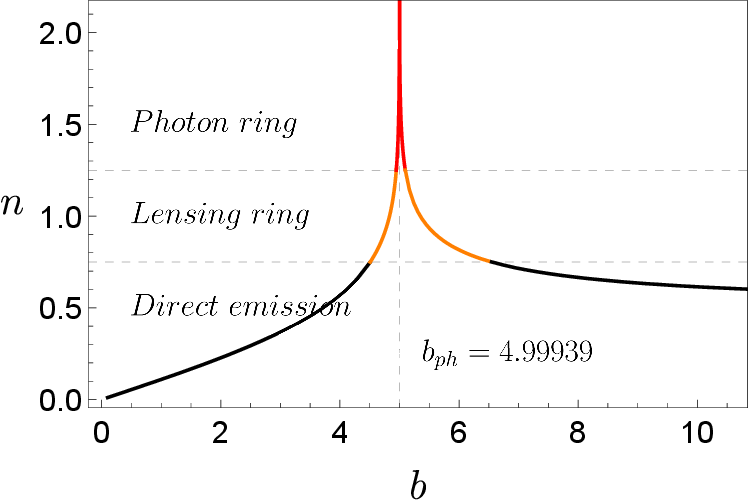}
    \end{subfigure}
    \begin{subfigure}[t]{.3\linewidth}
        \includegraphics[width=5cm]{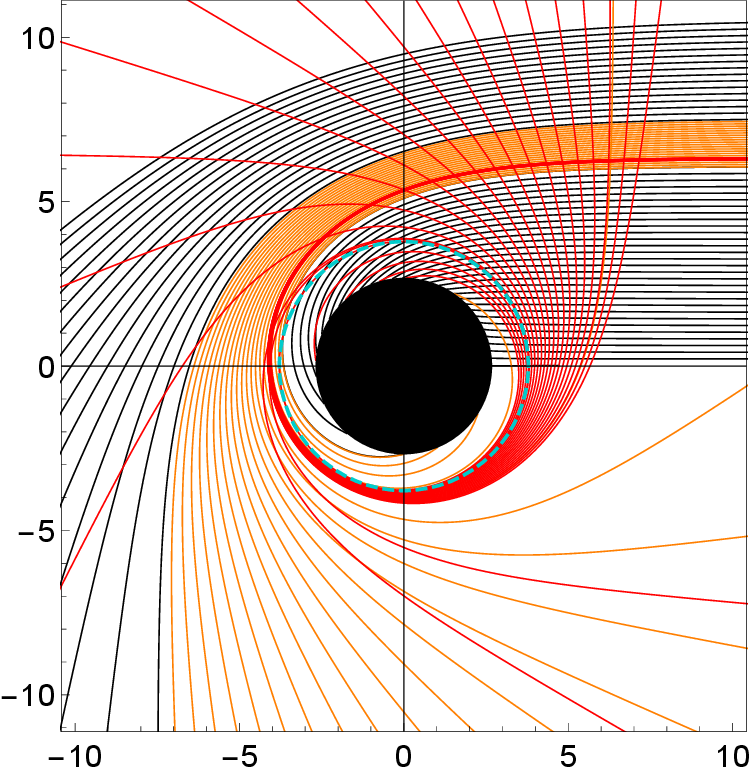}
        \caption{$\Tilde{P}=0.5$}
    \end{subfigure}
    \begin{subfigure}[t]{.3\linewidth}
        \includegraphics[width=5cm]{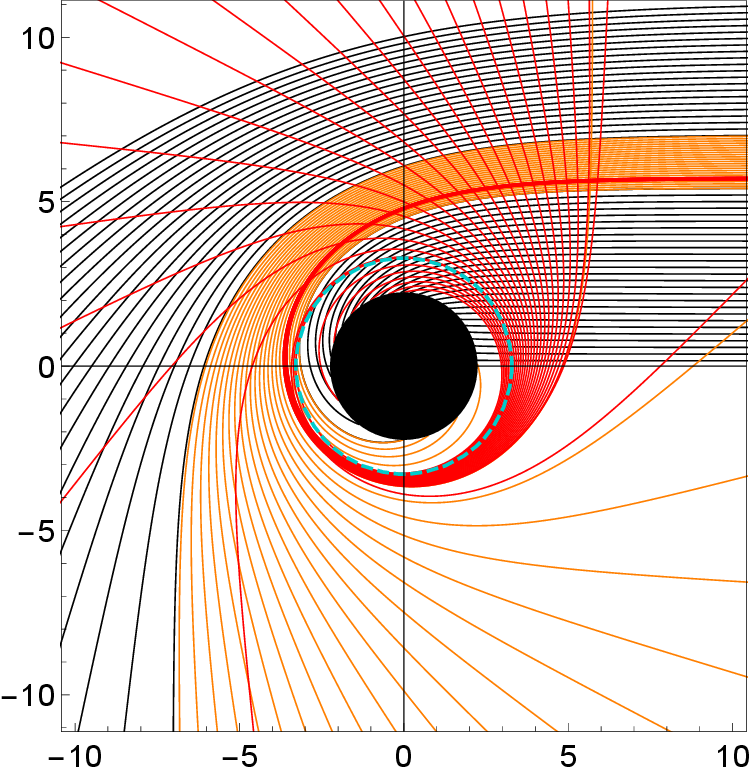}
        \caption{$\Tilde{P}=1$}
    \end{subfigure}
    \begin{subfigure}[t]{.3\linewidth}
        \includegraphics[width=5cm]{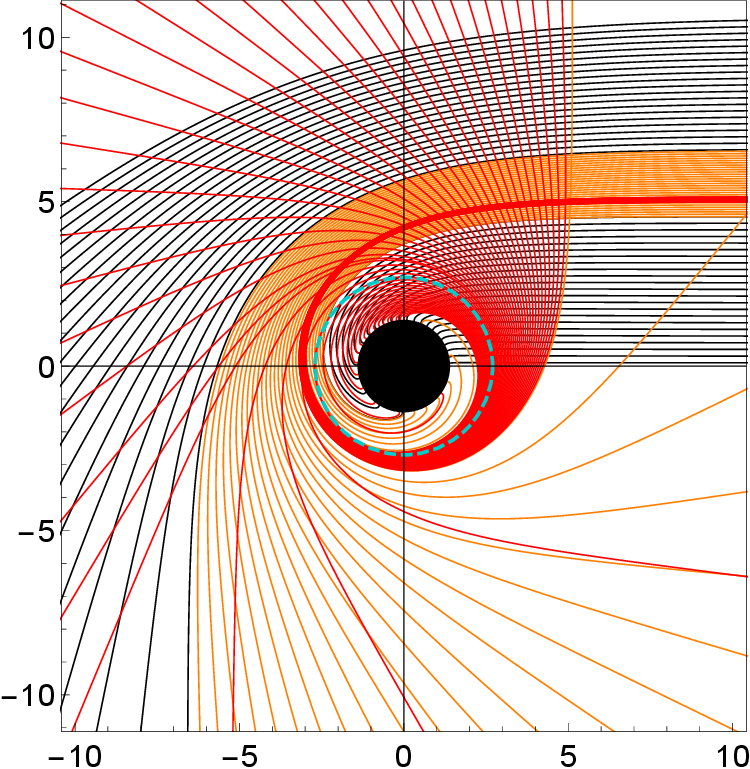}
        \caption{$\tilde{P}=\tilde{P}_{ext}(1.2522)$}
    \end{subfigure}
    \caption{The number of photon orbits $n$ as a function of impact parameter $b$ for fixed $\lambda=0.5$ and $\Tilde{D}=-0.8$ for three distinct values of $\Tilde{P}=0.5,1$ and $1.2522$ (top row, left to right). The trajectories of photon with direct emission (black), lensing ring (orange) and photon ring (red) for the same set of parameters (bottom row). The cyan dashed circle denotes the $r_{ph}$. \justifying   }
 \label{fig:raytrace3}
\end{figure}

Since the solution \eqref{bh} depends on three parameters $\Tilde{D},\Tilde{P}$ and $\lambda$, therefore we shall investigated $n$ vs $b$ and photon trajectories plots in Fig.~\ref{fig:raytrace1}--\ref{fig:raytrace3}. In these figures, we vary $\lambda,\Tilde{D}$ and $\Tilde{P}$ respectively. We denote direct, lensed and photon ring by black, orange and red curves in these plots. The horizon radius $\Tilde{r}_+$ is illustrated as a black circle at the center of the plot and the photon radius $r_{ph}$ is marked by cyan dashed line.

In Fig.~\ref{fig:raytrace1}, we display the number of photon orbits $n$ versus impact parameter $b$ for fixed $\Tilde{D}=-1$ and $\Tilde{P}=0.5$. 
As $\lambda$ increases, the peak to the right meaning that the critical impact parameter $b_{ph}$ and radius of unstable circular orbit decease. 
However, when $\lambda$ reaches $\sqrt{3}$, we observe that $b_{ph}$ slightly increases. It is noteworthy that our choice of setting $\lambda=\sqrt{3}$ is motivated by the fact that the action \eqref{action} describes the original five dimensional Kaluza-Klien theory with dimensional reduction \cite{Galtsov:1995mb}. 
Numerical values of $b$ corresponding to each range of $n$ can be found in Table~\ref{Tab:II}. 
Moreover, trajectories of photon around spherical objects in the EMD theory are shown in the Euclidean polar coordinates. 
It is observed that as $\lambda$ increases the size of horizon radius is larger. 
This is consistent with the left plot of Fig.~\ref{fig:extremalLines} and data listed in Table~\ref{Tab:I}. 
In addition, the distance between the horizon and photon sphere becomes less separated as $\lambda$ gets bigger. We also find that photon sphere's width decreases with $\lambda$ results in narrower brightness area. 
This can be seen in Table~\ref{Tab:II}.

We explore the effect of the dilaton flux $\Tilde{D}$ on $n$ versus $b$ and photon trajectories plots in Fig.~\ref{fig:raytrace2}. Increasing dilaton flux $\Tilde{D}$ results in peak shifting to the right of the $n$ versus $b$ plot. However, $b_{ph}$ marginally decreases as one approaches $\Tilde{D}=0.5$. The outer horizon radius and the photon sphere radius decrease with $\Tilde{D}$. Remark that, at $\Tilde{D}=0.5$, the horizon radius decreases significantly. Since $\Tilde{r}_+$ drops more rapidly than $r_{ph}$, therefore, the distance between $\Tilde{r}_+$ and $r_{ph}$ becomes larger as $\Tilde{D}$ increases. This can be seen evidently in the bottom panel of this figure. However, the photon sphere's width barely changes with the integrated dilaton flux.

In Fig.~\ref{fig:raytrace3}, variation of $\Tilde{P}$ on the number of photon orbits around gravitating object in the EMD theory and characteristic photon trajectories are explored. As magnetic charge increases, it is observed that $\Tilde{r}_+$ and $r_{ph}$ both decrease. As one approaches extremal value $\Tilde{P}_{ext}=1.2522$, the outer horizon radius drops swiftly as shown in Fig.~\ref{fig:extremalLines} (right panel). In $n$ versus $b$ plot, the peak shifts toward the left as $\Tilde{P}$ is larger indicating that the critical impact parameter is smaller with magnetic charge. Lastly, according to Table~\ref{Tab:II}, we observe that the photon sphere's width is broader with increasing in $\Tilde{P}$. Interestingly, the width is the largest at the extremal limit. This implies that the extremal case has the largest brightness region. This is confirmed by the relation between Lyapunov exponent and the photon sphere's width discussed in Appendix \ref{app:lypunov}. From Fig.~\ref{fig:newlypunov}, we observe that the smallest Lyapunov exponent occurs at the extremal value $\Tilde{P}_{ext}$. This indicates that, at the extremal limit, the photon sphere's width has the largest size.

\newpage

\newpage

\newpage

\begin{table}[htbp]
\caption{The region of impact parameter that represented the profile of direct emission, lensing ring, photon ring and critical impact parameter each variation case for Fig.~\ref{fig:raytrace1}--\ref{fig:raytrace3}.\justifying}

\begin{tabular}{|c|c|c|c|}
  \cline{1-4}
  \multirowcell{2}{Set of parameters} & \multicolumn{3}{c|}{$\Tilde{D}=-1~ \Tilde{P}=0.5$} \\
  \cline{2-4}
  \multicolumn{1}{|c|}{} & \eqmakebox[H]{$\lambda=0.5$} & \eqmakebox[H]{$\lambda=1$} & \eqmakebox[H]{$\lambda=\sqrt{3}$}  \\ \hline
  \multirowcell{2}{Direct emission} & $b<6.8521$ & $b<4.5999$ & $b<4.6448$ \\
  \cline{2-4}
   & $b>8.4944$ & $b>5.7542$ & $b>5.7834$ \\ \hline
  \multirowcell{2}{Lensing ring} & $6.8521<b<7.1001$ & $4.5999<b<4.7396$ & $4.6448<b<4.7830$ \\
  \cline{2-4}
   & $7.1597<b<8.4944$ & $4.7811<b<5.7542$ & $4.8230<b<5.7834$ \\ \hline
  Photon ring & $7.1001<b<7.1597$ & $4.7396<b<4.7811$ & $4.7830<b<4.8230$ \\ \hline
  Photon sphere width & $0.0596$ & $0.0415$ & $0.0400$ \\ \hline 
\end{tabular}

\vspace{5mm}

\begin{tabular}{|c|c|c|c|}
  \cline{1-4}
  \multirowcell{2}{Set of parameters} & \multicolumn{3}{c|}{$\lambda=1.5~ \Tilde{P}=0.1$} \\
  \cline{2-4}
  \multicolumn{1}{|c|}{} & \eqmakebox[H]{$\Tilde{D}=-1.5$} & \eqmakebox[H]{$\Tilde{D}=-0.5$} & \eqmakebox[H]{$\Tilde{D}=0.5$} \\ \hline
  \multirowcell{2}{Direct emission} & $b<4.3791$ & $b<4.9517$ & $b < 4.9451$ \\
  \cline{2-4}
   & $b>5.5445$ & $b>6.1007$ & $b > 6.0976$ \\ \hline
  \multirowcell{2}{Lensing ring} & $4.3791<b<4.4595$ & $4.9517 < b < 5.1163$ & $4.9451 < b < 5.1106$ \\
  \cline{2-4}
   & $4.5047<b<5.5445$ & $5.1563 < b < 6.1007$ & $5.1510 < b < 6.0976$ \\ \hline
  Photon ring & $4.4595<b<4.5047$ & $5.1163 < b < 5.1563 $ & $5.1106 < b < 5.1510$ \\ \hline
  Photon sphere width & $0.0452$ & $0.0400$ & $0.0404$ \\ \hline
\end{tabular}

\vspace{5mm}

\begin{tabular}{|c|c|c|c|}
  \cline{1-4}
  \multirowcell{2}{Set of parameters} & \multicolumn{3}{c|}{$\lambda=0.5~ \Tilde{D}=-0.8$} \\
  \cline{2-4}
  \multicolumn{1}{|c|}{} & \eqmakebox[H]{$\Tilde{P}=0.5$} & \eqmakebox[H]{$\Tilde{P}=1$} & \eqmakebox[H]{$\Tilde{P}_{ext}=1.2522$}  \\ \hline
  \multirowcell{2}{Direct emission} & $b<6.0190$ & $b<5.3615$ & $b<4.5093$ \\
  \cline{2-4}
   & $b>7.4785$ & $b>6.9853$ & $b>6.5487$ \\ \hline
  \multirowcell{2}{Lensing ring} & $6.0190<b<6.2380$ & $5.3615<b<5.6243$ & $4.5093<b<4.9460$ \\
  \cline{2-4}
   & $6.2914<b<7.4785$ & $5.6995<b<6.9853$ & $5.0976<b<6.5487$ \\ \hline
  Photon ring & $6.2380<b<6.2914$ & $5.6243<b<5.6995$ & $4.9460<b<5.0976$ \\ \hline
  Photon sphere width & $0.0534$ & $0.0752$ & $0.1516$ \\ \hline
\end{tabular}
\label{Tab:II}
\end{table}

\section{Shadow}\label{sec:shadow}

\begin{figure}[htbp]
    \centering
    \begin{subfigure}[t]{.3\linewidth}
        \includegraphics[scale=.47]{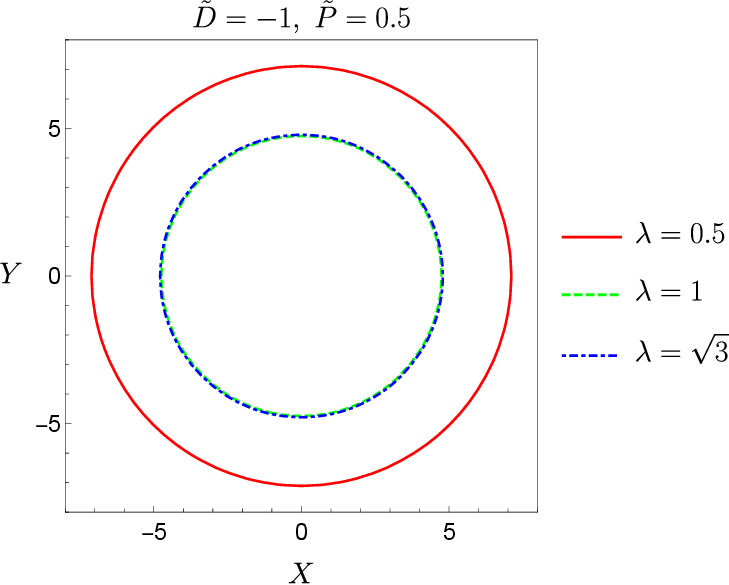}
    \end{subfigure}
    \hspace{5mm}
    \begin{subfigure}[t]{.3\linewidth}
        \includegraphics[scale=.47]{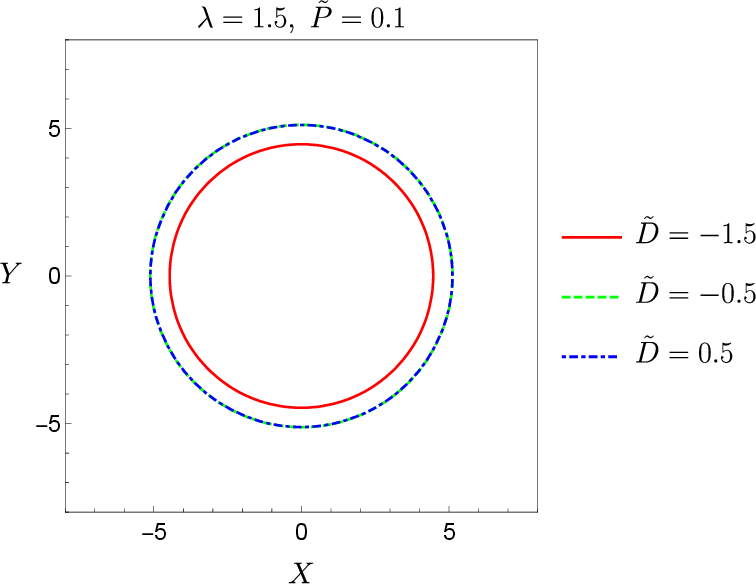}
    \end{subfigure}
    \hspace{5mm}
    \begin{subfigure}[t]{.3\linewidth}
        \includegraphics[scale=.47]{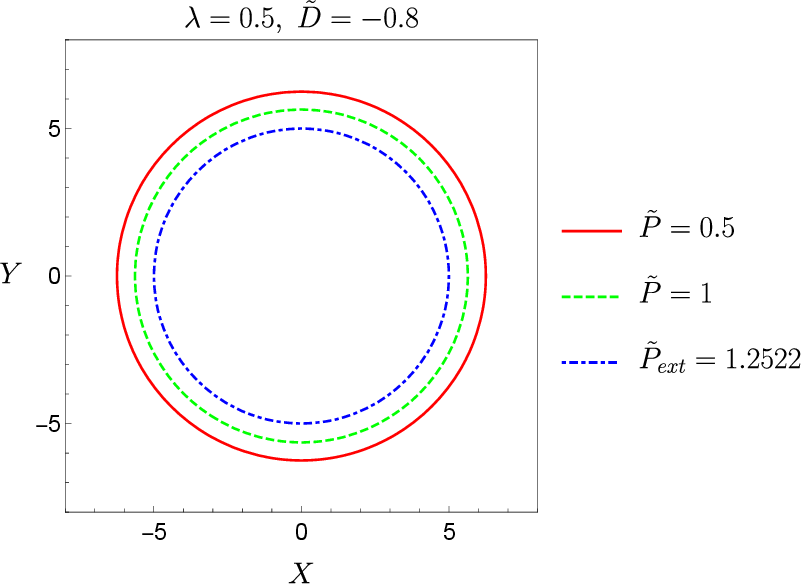}
    \end{subfigure}
    
    \begin{subfigure}[t]{.3\linewidth}
        \includegraphics[scale=.47]{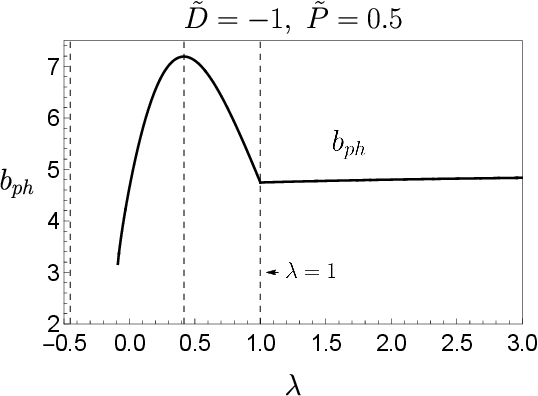}
    \end{subfigure}
    \hspace{5mm}
    \begin{subfigure}[t]{.3\linewidth}
        \includegraphics[scale=.47]{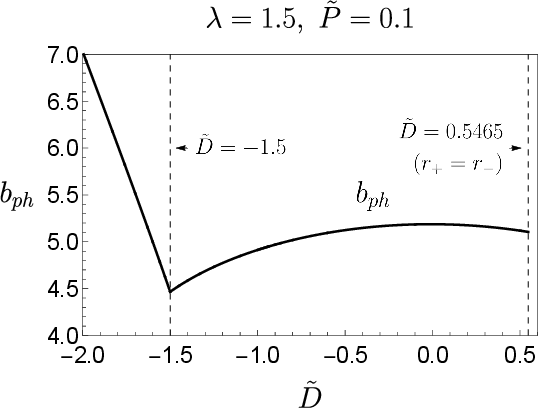}
    \end{subfigure}
    \hspace{5mm}
    \begin{subfigure}[t]{.3\linewidth}
        \includegraphics[scale=.47]{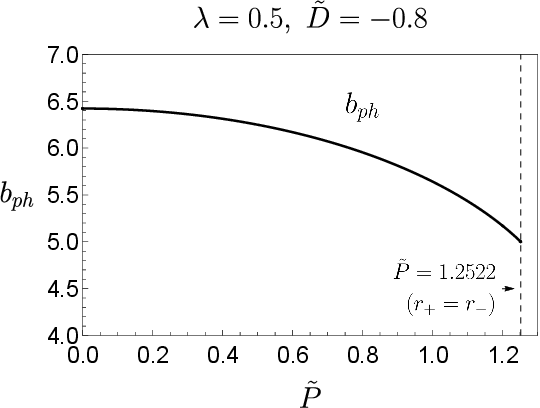}
    \end{subfigure}
    \caption{The shadow radius $R_s$ of $q$-deformed object for different values of parameters $\tilde{D}$, $\tilde{P}$ and $\lambda$. For comparison between the shadow radius and the critical impact parameter, in the bottom row, we show the critical impact parameter $b_{ph}$ as a function with respect to parameters that correspond to the top row. In the left panel is the variation of parameter $\lambda$ with fixed $\tilde{D}=-1$ and $\tilde{P}=0.5$. The middle panel is the variation of parameter $\tilde{D}$ with fixed $\lambda=1.5$ and $\tilde{P}=0.1$. The right panel shows the variation of parameter $\tilde{P}$ with fixed $\lambda=0.5$ and $\tilde{D}=-0.8$. \justifying}
     \label{fig:Rs}
\end{figure}

In this section, we explore how $\lambda,\Tilde{D}$ and $\Tilde{P}$ influence the shadow of a spherically symmetric $q$-deformed object in the EMD gravity. 
More precisely, we visualize how these parameters affect the shape and size of the shadow. 
We place a light source at some distance from the central object such that an incoming light ray can be treated as a parallel ray. A static observer has only one non-vanishing component of four velocities i.e., $u^{a}=\{1,0,0,0\}$. A shadow image is usually depicted on a celestial plane $(X,Y)$ \cite{Vazquez:2003zm,PhysRevD.90.024073}. The celestial coordinates are given by   
\begin{align}
    X &= -\lim_{r_\ast \to \infty} r_\ast^2 \sin \theta_O \frac{d\phi}{dr}, \label{Xcoor} \\
    Y &= \lim_{r_\ast \to \infty} r_\ast^2 \frac{d\theta}{dr}, \label{Ycoor}
\end{align}
where the distance between the observer and the central object is denoted by $r_\ast$. The inclination angle between the line of sight of the observer and the normal to the celestial plane is represented by the angular coordinate $\theta_O$. On an equatorial plane $\theta_O=\pi/2$, the radius of shadow is \cite{Promsiri:2023rez}
\begin{align}
    R_s &= \sqrt{X^2+Y^2}.\label{eq39}
\end{align}
To evaluate celestial coordinates $X$ and $Y$, we firstly need to determine $d\phi/dr$ and $d\theta/dr$. From \eqref{eq9},\eqref{eq32} and \eqref{eq33}, these can be explicitly obtained
\begin{align}
    \frac{d\phi}{dr} &= L\csc^2\theta \sqrt{\frac{fg}{\mathcal{R}}}, \\
    \frac{d\theta}{dr} &= \sqrt{\Theta}\sqrt{\frac{fg}{\mathcal{R}}}.
\end{align}
Then, we substitute these two relations back into \eqref{Xcoor} and \eqref{Ycoor}. At $\theta_O=\pi/2$, we get $X=-L/E$ and $Y=\sqrt{\mathcal{Q}}/E$. Therefore, from \eqref{eq32}, we obtain 
\begin{align}
    \left(\frac{dr}{d\tilde{\sigma}}\right)^2&=\frac{1}{f(r)g(r)b^2}\left[1-b^2\frac{f(r)}{h(r)}\left(1+\frac{\mathcal{Q}}{L^2}\right)\right], \label{eq43}
\end{align}
for null geodesics. The circular photon orbit equation at $r=r_{ph}$ implies 
\begin{align}
\frac{1}{b_{ph}^2} &= \frac{f(r_{ph})}{h(r_{ph})}\left(1+\frac{\mathcal{Q}}{L^2}\right), \nonumber \\
    \frac{f(r_{ph})}{h(r_{ph})} &= \frac{E^2}{L^2+\mathcal{Q}}, \nonumber \\
    &= \frac{1}{R_s^2}.
\end{align}
Therefore, the appearance shadow radius can be explicitly illustrated by the critical impact parameter $b_{ph}$. 
This is demonstrated in the top row of Fig.~\ref{fig:Rs} where the shadow radius $R_s$ is plotted in celestial coordinates. 


To clarify the non-monotonic tendency of the critical impact parameter as a function of parameters $\lambda$, $\tilde{D}$ and $\tilde{P}$ in the bottom row of Fig.~\ref{fig:Rs}, which corresponds to the shadow region in the top row. In the left panel, we display the critical impact parameter as a function of $\lambda$ with fixed $\tilde{D}=-1$ and $\tilde{P}=0.5$. 
At small $\lambda$, i.e., $\lambda<1$, $b_{ph}$ initially increases with $\lambda$. After it reaches its maximum at $\lambda \approx 0.5$, the $b_{ph}$ then decreases. In contrast when $\lambda>1$, the $b_{ph}$ moderately increases with $\lambda$.
In the middle panel, the critical impact parameter behavior is shown for varying $\tilde{D}$ with fixed $\lambda=1.5$ and $\tilde{P}=0.1$. The impact parameter linearly decreases on $-\infty<\tilde{D}\leq-1.5$, then increasing when $-1.5\leq\tilde{D}\leq -0.01$. As $-0.01<\tilde{D}\leq 0.5465$, the critical impact parameter slowly decreases until it reaches its extremal limit at $\tilde{D}=0.5465$. 
With these results, the critical impact parameter on this variation has the smallest value of $b_{ph}$ at $\tilde{D}=-1.5$. 
For the right panel, the trend of the critical impact parameter is illustrated for varying $\tilde{P}$ with fixed $\lambda=0.5$ and $\tilde{D}=-0.8$. As can be seen, the impact parameter monotonically decreases as $\tilde{P}$ increases.

\section{Optical images of q-deformed objects surrounded by thin accretion disk}\label{sec:images}

In this section, we investigate the observational features of a static, spherically symmetric object in EMD gravity. We consider a scenario where the object is illuminated by an optically thin accretion disk situated on the equatorial plane. 
A distant observer is positioned at an inclination angle of $90^\circ$, such that the line of sight is perpendicular to the equatorial plane (see, e.g., \cite{Promsiri:2023rez}).
Utilizing a backward ray-tracing technique, we track null geodesics from the observer's screen back toward the emitting disk. 
For a given impact parameter $b$, a geodesic may intersect the disk multiple times; the first, second, and subsequent intersections generate the direct image, the lensing ring, and the photon rings.
Each $m$th intersection contributes to the total observed intensity, $I_{obs}$, which is determined by the sum of the local emissivities from all disk crossings.
Since the emission intensity $I_{em}$ is a function of the radial coordinate $r$, it is essential to determine the radial position $r_m(b)$ for each $m$th intersection. 
The mapping of $r_m(b)$ is defined as the $m$th order transfer function, which encodes how the background geometry organizes image features on the observer's screen and provides the bridge between the spacetime, the disk model, and the resulting brightness profile.

\subsection{Transfer function}

To determine the image perceived by a distant observer, we analyze the propagation of specific intensity along the light ray. According to the Liouville's theorem, the phase space density of photons is conserved during their propagation along null geodesics . 
In curved spacetime, this principle dictates that $I_\nu/\nu^3$ remains constant along the photon's trajectory \cite{Luminet:1979nyg,schneider1992gravitational}.
Moreover, the local frequencies of emitted photons $\nu_{em}$ and those received photons $\nu_{obs}$ by a distant observer differ due to the gravitational redshift, characterized by the factor $z=\nu_{obs}/\nu_{em}=\sqrt{f(r)}$.
Consequently, we have $I_{\nu_{obs}}^{obs}(r) = z^3 I_{\nu_{em}}^{em}(r)$, where $I_{\nu_{obs}}^{obs}$ and $I_{\nu_{em}}^{em}$ represent the observed and emitted specific intensities, respectively.
The total intensity $I_{obs}$ is obtained by integrating this expression over the entire frequency spectrum as follows
\begin{align}
    I_{obs}&=\int I^{obs}_{\nu_{obs}} d\nu_{obs}
    =z^4 \int I^{em}_{\nu_{em}}d\nu_{em}
    =z^4 I_{em}. \nonumber
\end{align}

\begin{figure}[htbp]
    \begin{subfigure}[t]{0.3\textwidth}
        \includegraphics[width = \linewidth]{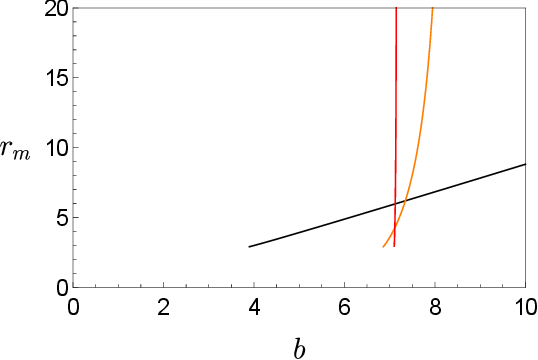}
        \caption{$\lambda=0.5$}
    \end{subfigure}
    \begin{subfigure}[t]{0.3\textwidth}
        \includegraphics[width = \linewidth]{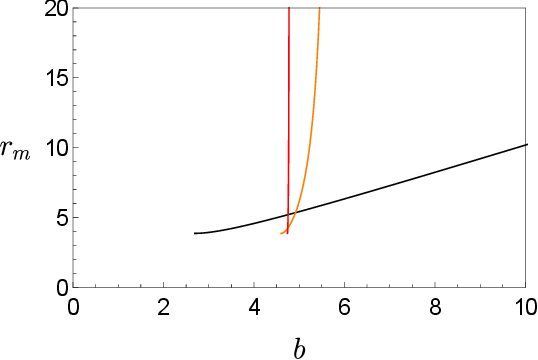}
        \caption{$\lambda=1$}
    \end{subfigure}
    \begin{subfigure}[t]{0.3\textwidth}
        \includegraphics[width = \linewidth]{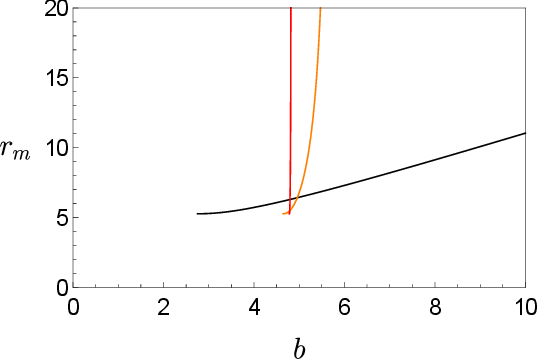}
        \caption{$\lambda=\sqrt{3}$}
    \end{subfigure}
    \begin{subfigure}[t]{.3\textwidth}
        \includegraphics[width = \linewidth]{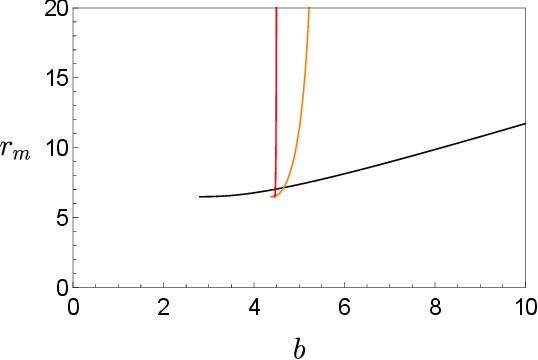}
        \caption{$\tilde{D}=-1.5$}
    \end{subfigure}
    \begin{subfigure}[t]{.3\textwidth}
        \includegraphics[width = \linewidth]{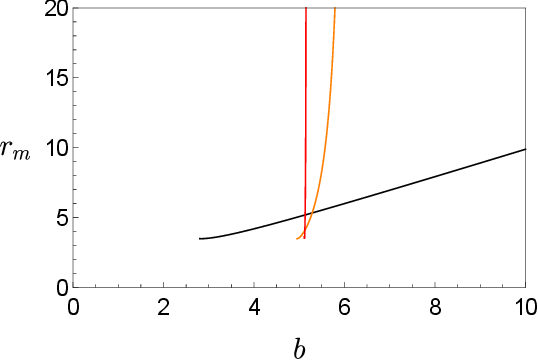}
        \caption{$\tilde{D}=-0.5$}
    \end{subfigure}
    \begin{subfigure}[t]{.3\textwidth}
        \includegraphics[width = \linewidth]{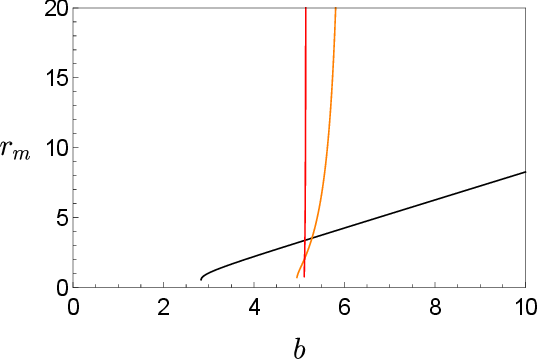}
        \caption{$\tilde{D}=0.5$}
    \end{subfigure}
    \begin{subfigure}[t]{.3\textwidth}
        \includegraphics[width = \linewidth]{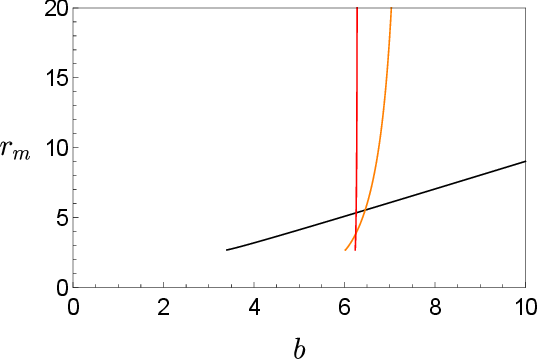}
        \caption{$\tilde{P}=0.5$}
    \end{subfigure}
    \begin{subfigure}[t]{.3\textwidth}
        \includegraphics[width = \linewidth]{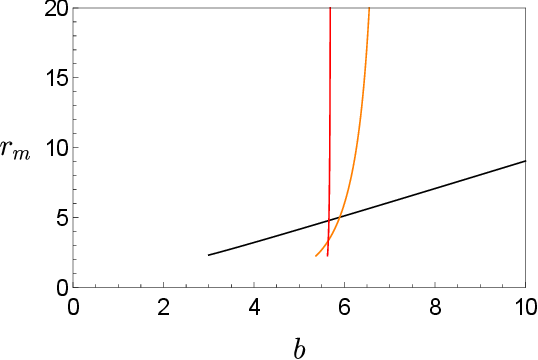}
        \caption{$\tilde{P}=1$}
    \end{subfigure}
    \begin{subfigure}[t]{.3\textwidth}
        \includegraphics[width = \linewidth]{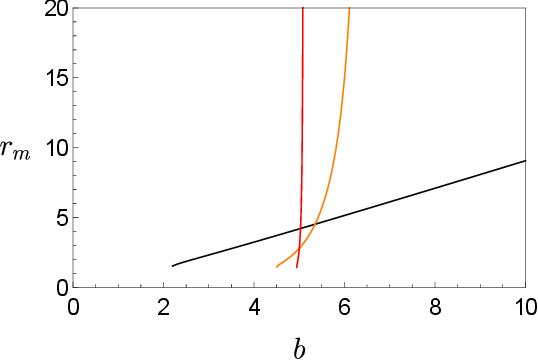}
        \caption{$\tilde{P}_{ext}=1.2522$}
    \end{subfigure}
    \caption{The transfer functions $r_m$ against impact parameter $b$. Top row: $\tilde{D}=-1$ and $\tilde{P}=0.5$ for $\lambda=0.5,1,\sqrt{3}$ Middle row: $\lambda=1.5$ and $\tilde{P}=0.1$ for $\tilde{D}=-1.5,-0.5,0.5$ Bottom row:  $\lambda=0.5$ and $\tilde{D}=-0.8$ for $\tilde{P}=0.5,1,1.2522$. The black, orange and red curves represent direct emission, lensing ring and photon ring profile, respectively \justifying}
    \label{fig:transfer}
\end{figure}

As a photon travels along its trajectory around the compact object, it may intersect the accretion disk multiple times due to strong gravitational lensing. 
Assuming the disk is optically thin, the photon accumulates intensity at each intersection without being absorbed. 
Consequently, the total observed intensity $I_{obs}(b)$ at a given impact parameter $b$ is obtained by summing the contributions from all such crossings as follows
\begin{align}
    I_{obs}(b)=\sum_{m}f(r)^2 I_{em}(r)\vert_{r=r_m(b)},
\end{align}
where $r_m(b)$ represents the $m$th order transfer function. The transfer functions are obtained by numerically solving the geodesic equation in \eqref{eqU}.
For a given $b$, the radial position of each intersection is determined by the variable $1/u(\phi)$ evaluated at $\phi=\pi/2,3\pi/2$ and $5\pi/2$.
The behavior of $r_m(b)$ for $q-$ deformed solutions is illustrated in Fig.~\ref{fig:transfer}, where the curves for $m=1$ (black), $m=2$ (orange), and $m=3$ (red) represent the direct emission, the lensing ring and the photon ring, respectively.
Notably, the slope of the transfer function, $dr_m/db$, serves as a measure of the demagnification of the image. 
For direct emission ($m=1$), the slope is approximately unity ($dr_1/db \approx 1$), implying that the impact parameter on the observer's screen corresponds almost directly to the radial coordinate on the disk. 
In contrast, the lensing and photon rings ($m \geq 2$) exhibit extremely steep slopes ($dr_m/db \gg 1$). 
Such steepness indicates that a wide range of radial coordinates on the accretion disk is compressed into a very narrow range of impact parameters, resulting in a highly demagnified and thin ring-like appearance. 
Conversely, a slope with $dr_m/db < 1$ denotes a magnification effect, where the emission profile appears broader in the observed image.

As shown in the top row in Fig.~\ref{fig:transfer}, increasing $\lambda$ results in a steeper slope for each $r_m(b)$ curve.
While direct emission maintains a slope near unity, the lensing and photon rings exhibit extremely steep slopes. 
With this high demagnification of the higher-order transfer function, the resulting images display extremely thin subring structures on the observer's screen.
By varying the scalar charge $\tilde{D}$, a distinctive behavior appears when $\tilde{D}$ is negative. 
In this regime, the first transfer function develops a slope smaller than unity, enhancing a direct image more evidently.
As $\tilde{D}$ increases, the slope for each $r_m(b)$ is steeper, as shown in the middle row in Fig.~\ref{fig:transfer}.
Notably, as $\tilde{D}$ increases, the starting values of $r_m$ decrease, allowing the observer to see images from deeper within the disk.
Increasing $\tilde{P}$ leads to a decrease in the slopes of $r_m(b)$, which reduces the demagnification of higher-order images. 
Consequently, sub-ring structures become more visible in the resulting images.

\subsection{Thin accretion disk profiles}

In this work, we model the emission profile of optically thin accretion disk by using the Gralla-Lupsasca-Marrone (GLM) model \cite{PhysRevD.102.124004}. The GLM model is constructed from a general relativistic magnetohydrodynamics (GRMHD) simulation for an accretion disk \cite{Johnson:2019ljv, Chael:2021rjo, Vincent:2022fwj}. 

The emitted intensity profile of the GLM model is 
\begin{eqnarray}
    I_\text{em}(r;\gamma ,\alpha , \beta)=\frac{\text{exp}\left[ -\frac{1}{2}\left(\gamma +\text{arcsinh}\left( \frac{r-\alpha}{\beta}\right) \right)^2\right]}{\sqrt{(r-\alpha)^2+\beta^2}}.
\end{eqnarray}
The intensity profile $I_{em}$ is described by three parameters. First, $\alpha$ dictates how the peak of $I_m$ has shifted from a central position. Second, $\beta$ relates to the width of the profile and typically is given in a unit of mass $M$. Third, $\gamma$ controls the rate at which the intensity inclines to its peak from infinity. 

\begin{figure}[htbp]
    \begin{subfigure}[t]{.3\textwidth}
        \includegraphics[width = \linewidth]{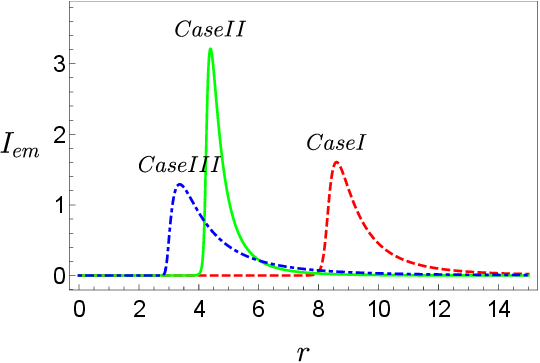}
        \caption{$\lambda=0.5$}
    \end{subfigure}
    \begin{subfigure}[t]{.3\linewidth}
        \includegraphics[width = \linewidth]{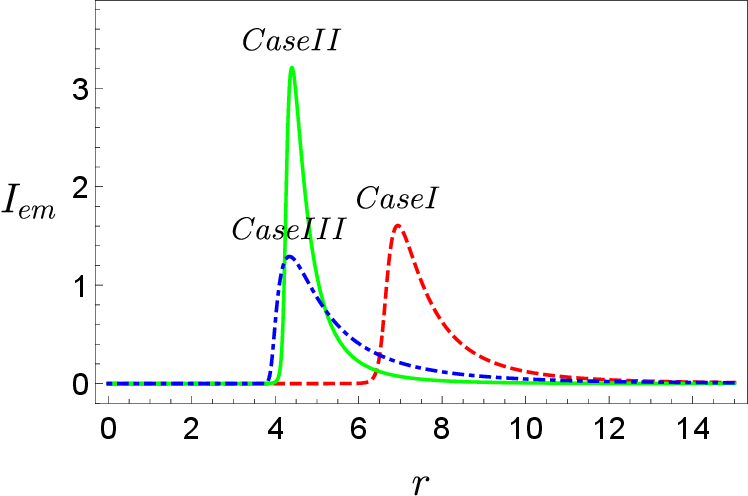}
        \caption{$\lambda=1$}
    \end{subfigure}
    \begin{subfigure}[t]{.3\linewidth}
        \includegraphics[width = \linewidth]{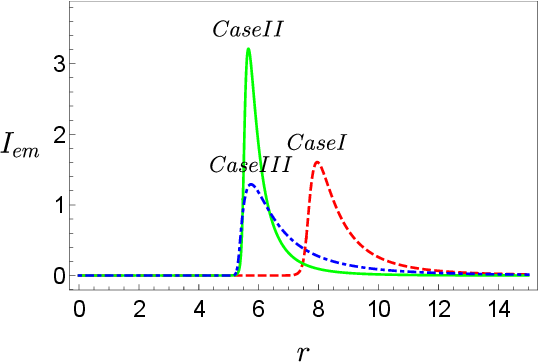}
        \caption{$\lambda=\sqrt{3}$}
    \end{subfigure}
    \begin{subfigure}[t]{.3\linewidth}
        \includegraphics[width = \linewidth]{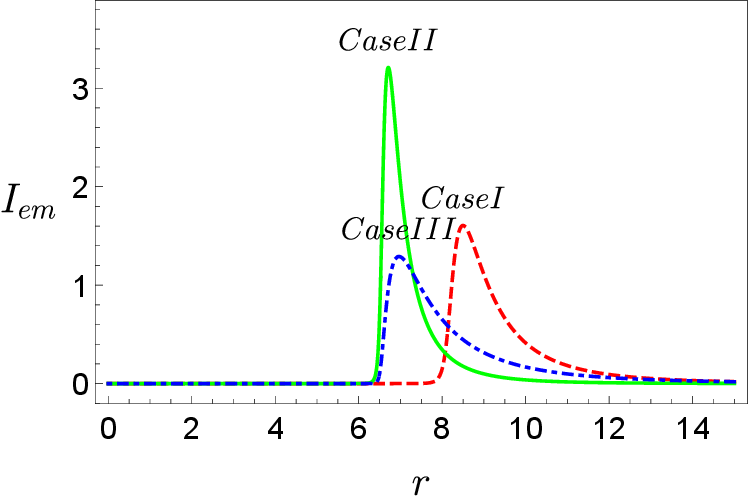}
        \caption{$\tilde{D}=-1.5$}
    \end{subfigure}
    \begin{subfigure}[t]{.3\linewidth}
        \includegraphics[width = \linewidth]{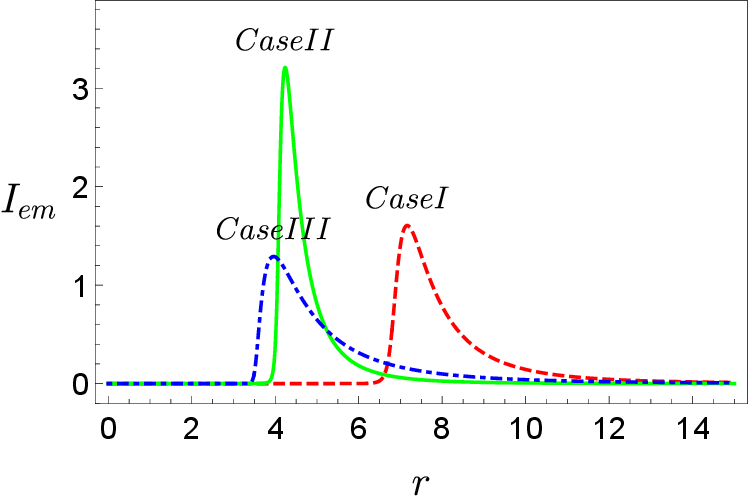}
        \caption{$\tilde{D}=-0.5$}
    \end{subfigure}
    \begin{subfigure}[t]{.3\linewidth}
        \includegraphics[width = \linewidth]{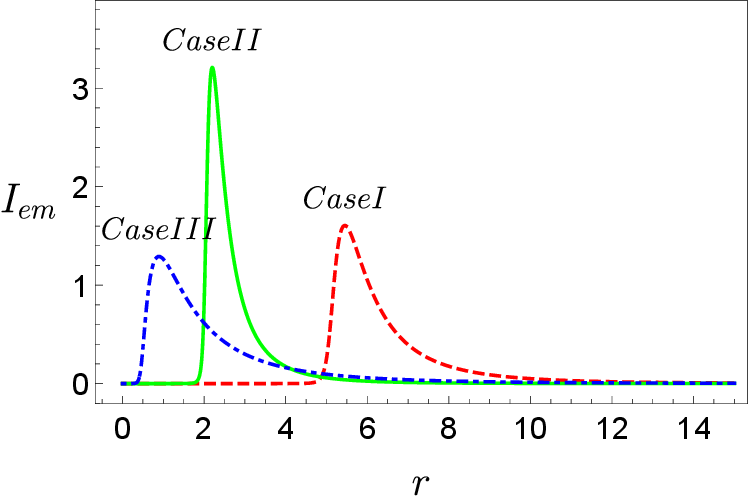}
        \caption{$\tilde{D}=0.5$}
    \end{subfigure}
    \begin{subfigure}[t]{.3\linewidth}
        \includegraphics[width = \linewidth]{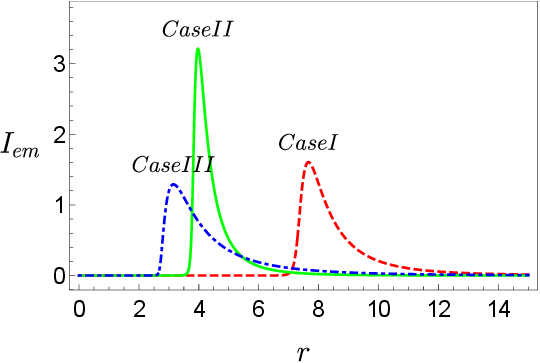}
        \caption{$\tilde{P}=0.5$}
    \end{subfigure}
    \begin{subfigure}[t]{.3\linewidth}
        \includegraphics[width = \linewidth]{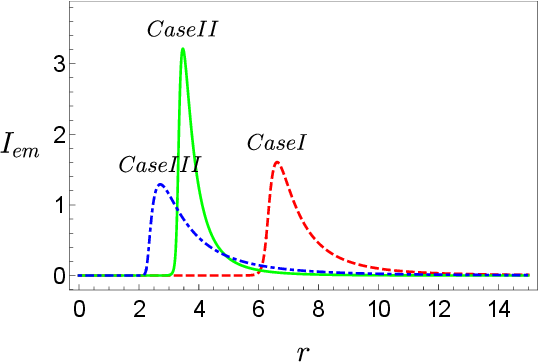}
        \caption{$\tilde{P}=1$}
    \end{subfigure}
    \begin{subfigure}[t]{.3\linewidth}
        \includegraphics[width = \linewidth]{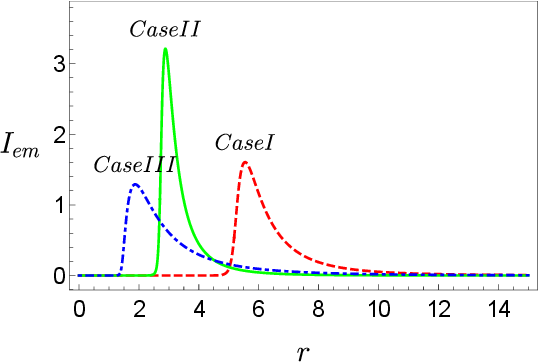}
        \caption{$\tilde{P}_{ext}=1.2522$}
    \end{subfigure}
    \caption{The emission profile ($I_{em}$) of thin accretion disk in \textbf{CaseI}, \textbf{CaseII} and \textbf{CaseIII}. The variation of $\lambda,\tilde{D}$ and $\tilde{P}$ are shown in the top ($\tilde{D}=-1$ and $\tilde{P}=0.5$), middle ($\lambda=1.5$ and $\tilde{P}=0.1$.) and bottom ($\lambda=0.5$ and $\tilde{D}=-0.8$) rows, respectively. The red-dashed line, green-solid line and blue-dot-dashed line illustrate \textbf{CaseI}, \textbf{CaseII} and \textbf{CaseIII}, respectively.\justifying}
    \label{fig:Iem}
\end{figure}

In our work, we particularly focus in three specific cases of the GLM profile \cite{Promsiri:2023rez}. The profiles are distinguished by the location of their maxima. They are 
\begin{itemize}
    \item \textbf{CaseI}: $\gamma =-2, \alpha =r_\text{ISCO}$ and $\beta =M/4$. This case has a peak around the $r_{ISCO}$ region. We provide the details calculation of $r_{ISCO}$ in Appendix~\ref{app:isco}.
    \item \textbf{CaseII}: $\gamma =-2, \alpha =r_{ph}$ and $\beta =M/8$. This case has a peak around the photon ring radius $r_{ph}$, which implies that the accretion is located around the photon ring.
    \item \textbf{CaseIII}: $\gamma =-3, \alpha =r_+$ and $\beta =M/8$. This case has a peak around the horizon radius $r_+$. 
\end{itemize}
The emitted intensity profile of these three cases of the GLM model, i.e., \textbf{Case~I} (red dashed line), \textbf{Case~II} (green solid line) and \textbf{Case~III} (blue dot-dashed line), are shown in Fig.~\ref{fig:Iem}. Remark that, We observe that the \textbf{Case~II} intensity profile has the highest maximum value, which indicates that this case has the brightest images.

\newpage

\begin{figure}
    \centering
    \begin{subfigure}[t]{.3\linewidth}
            \includegraphics[width=5.4cm]{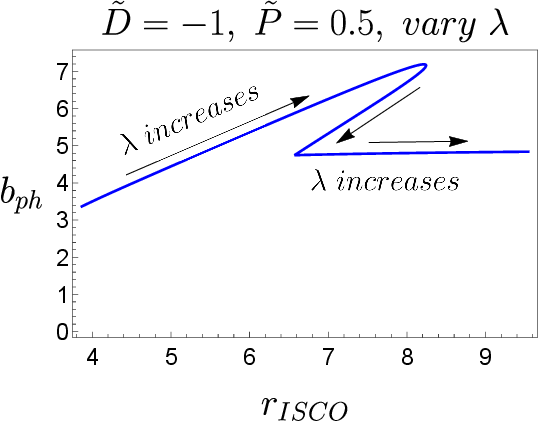}
    \end{subfigure}
    \hspace{6mm}
    \begin{subfigure}[t]{.3\linewidth}
        \includegraphics[width=5.4cm]{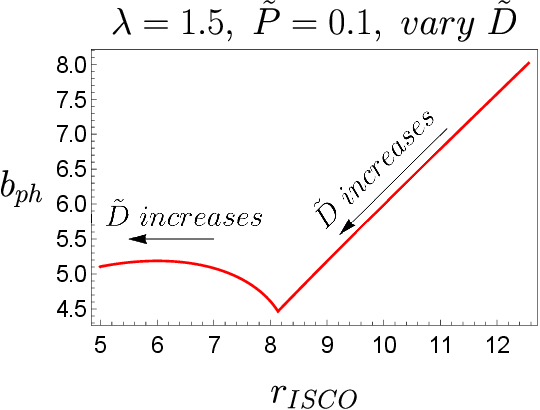}
    \end{subfigure}
    \hspace{6 mm}
    \begin{subfigure}[t]{.3\linewidth}
        \includegraphics[width=5.4cm]{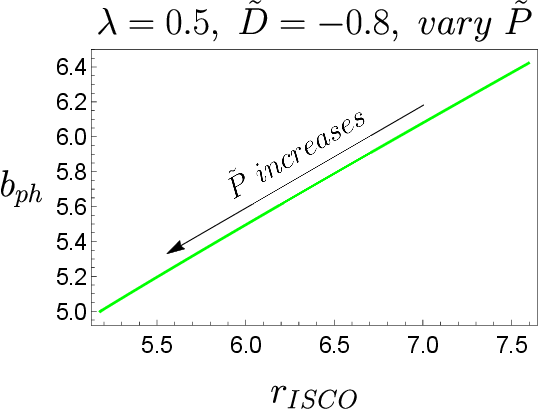}
    \end{subfigure}

    \begin{subfigure}[t]{.3\linewidth}
        \includegraphics[width=5.4cm]{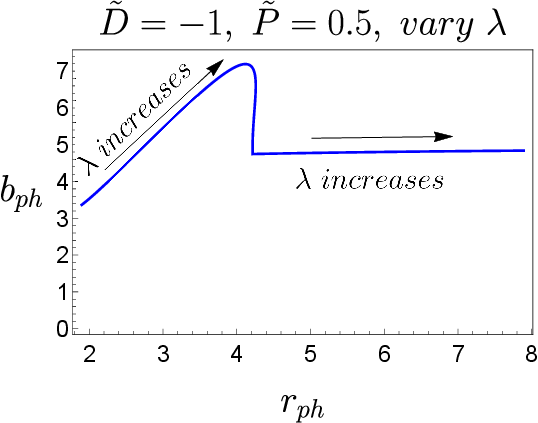}
    \end{subfigure}
    \hspace{6mm}
    \begin{subfigure}[t]{.3\linewidth}
        \includegraphics[width=5.4cm]{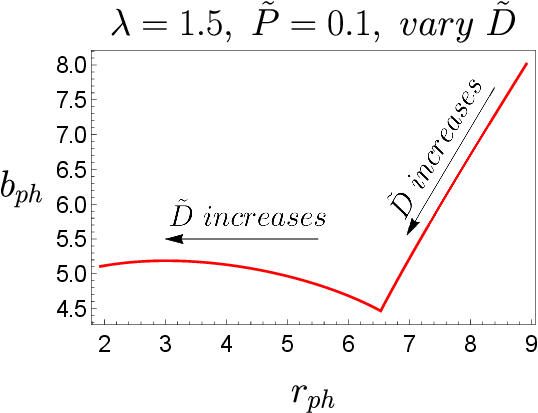}
    \end{subfigure}
    \hspace{6 mm}
    \begin{subfigure}[t]{.3\linewidth}
        \includegraphics[width=5.4cm]{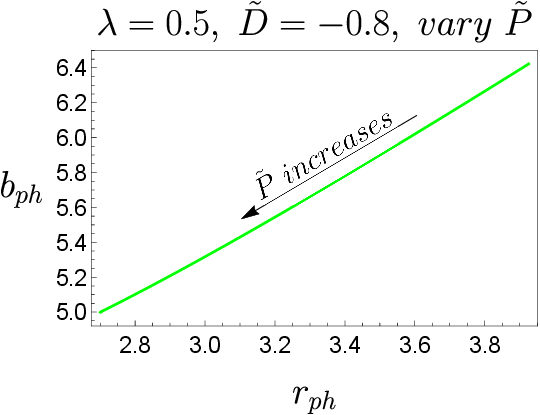}
    \end{subfigure}

    \begin{subfigure}[t]{.3\linewidth}
        \includegraphics[width=5.4cm]{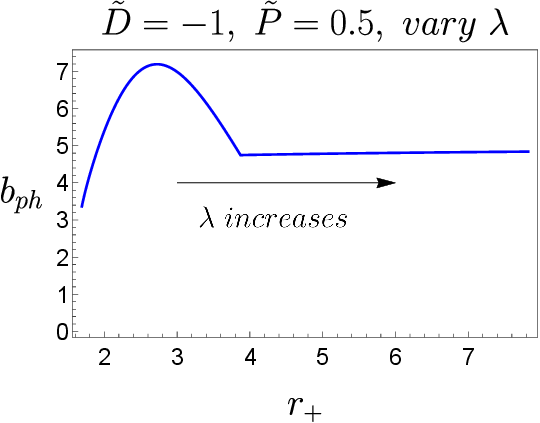}
    \end{subfigure}
    \hspace{6mm}
    \begin{subfigure}[t]{.3\linewidth}
        \includegraphics[width=5.4cm]{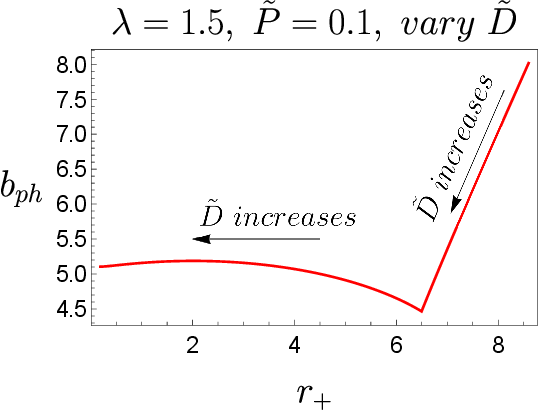}
    \end{subfigure}
    \hspace{6 mm}
    \begin{subfigure}[t]{.3\linewidth}
        \includegraphics[width=5.4cm]{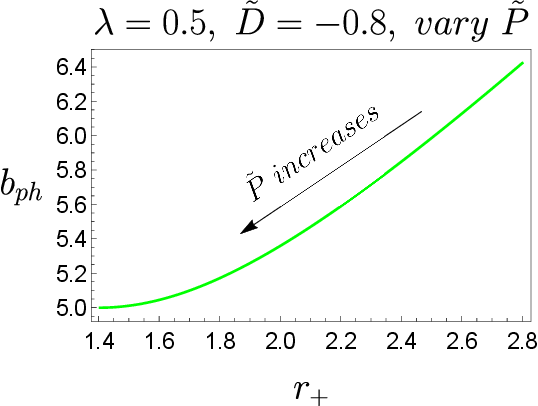}
    \end{subfigure}
    \caption{The critical impact parameter $b_{ph}$ as a function of $r_{ISCO}$ (top row), photon ring radius (middle row) $r_{ph}$ and horizon $r_+$ (bottom row) for different variations of parameters $\lambda,~\tilde{D},$ and $\tilde{P}$. The black arrows indicate an increasing direction of each respective parameters\justifying} 
    \label{fig:ISCO_bc}
\end{figure}

\newpage

Figure~\ref{fig:ISCO_bc} illustrates how the characteristic radii of the background $r_{ISCO}$ (top row), the photon ring radius $r_{ph}$ (middle) and the horizon radius $r_+$ (bottom) map onto the critical impact parameter $b_{ph}$ in each GLM case by varying parameters $\lambda, \tilde{D}$ and $\tilde{P}$. 
Note that the black arrows indicate the direction of increasing $\lambda, \tilde{D}$ and $\tilde{P}$.
Since $b_{ph}$ determines the shadow boundary seen by a distance observer, these trends encode how spacetime geometry controls the image of $q$-deformed objects surrounded by a thin accretion disk in the GLM model.
Two qualitative features are noteworthy, namely $r_{ISCO}$, $r_{ph}$ and $r_+$ vary non-monotonically with $b_{ph}$ when $\lambda$ and $\tilde{D}$ increase as observed in the first and second column of Fig.~\ref{fig:ISCO_bc}.
By comparing with the bottom row in Fig.~\ref{fig:Rs}, we find the following dependence on the dilaton coupling. For $0 \lesssim \lambda \lesssim 0.5$, we observe that the $b_{ph},r_{ISCO}, r_{ph}$ and $r_+$ are all increasing with $\lambda$. For $0.5\lesssim \lambda \lesssim1$, $b_{ph}$, $r_{ISCO}$ and $r_{ph}$ decrease, while $r_+$ remains increasing. For $\lambda > 1$, $r_{ISCO},\, r_{ph}$ and $r_+$ increase and the shadow size remains nearly the same as it changes very slightly.
With fixed $\lambda=1.5$ and $\tilde{P}=0.1$, the dependence on $\tilde{D}$ is as follows.
The scalar charge increases from negative to positive values from right to left as shown in the middle panel in Fig.~\ref{fig:ISCO_bc}.
For the $-2\lesssim \tilde{D}\lesssim -1.5$ region, $b_{ph}$ decreases with $r_{ISCO}$, $r_{ph}$ and $r_+$ and $b_{ph}$ has a minimum value at $\tilde{D}\approx -1.5$ that corresponds to the curve in the bottom row in the middle column of Fig.~\ref{fig:Rs}.
For $-1.5 \lesssim \tilde{D}\lesssim 0.5465$, $b_{ph}$ starts to increases and then slowly decreases over some range of $\tilde{D}$ before the extremal limit is reached, while $r_{ISCO}$, $r_{ph}$ and $r_+$ decrease in this range.
For increasing $\tilde{P}$, we find that $r_{ISCO}$, $r_{ph}$ and $r_+$ monotonically decrease with $b_{ph}$ as shown tn the third column of Fig.~\ref{fig:ISCO_bc}.

\newpage

\subsection{Images}

\begin{figure}[htbp]
 	     \begin{subfigure}[t]{.24\textwidth}
 		         \includegraphics[width = \textwidth]{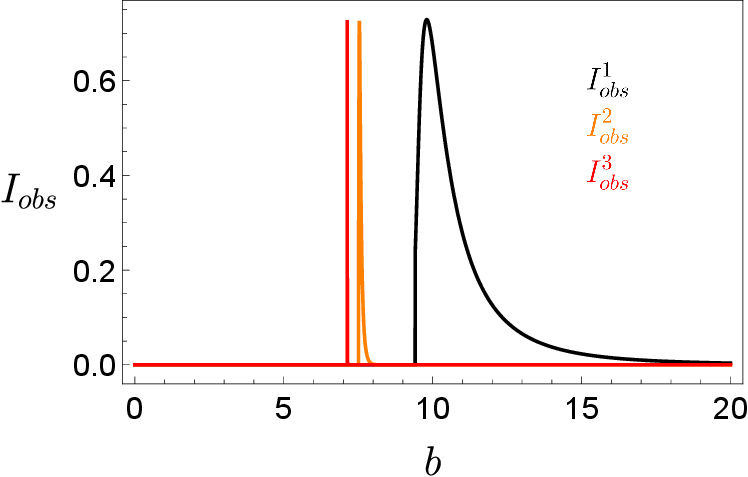}
 		     \end{subfigure}
 	     \begin{subfigure}[t]{.24\textwidth}
 		         \includegraphics[width = \textwidth]{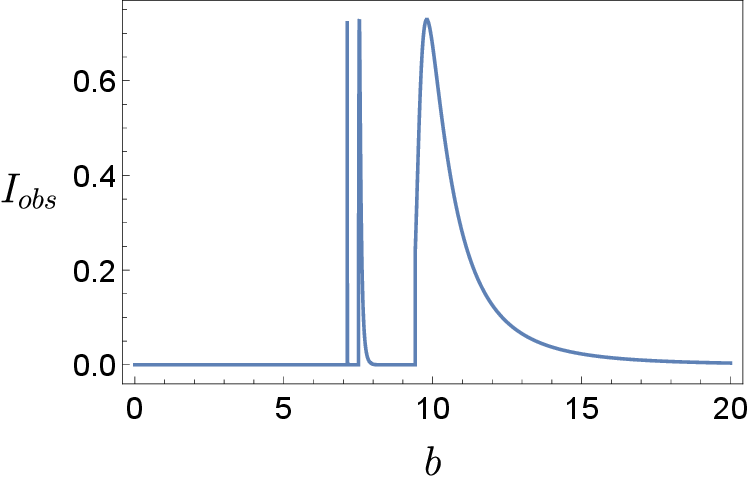}
 		     \end{subfigure}
 	     \begin{subfigure}[t]{.24\textwidth}
 		         \includegraphics[width = \textwidth]{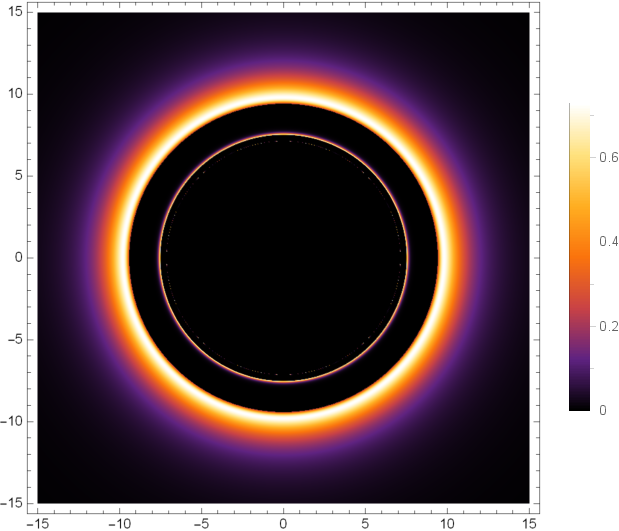}
 		     \end{subfigure}
 	     \begin{subfigure}[t]{.24\textwidth}
 		         \includegraphics[width = \textwidth]{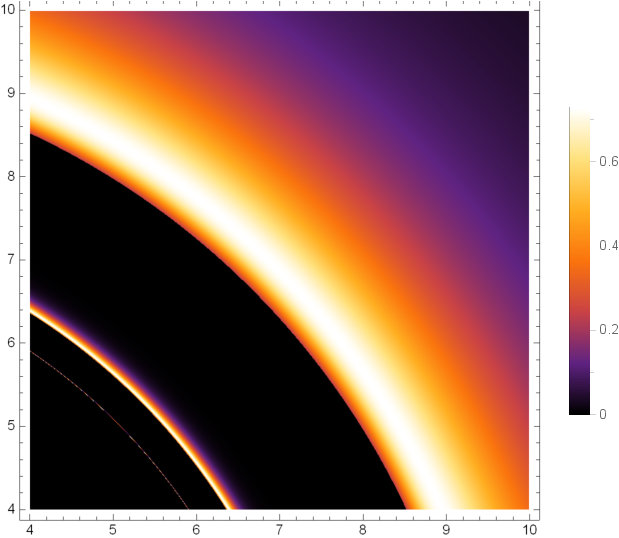}
 		     \end{subfigure}
 
      \begin{subfigure}[t]{.24\textwidth}
 	         \includegraphics[width = \textwidth]{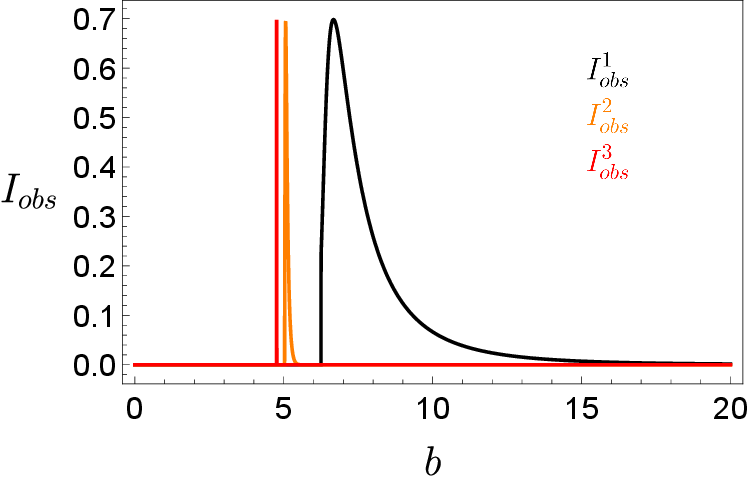}
 	     \end{subfigure}
      \begin{subfigure}[t]{.24\textwidth}
 	         \includegraphics[width = \textwidth]{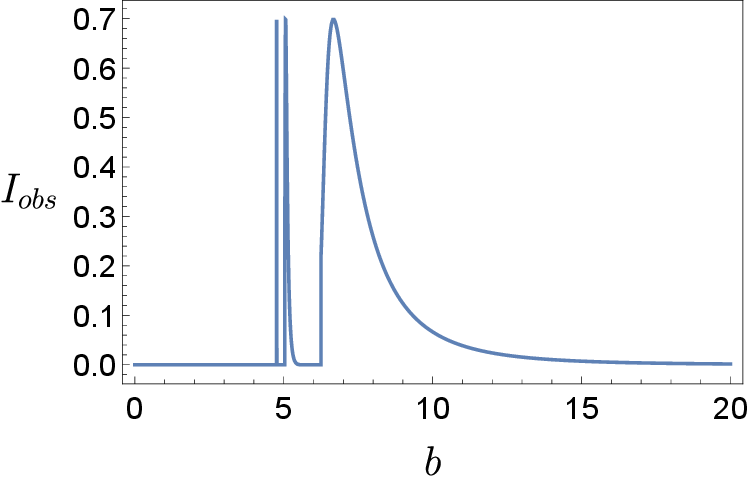}
 	     \end{subfigure}
      \begin{subfigure}[t]{.24\textwidth}
 	         \includegraphics[width = \textwidth]{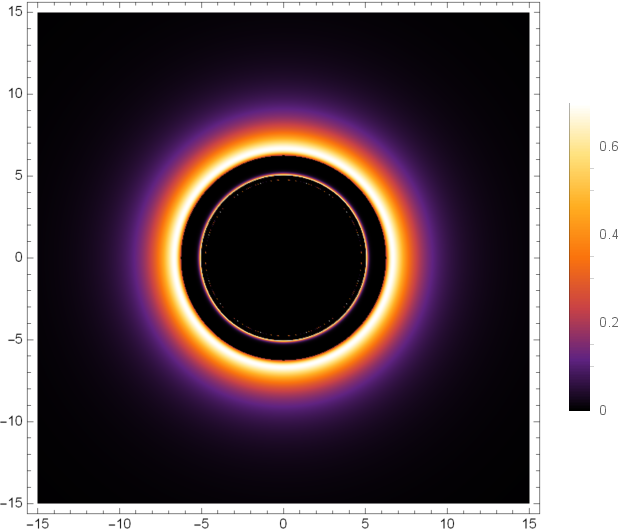}
 	     \end{subfigure}
      \begin{subfigure}[t]{.24\textwidth}
 	         \includegraphics[width = \textwidth]{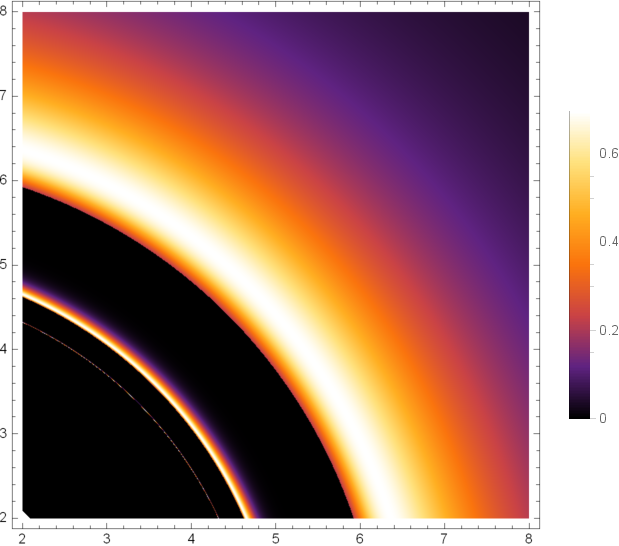}
 	     \end{subfigure}
 
      \begin{subfigure}[t]{.24\textwidth}
 	         \includegraphics[width = \textwidth]{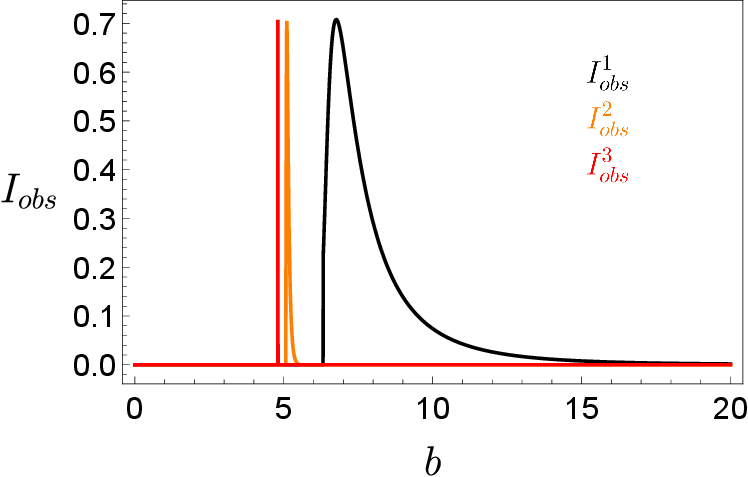}
 	     \end{subfigure}
      \begin{subfigure}[t]{.24\textwidth}
 	         \includegraphics[width = \textwidth]{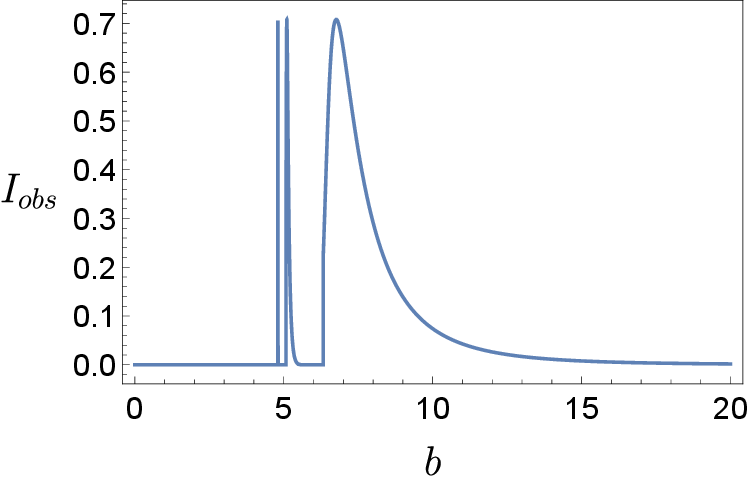}
 	     \end{subfigure}
      \begin{subfigure}[t]{.24\textwidth}
 	         \includegraphics[width = \textwidth]{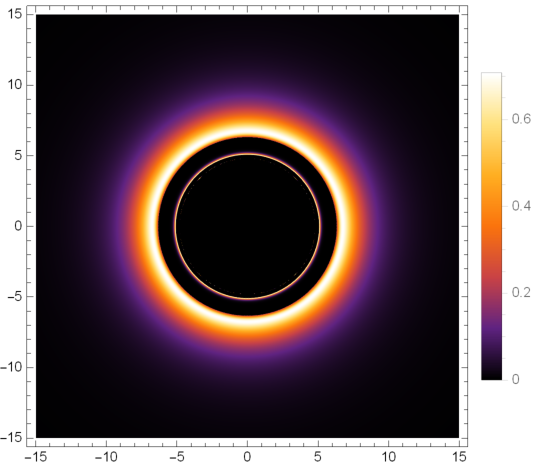}
 	     \end{subfigure}
      \begin{subfigure}[t]{.24\textwidth}
 	         \includegraphics[width = \textwidth]{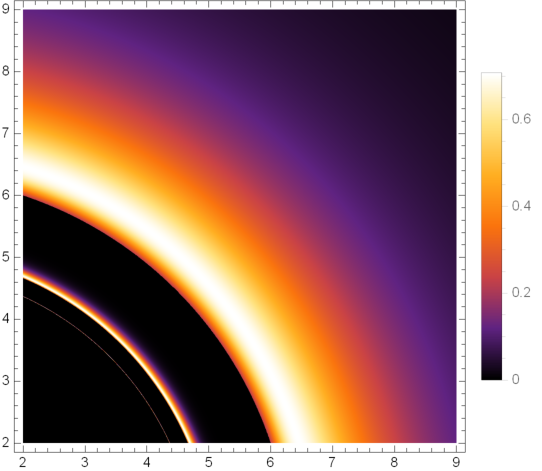}
 	     \end{subfigure}
      \caption{Optical appearance of $q$-deformed objects illuminated by thin accretion disk profile in \textbf{Case~I} of the GLM model. 
      From top to bottom row, the dilaton coupling constant is varies as $\lambda=0.5,1,\sqrt{3}$ with fixed $\tilde{D}=-1$ and $\tilde{P}=0.5$, respectively.
      The first column shows the individual observed intensity contributions from the direct (black), lensing ring (orange) and photon ring (red) regions. 
      The second column presents the total observed intensity. 
      The third column displays the resulting image of a light source near the $q$-deformed object and the close up images are shown in the last column. \justifying}
   \label{fig:shdwvaryLmodel1}
  \end{figure}

\begin{figure}[htbp]
      \begin{subfigure}[t]{.24\textwidth}
 	         \includegraphics[width = \textwidth]{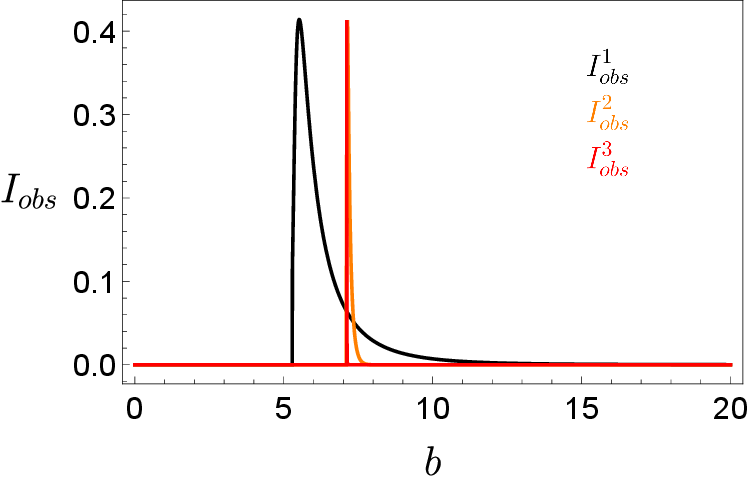}
 	     \end{subfigure}
      \begin{subfigure}[t]{.24\textwidth}
 	         \includegraphics[width = \textwidth]{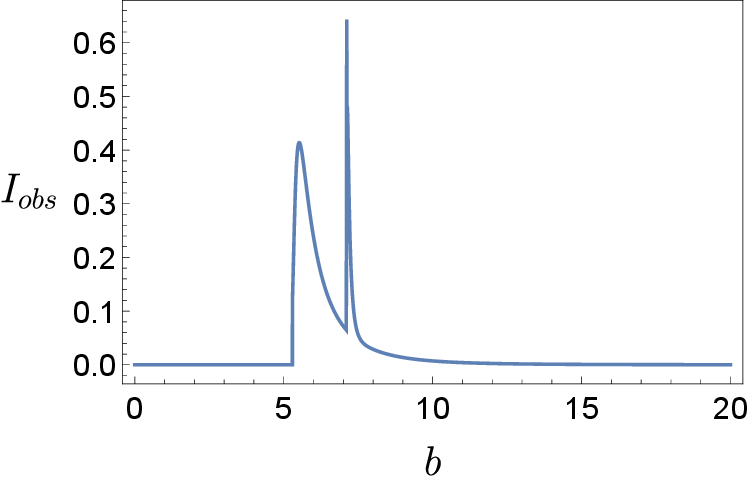}
 	     \end{subfigure}
      \begin{subfigure}[t]{.24\textwidth}
 	         \includegraphics[width = \textwidth]{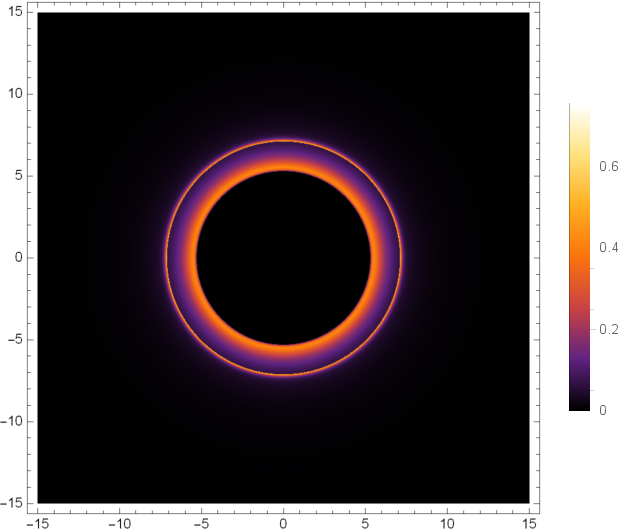}
 	     \end{subfigure}
      \begin{subfigure}[t]{.24\textwidth}
 	         \includegraphics[width = \textwidth]{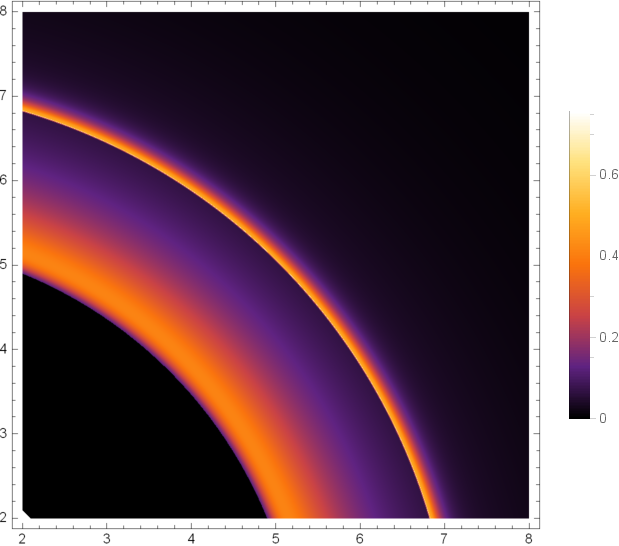}
 	     \end{subfigure}
 
      \begin{subfigure}[t]{.24\textwidth}
 	         \includegraphics[width = \textwidth]{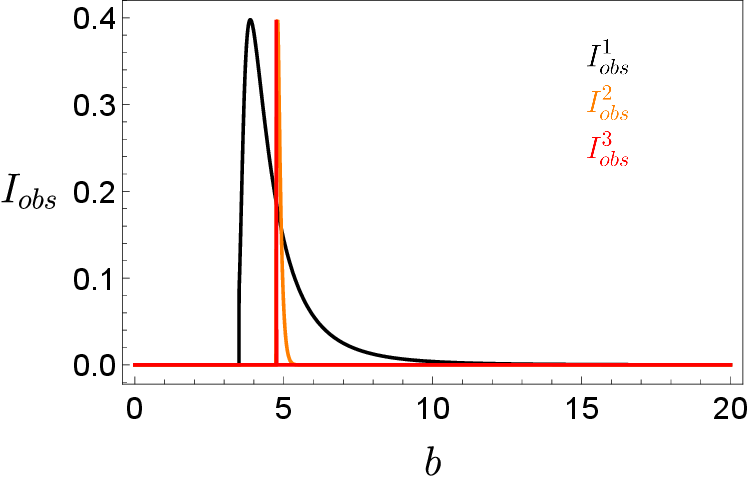}
 	     \end{subfigure}
      \begin{subfigure}[t]{.24\textwidth}
 	         \includegraphics[width = \textwidth]{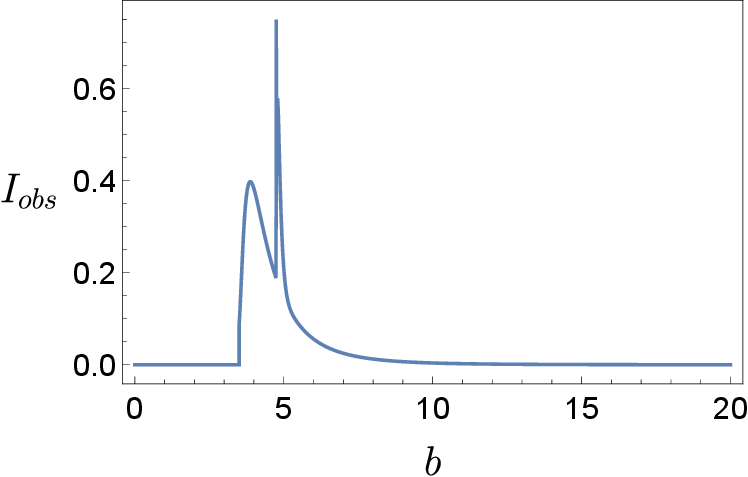}
 	     \end{subfigure}
      \begin{subfigure}[t]{.24\textwidth}
 	         \includegraphics[width = \textwidth]{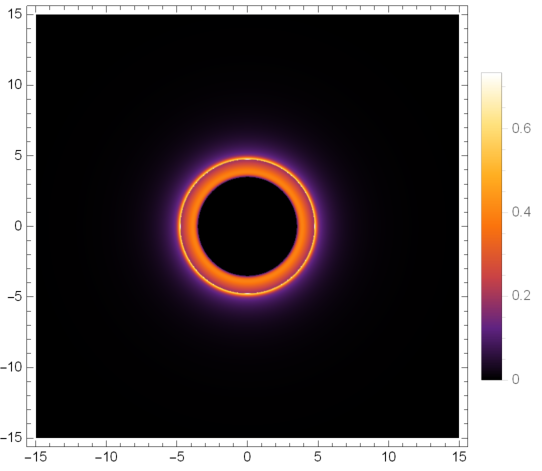}
 	     \end{subfigure}
      \begin{subfigure}[t]{.24\textwidth}
 	         \includegraphics[width = \textwidth]{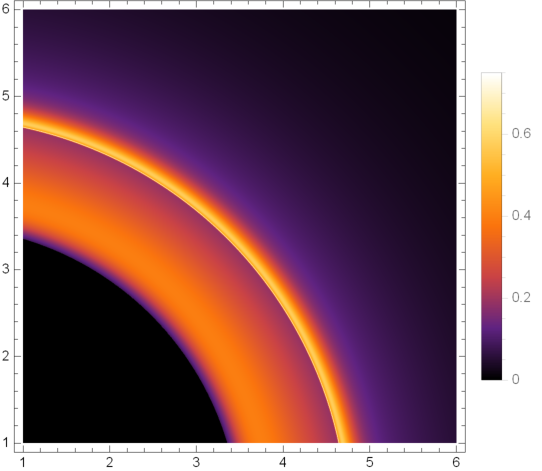}
 	     \end{subfigure}
 
      \begin{subfigure}[t]{.24\textwidth}
 	         \includegraphics[width = \textwidth]{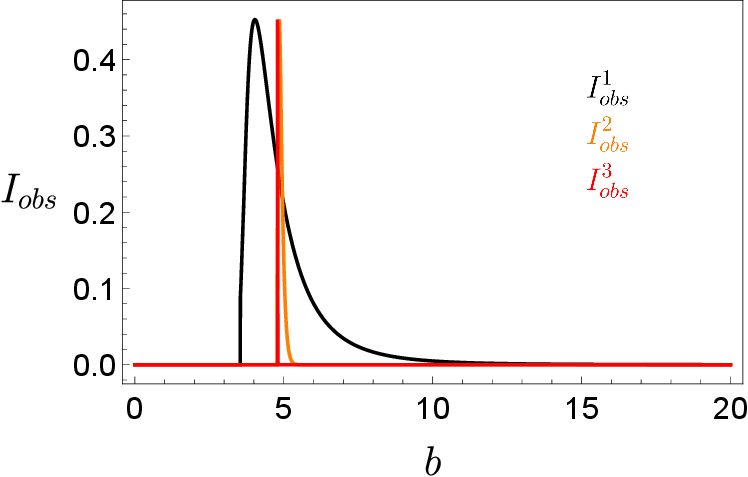}
 	     \end{subfigure}
      \begin{subfigure}[t]{.24\textwidth}
 	         \includegraphics[width = \textwidth]{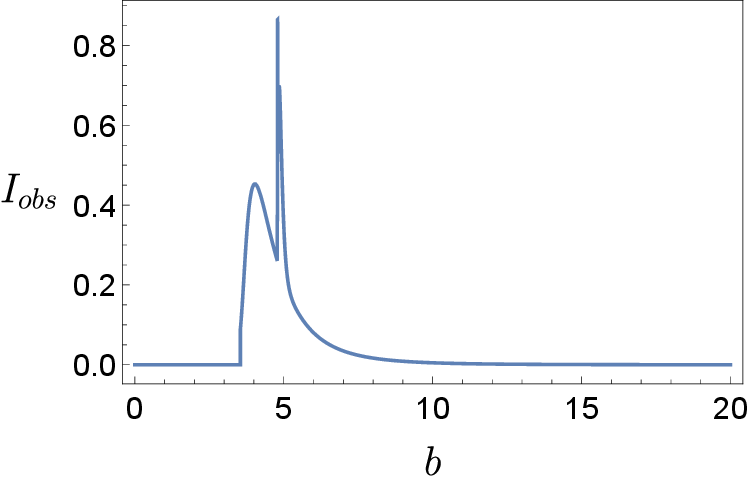}
 	     \end{subfigure}
      \begin{subfigure}[t]{.24\textwidth}
 	         \includegraphics[width = \textwidth]{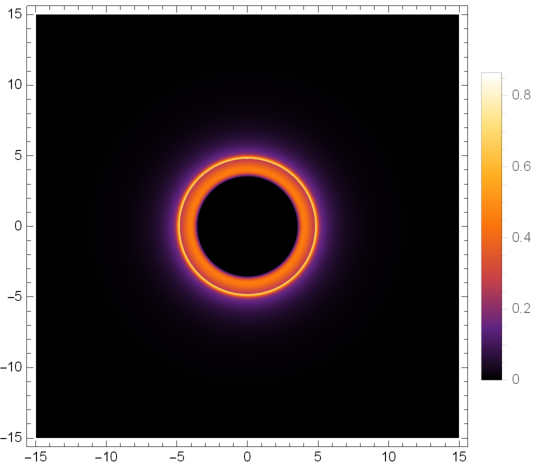}
 	     \end{subfigure}
      \begin{subfigure}[t]{.24\textwidth}
 	         \includegraphics[width = \textwidth]{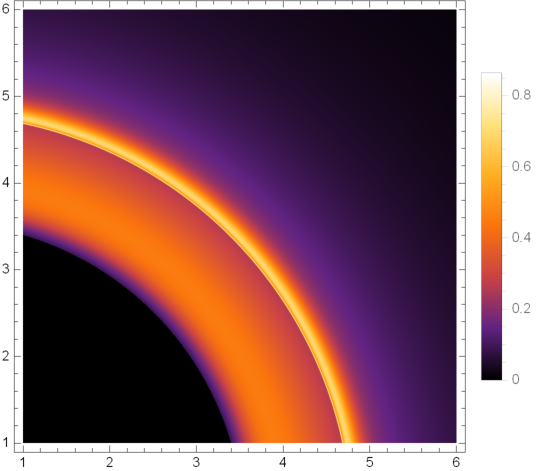}
 	     \end{subfigure}
      \caption{Optical appearance of $q$-deformed objects illuminated by thin accretion disk profile in \textbf{Case~II} of the GLM model. 
      From top to bottom row, the dilaton coupling constant is varies as $\lambda=0.5,1,\sqrt{3}$ with fixed $\tilde{D}=-1$ and $\tilde{P}=0.5$, respectively.
      The first column shows the individual observed intensity contributions from the direct (black), lensing ring (orange) and photon ring (red) regions. 
      The second column presents the total observed intensity. 
      The third column displays the resulting image of a light source near the $q$-deformed object and the close up images are shown in the last column. \justifying}
    \label{fig:shdwvaryLmodel2}
  \end{figure}

\begin{figure}[htbp]
      \begin{subfigure}[t]{.24\textwidth}
 	         \includegraphics[width = \textwidth]{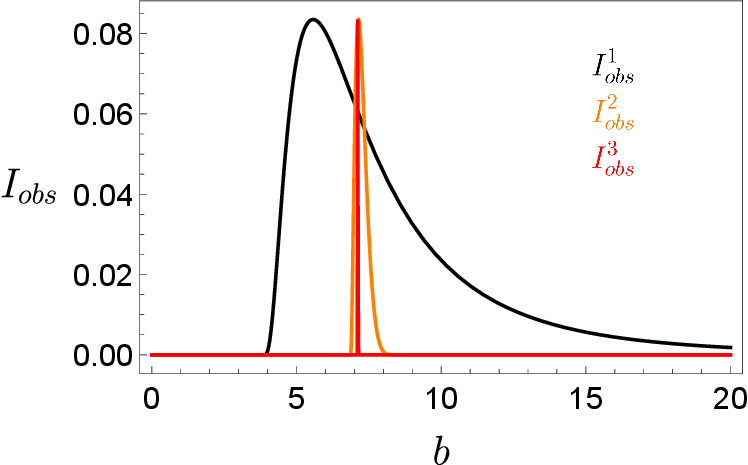}
 	     \end{subfigure}
      \begin{subfigure}[t]{.24\textwidth}
 	         \includegraphics[width = \textwidth]{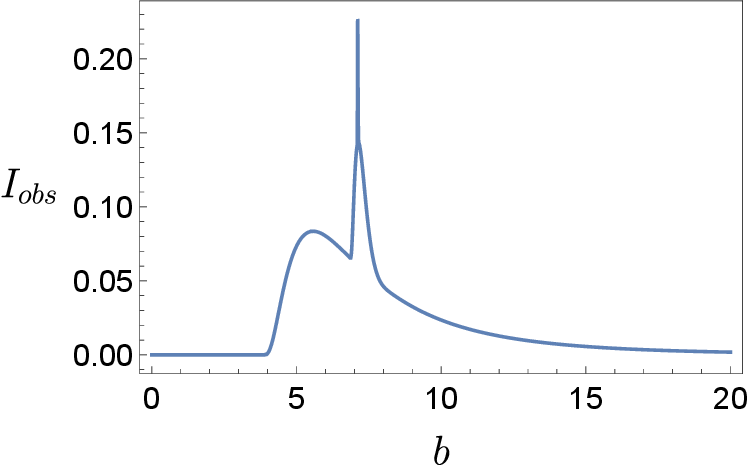}
 	     \end{subfigure}
      \begin{subfigure}[t]{.24\textwidth}
 	         \includegraphics[width = \textwidth]{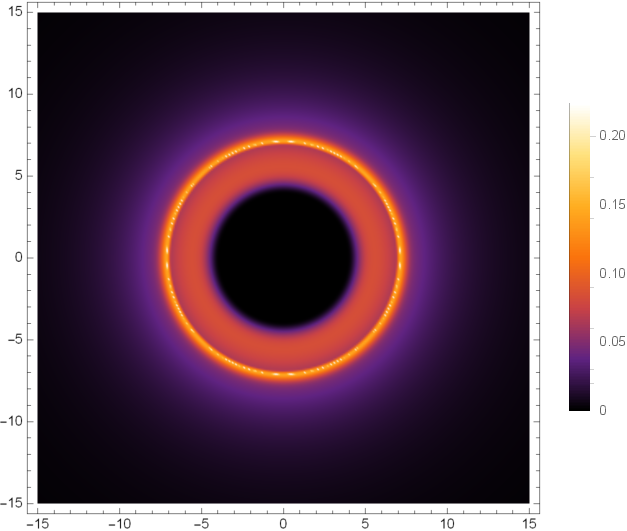}
 	     \end{subfigure}
      \begin{subfigure}[t]{.24\textwidth}
 	         \includegraphics[width = \textwidth]{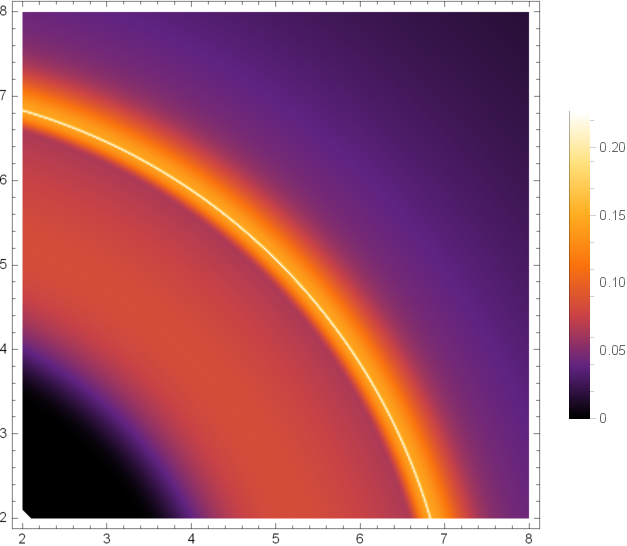}
 	     \end{subfigure}
 
      \begin{subfigure}[t]{.24\textwidth}
 	         \includegraphics[width = \textwidth]{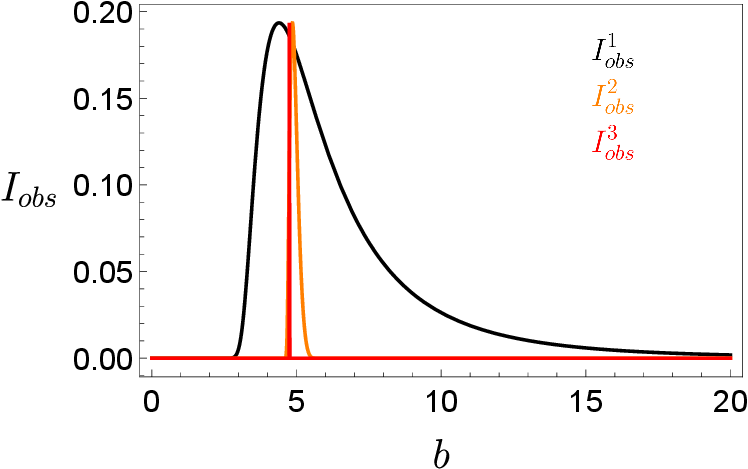}
 	     \end{subfigure}
      \begin{subfigure}[t]{.24\textwidth}
 	         \includegraphics[width = \textwidth]{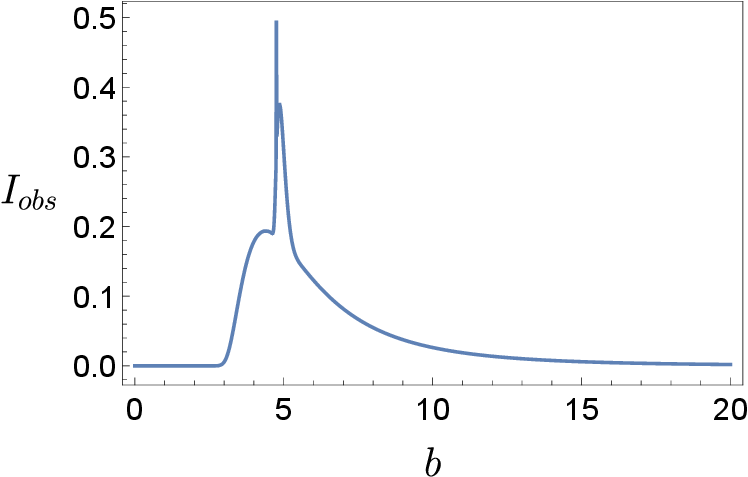}
 	     \end{subfigure}
      \begin{subfigure}[t]{.24\textwidth}
 	         \includegraphics[width = \textwidth]{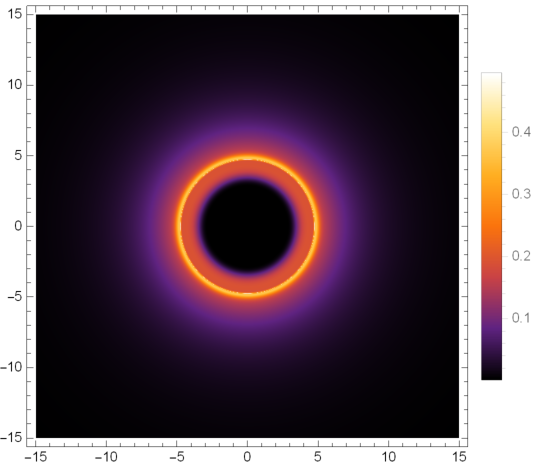}
 	     \end{subfigure}
      \begin{subfigure}[t]{.24\textwidth}
 	         \includegraphics[width = \textwidth]{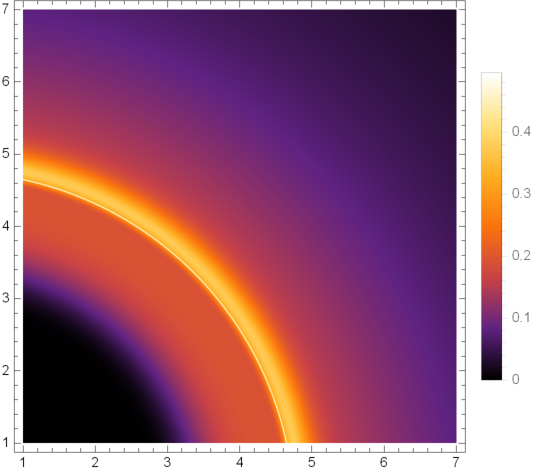}
 	     \end{subfigure}
 
      \begin{subfigure}[t]{.24\textwidth}
 	         \includegraphics[width = \textwidth]{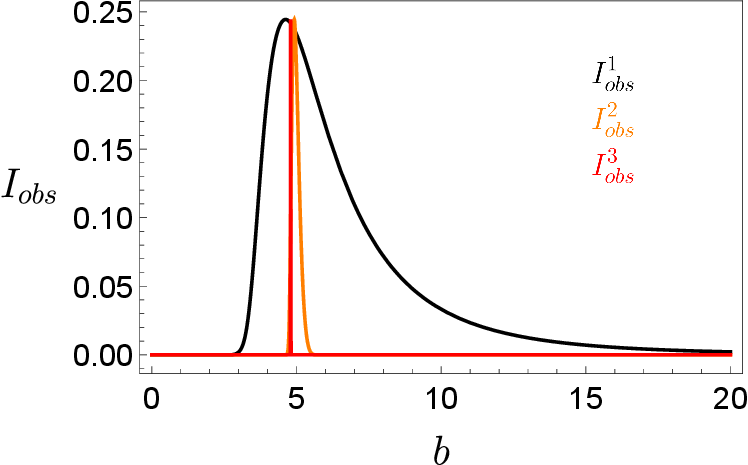}
 	     \end{subfigure}
      \begin{subfigure}[t]{.24\textwidth}
 	         \includegraphics[width = \textwidth]{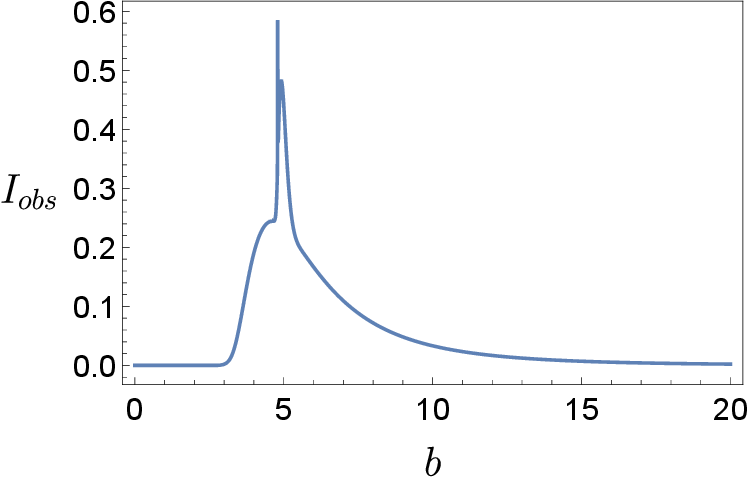}
 	     \end{subfigure}
      \begin{subfigure}[t]{.24\textwidth}
 	         \includegraphics[width = \textwidth]{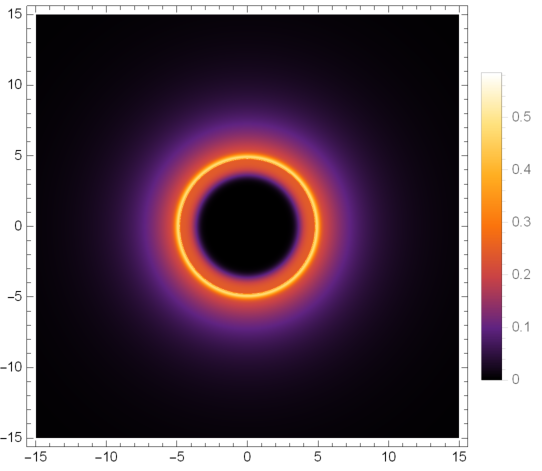}
 	     \end{subfigure}
      \begin{subfigure}[t]{.24\textwidth}
 	         \includegraphics[width = \textwidth]{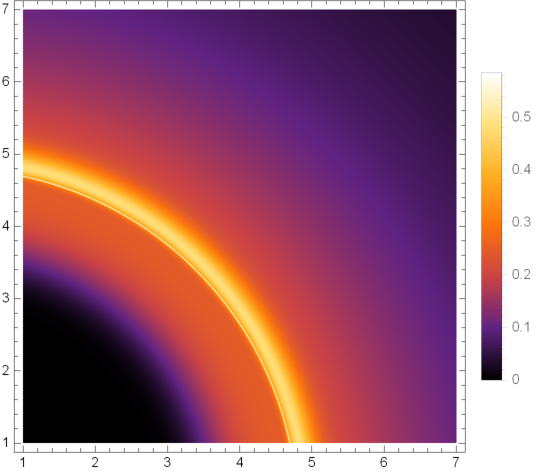}
 	     \end{subfigure}
      \caption{Optical appearance of $q$-deformed objects illuminated by thin accretion disk profile in \textbf{Case~III} of the GLM model. 
      From top to bottom row, the dilaton coupling constant is varies as $\lambda=0.5,1,\sqrt{3}$ with fixed $\tilde{D}=-1$ and $\tilde{P}=0.5$, respectively.
      The first column shows the individual observed intensity contributions from the direct (black), lensing ring (orange) and photon ring (red) regions. 
      The second column presents the total observed intensity. 
      The third column displays the resulting image of a light source near the $q$-deformed object and the close up images are shown in the last column. \justifying}
    \label{fig:shdwvaryLmodel3}
  \end{figure}

\begin{figure}[htbp]
      \begin{subfigure}[t]{.24\textwidth}
 	         \includegraphics[width = \textwidth]{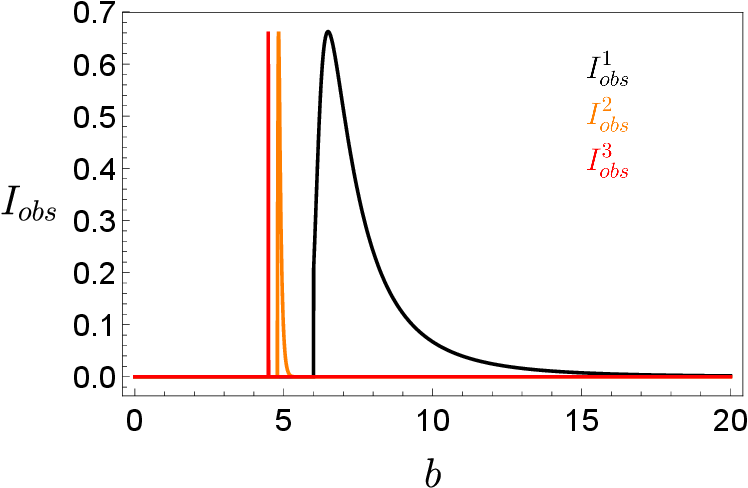}
 	     \end{subfigure}
      \begin{subfigure}[t]{.24\textwidth}
 	         \includegraphics[width = \textwidth]{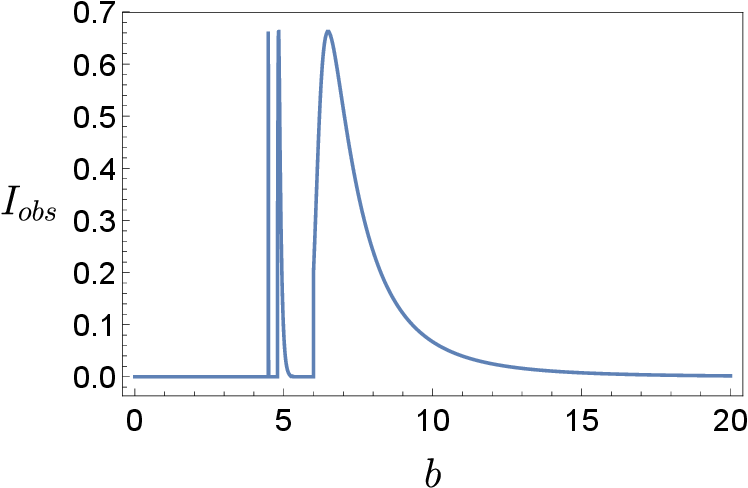}
 	     \end{subfigure}
      \begin{subfigure}[t]{.24\textwidth}
 	         \includegraphics[width = \textwidth]{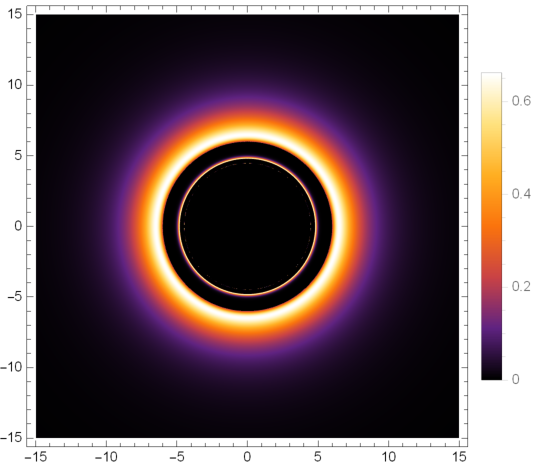}
 	     \end{subfigure}
      \begin{subfigure}[t]{.24\textwidth}
 	         \includegraphics[width = \textwidth]{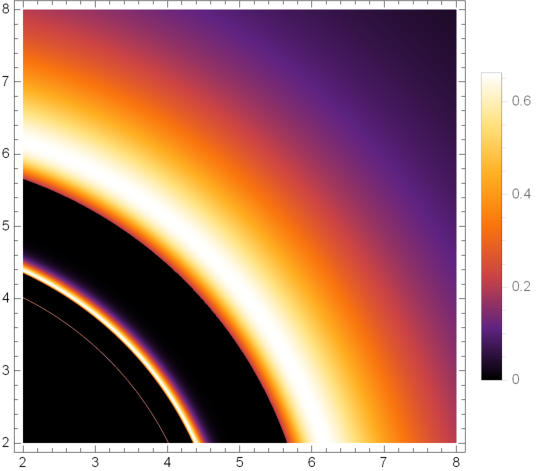}
 	     \end{subfigure}
 
      \begin{subfigure}[t]{.24\textwidth}
 	         \includegraphics[width = \textwidth]{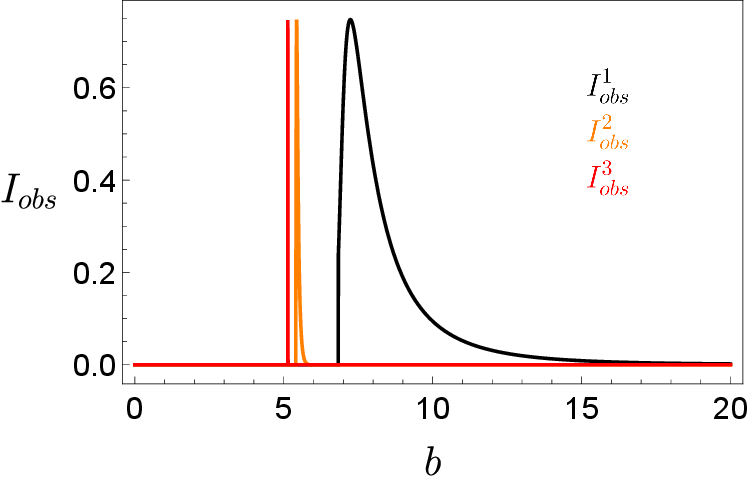}
 	     \end{subfigure}
      \begin{subfigure}[t]{.24\textwidth}
 	         \includegraphics[width = \textwidth]{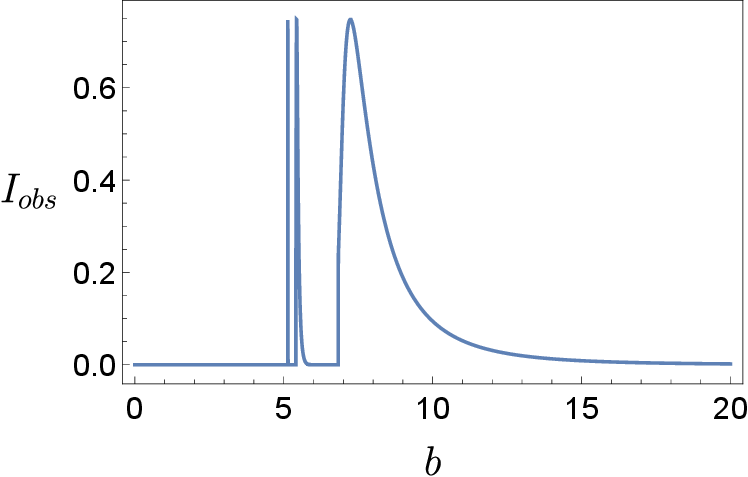}
 	     \end{subfigure}
      \begin{subfigure}[t]{.24\textwidth}
 	         \includegraphics[width = \textwidth]{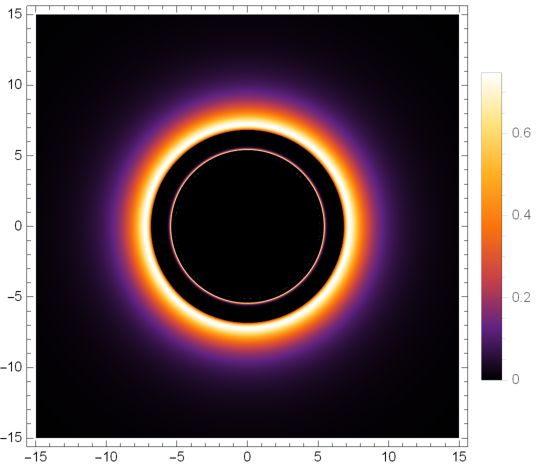}
 	     \end{subfigure}
      \begin{subfigure}[t]{.24\textwidth}
 	         \includegraphics[width = \textwidth]{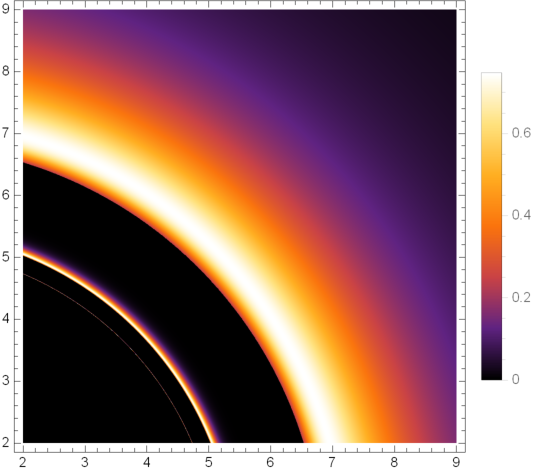}
 	     \end{subfigure}
 
      \begin{subfigure}[t]{.24\textwidth}
 	         \includegraphics[width = \textwidth]{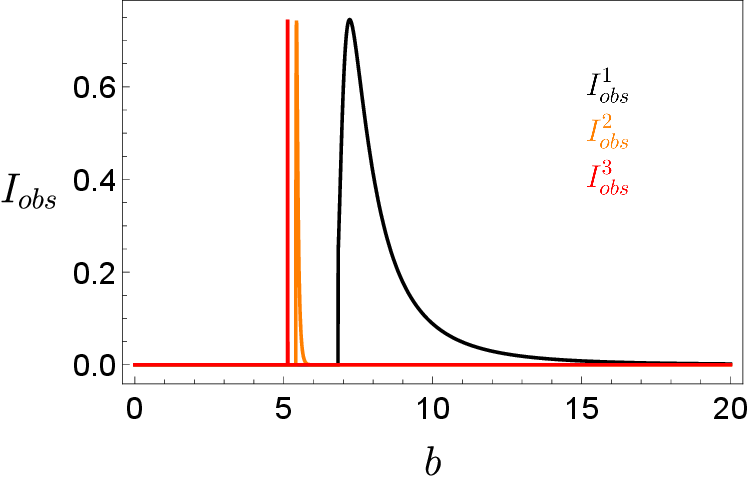}
 	     \end{subfigure}
      \begin{subfigure}[t]{.24\textwidth}
 	         \includegraphics[width = \textwidth]{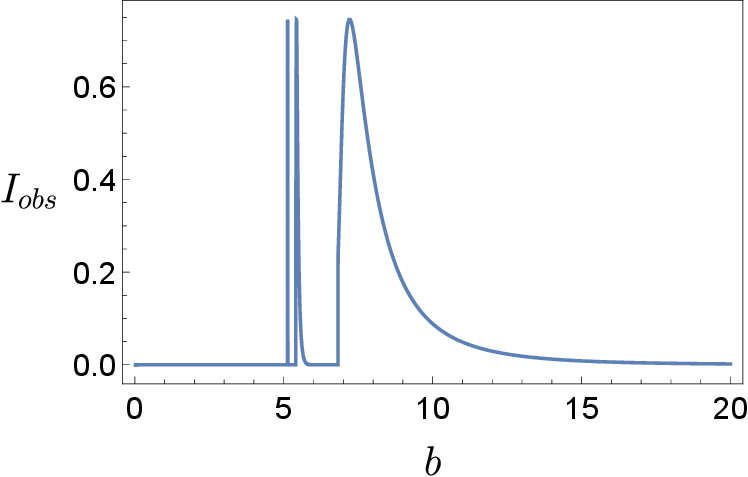}
 	     \end{subfigure}
      \begin{subfigure}[t]{.24\textwidth}
 	         \includegraphics[width = \textwidth]{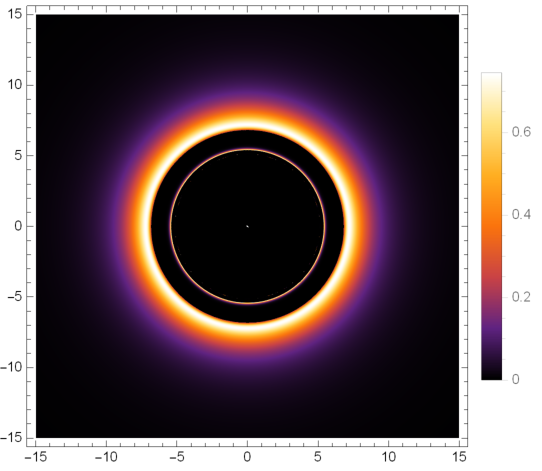}
 	     \end{subfigure}
      \begin{subfigure}[t]{.24\textwidth}
 	         \includegraphics[width = \textwidth]{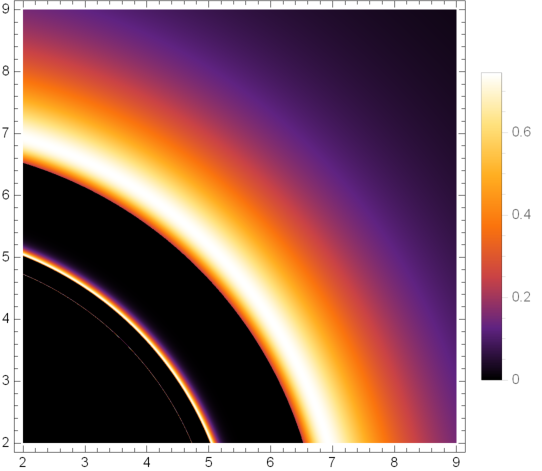}
 	     \end{subfigure}
      \caption{Optical appearance of $q$-deformed objects illuminated by thin accretion disk profile in \textbf{Case~I} of the GLM model. 
      From top to bottom row, the scalar charge is varies as $\tilde{D}=-1.5,-0.5$ and $0.5$ with fixed $\lambda=1.5$ and $\tilde{P}=0.1$, respectively.
      The first column shows the individual observed intensity contributions from the direct (black), lensing ring (orange) and photon ring (red) regions. 
      The second column presents the total observed intensity. 
      The third column displays the resulting image of a light source near the $q$-deformed object and the close up images are shown in the last column. \justifying}
    \label{fig:shdwvaryDmodel1}
  \end{figure}

\begin{figure}[htbp]
      \begin{subfigure}[t]{.24\textwidth}
 	         \includegraphics[width = \textwidth]{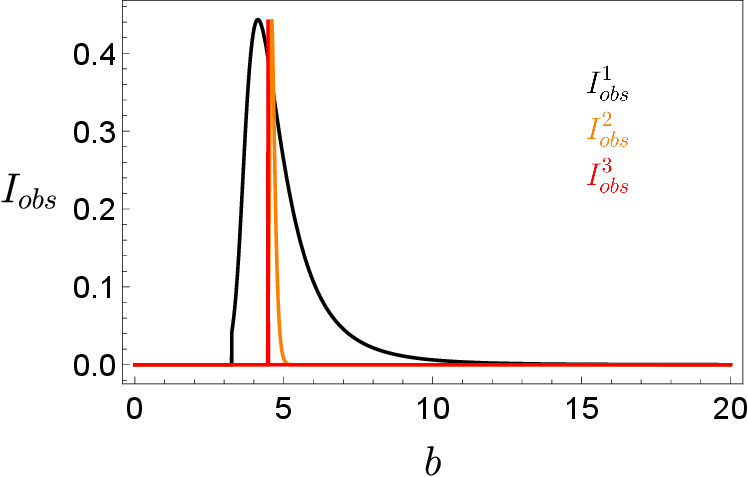}
 	     \end{subfigure}
      \begin{subfigure}[t]{.24\textwidth}
 	         \includegraphics[width = \textwidth]{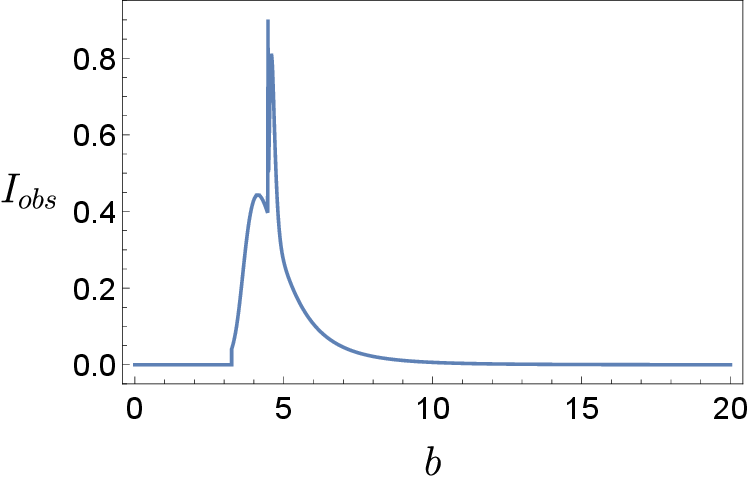}
 	     \end{subfigure}
      \begin{subfigure}[t]{.24\textwidth}
 	         \includegraphics[width = \textwidth]{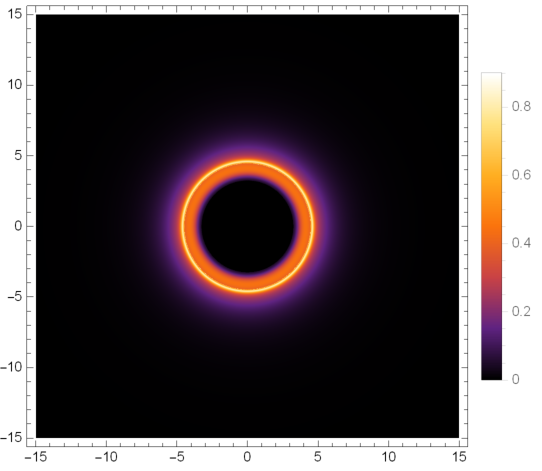}
 	     \end{subfigure}
      \begin{subfigure}[t]{.24\textwidth}
 	         \includegraphics[width = \textwidth]{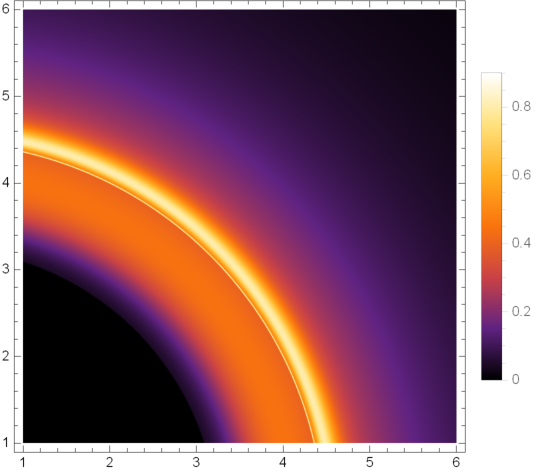}
 	     \end{subfigure}
 
      \begin{subfigure}[t]{.24\textwidth}
 	         \includegraphics[width = \textwidth]{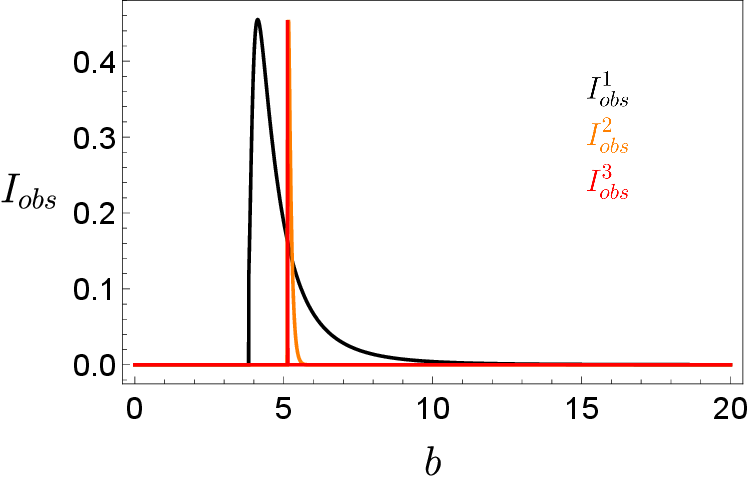}
 	     \end{subfigure}
      \begin{subfigure}[t]{.24\textwidth}
 	         \includegraphics[width = \textwidth]{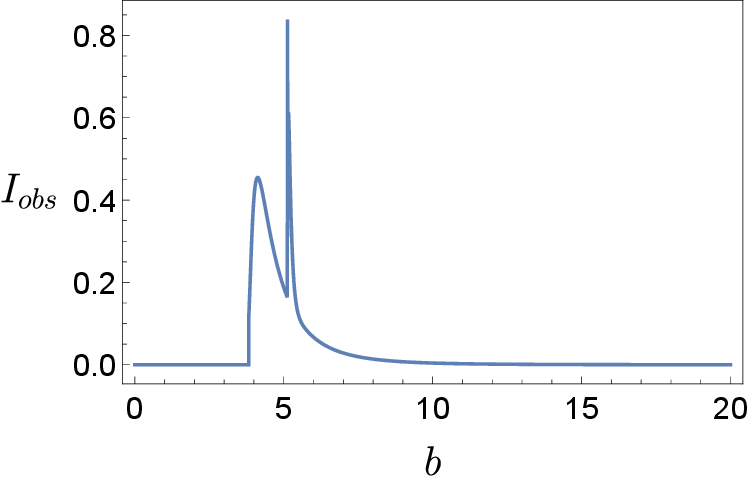}
 	     \end{subfigure}
      \begin{subfigure}[t]{.24\textwidth}
 	         \includegraphics[width = \textwidth]{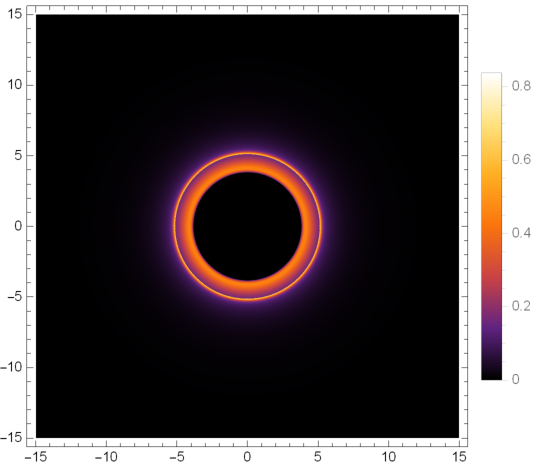}
 	     \end{subfigure}
      \begin{subfigure}[t]{.24\textwidth}
 	         \includegraphics[width = \textwidth]{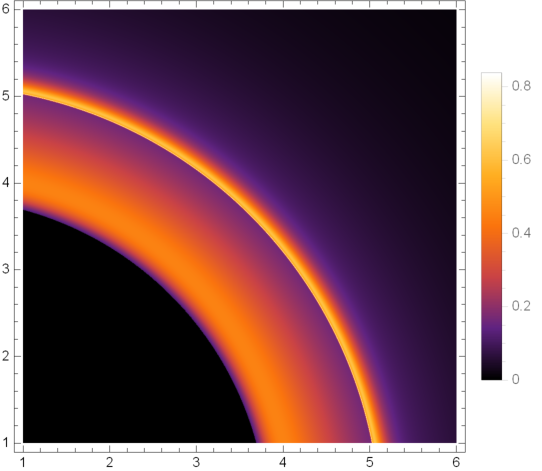}
 	     \end{subfigure}
 
      \begin{subfigure}[t]{.24\textwidth}
 	         \includegraphics[width = \textwidth]{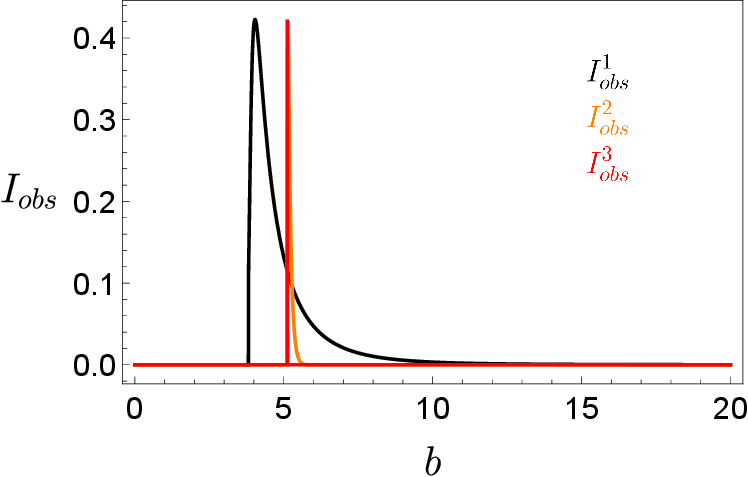}
 	     \end{subfigure}
      \begin{subfigure}[t]{.24\textwidth}
 	         \includegraphics[width = \textwidth]{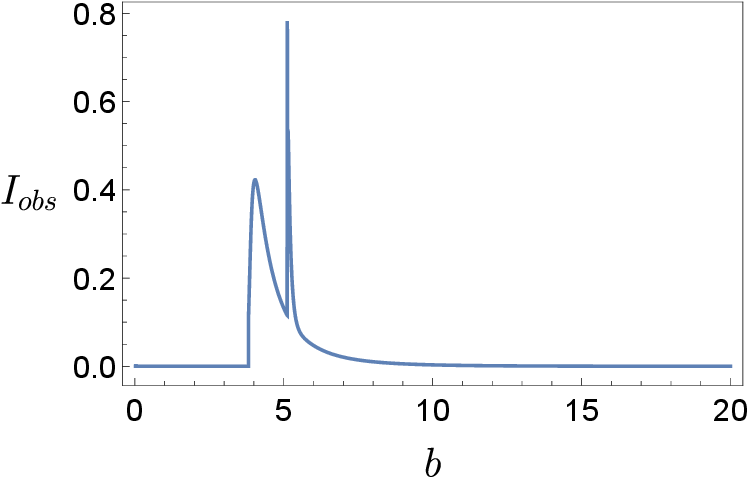}
 	     \end{subfigure}
      \begin{subfigure}[t]{.24\textwidth}
 	         \includegraphics[width = \textwidth]{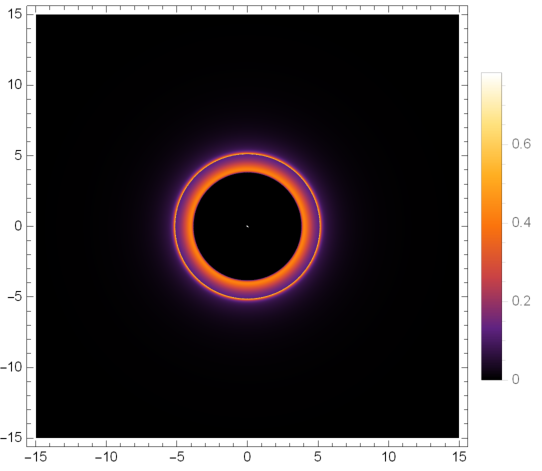}
 	     \end{subfigure}
      \begin{subfigure}[t]{.24\textwidth}
 	         \includegraphics[width = \textwidth]{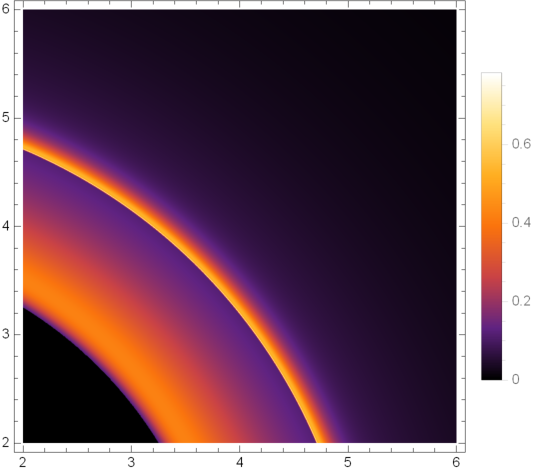}
 	     \end{subfigure}
      \caption{Optical appearance of $q$-deformed objects illuminated by thin accretion disk profile in \textbf{Case~II} of the GLM model. 
      From top to bottom row, the scalar charge is varies as $\tilde{D}=-1.5,-0.5$ and $0.5$ with fixed $\lambda=1.5$ and $\tilde{P}=0.1$, respectively.
      The first column shows the individual observed intensity contributions from the direct (black), lensing ring (orange) and photon ring (red) regions. 
      The second column presents the total observed intensity. 
      The third column displays the resulting image of a light source near the $q$-deformed object and the close up images are shown in the last column. \justifying}
    \label{fig:shdwvaryDmodel2}
  \end{figure}

\begin{figure}[htbp]
      \begin{subfigure}[t]{.24\textwidth}
 	         \includegraphics[width = \textwidth]{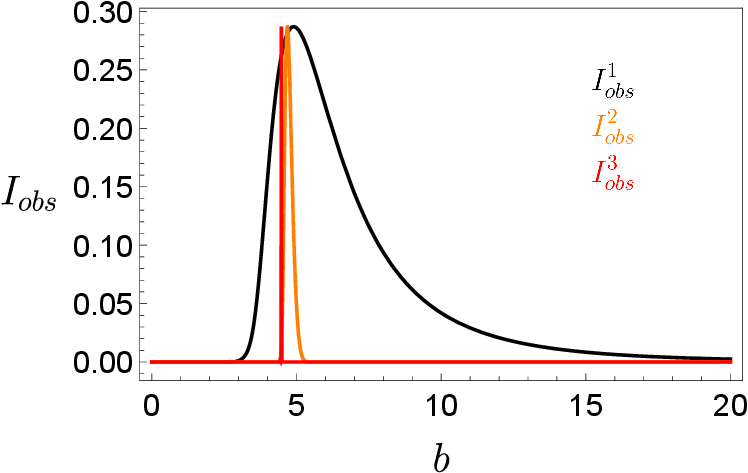}
 	     \end{subfigure}
      \begin{subfigure}[t]{.24\textwidth}
 	         \includegraphics[width = \textwidth]{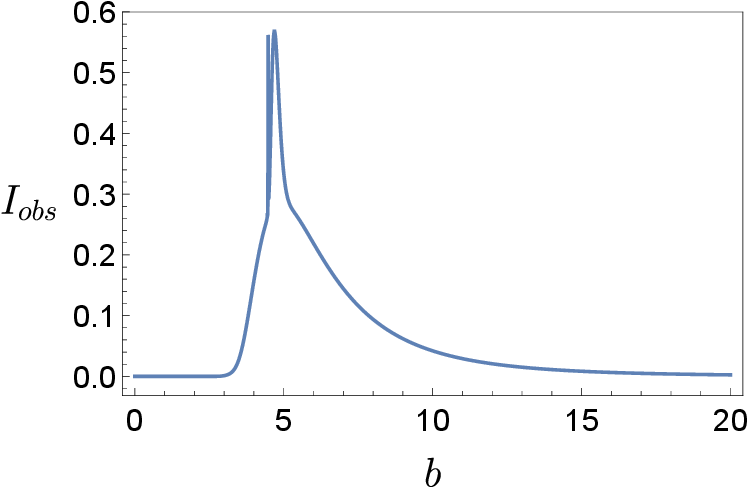}
 	     \end{subfigure}
      \begin{subfigure}[t]{.24\textwidth}
 	         \includegraphics[width = \textwidth]{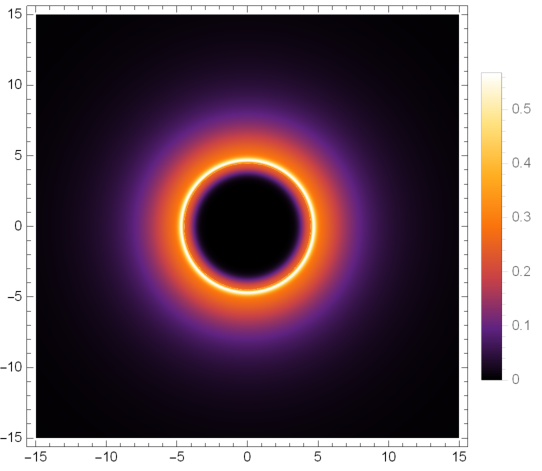}
 	     \end{subfigure}
      \begin{subfigure}[t]{.24\textwidth}
 	         \includegraphics[width = \textwidth]{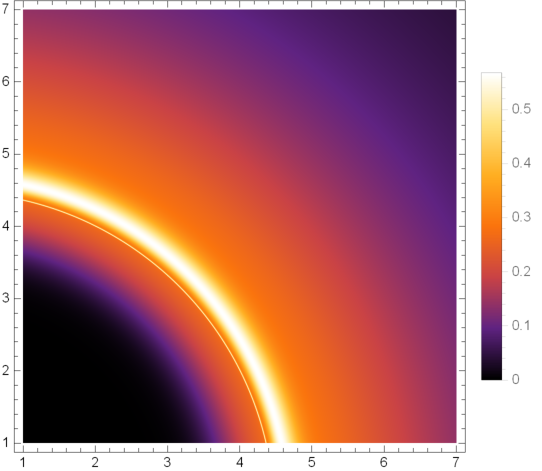}
 	     \end{subfigure}
 
      \begin{subfigure}[t]{.24\textwidth}
 	         \includegraphics[width = \textwidth]{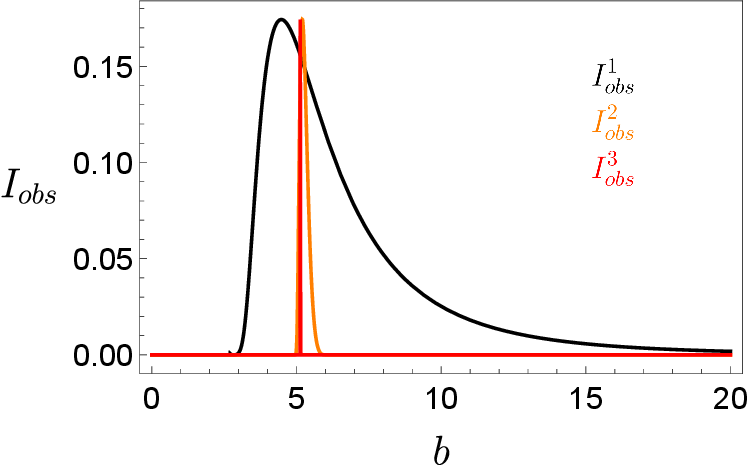}
 	     \end{subfigure}
      \begin{subfigure}[t]{.24\textwidth}
 	         \includegraphics[width = \textwidth]{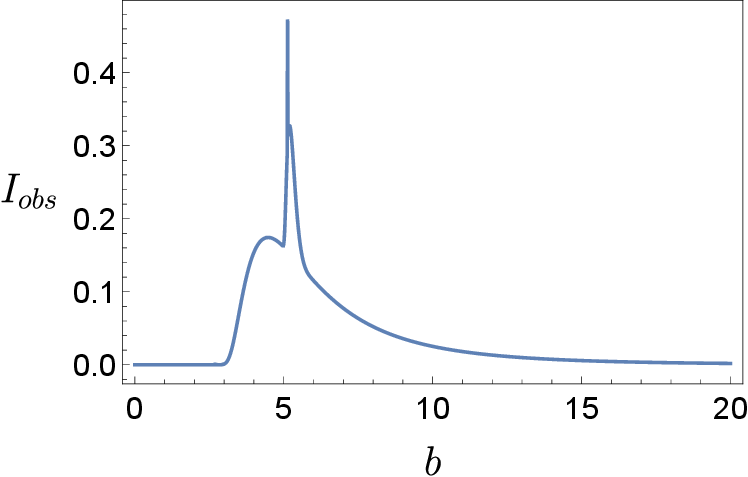}
 	     \end{subfigure}
      \begin{subfigure}[t]{.24\textwidth}
 	         \includegraphics[width = \textwidth]{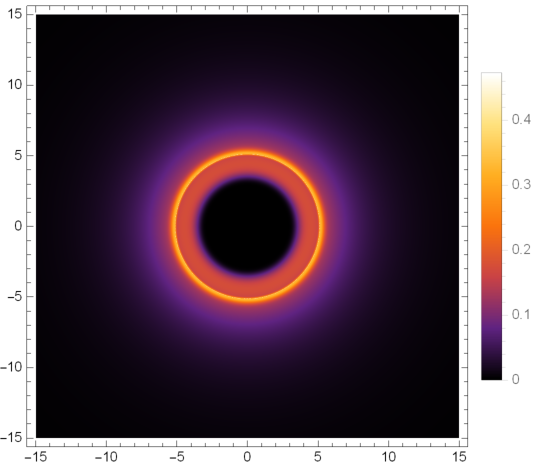}
 	     \end{subfigure}
      \begin{subfigure}[t]{.24\textwidth}
 	         \includegraphics[width = \textwidth]{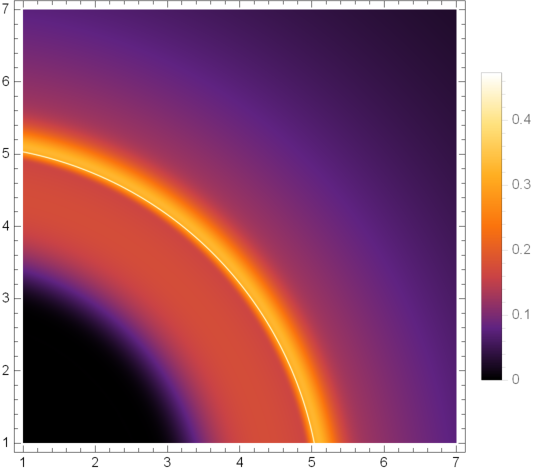}
 	     \end{subfigure}
 
      \begin{subfigure}[t]{.24\textwidth}
 	         \includegraphics[width = \textwidth]{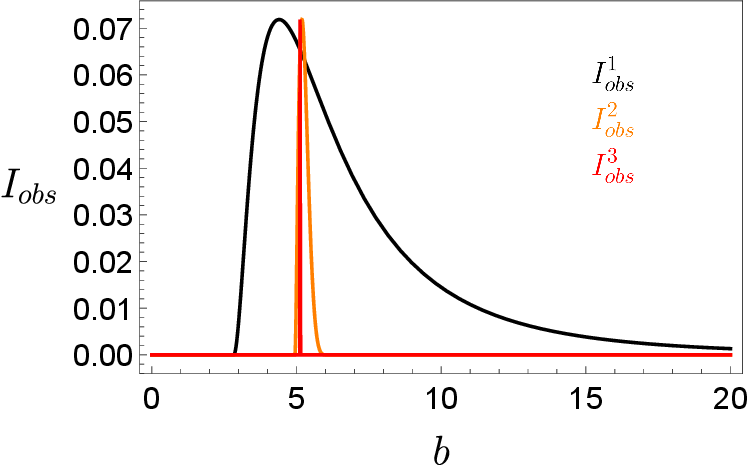}
 	     \end{subfigure}
      \begin{subfigure}[t]{.24\textwidth}
 	         \includegraphics[width = \textwidth]{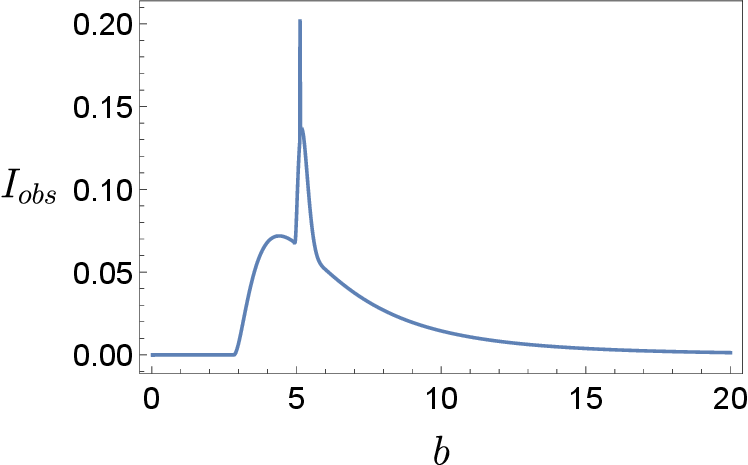}
 	     \end{subfigure}
      \begin{subfigure}[t]{.24\textwidth}
 	         \includegraphics[width = \textwidth]{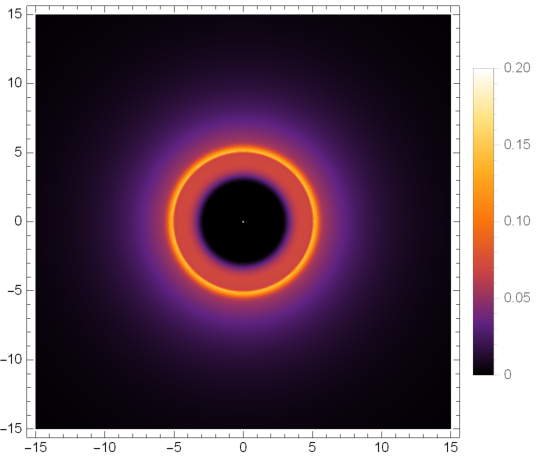}
 	     \end{subfigure}
      \begin{subfigure}[t]{.24\textwidth}
 	         \includegraphics[width = \textwidth]{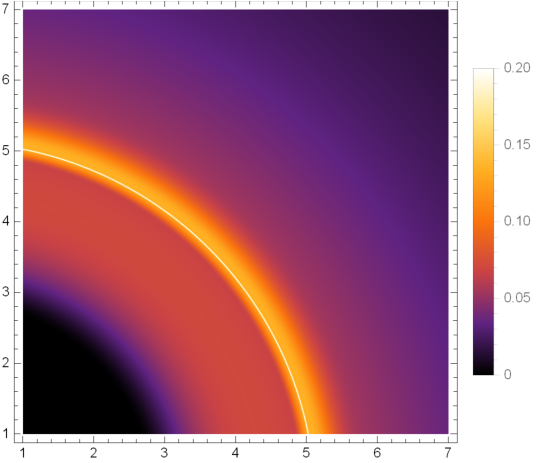}
 	     \end{subfigure}
      \caption{Optical appearance of $q$-deformed objects illuminated by thin accretion disk profile in \textbf{Case~III} of the GLM model. 
      From top to bottom row, the scalar charge is varies as $\tilde{D}=-1.5,-0.5$ and $0.5$ with fixed $\lambda=1.5$ and $\tilde{P}=0.1$, respectively.
      The first column shows the individual observed intensity contributions from the direct (black), lensing ring (orange) and photon ring (red) regions. 
      The second column presents the total observed intensity. 
      The third column displays the resulting image of a light source near the $q$-deformed object and the close up images are shown in the last column. \justifying}
    \label{fig:shdwvaryDmodel3}
  \end{figure}

\begin{figure}[htbp]
      \begin{subfigure}[t]{.24\textwidth}
 	         \includegraphics[width = \textwidth]{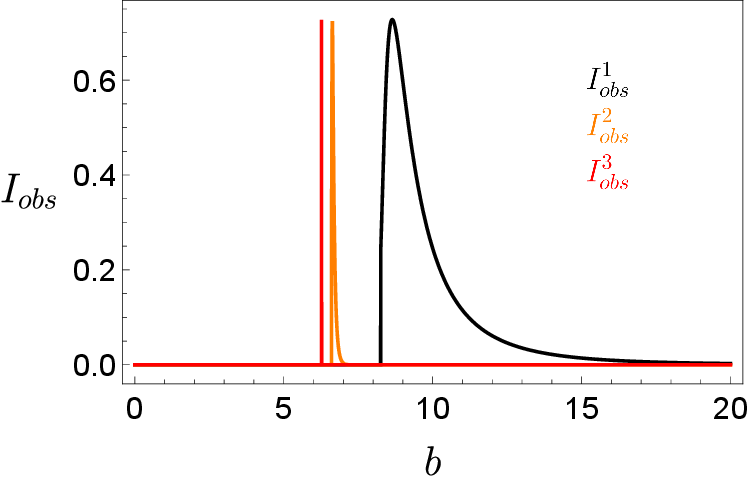}
 	     \end{subfigure}
      \begin{subfigure}[t]{.24\textwidth}
 	         \includegraphics[width = \textwidth]{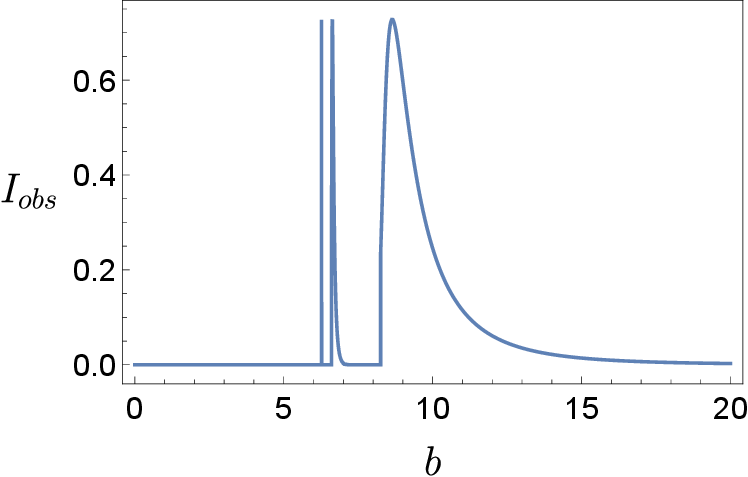}
 	     \end{subfigure}
      \begin{subfigure}[t]{.24\textwidth}
 	         \includegraphics[width = \textwidth]{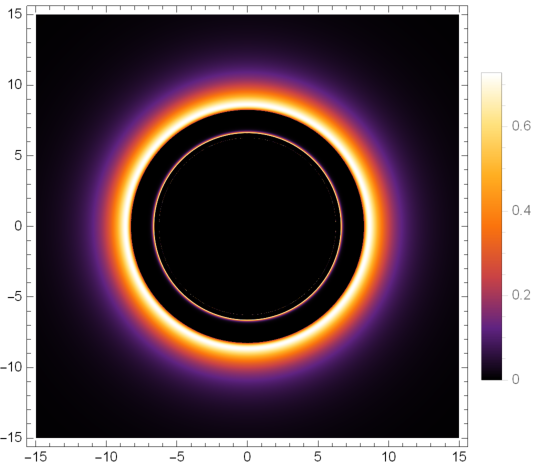}
 	     \end{subfigure}
      \begin{subfigure}[t]{.24\textwidth}
 	         \includegraphics[width = \textwidth]{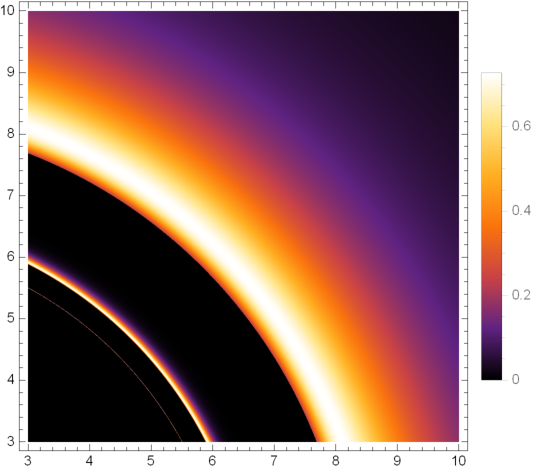}
 	     \end{subfigure}
 
      \begin{subfigure}[t]{.24\textwidth}
 	         \includegraphics[width = \textwidth]{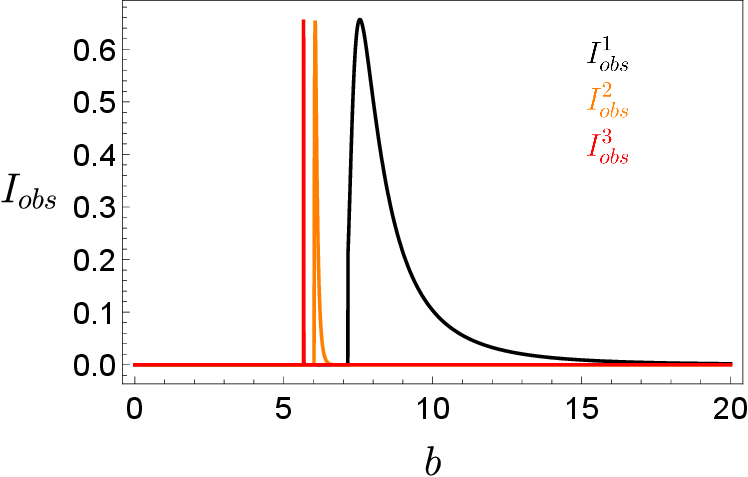}
 	     \end{subfigure}
      \begin{subfigure}[t]{.24\textwidth}
 	         \includegraphics[width = \textwidth]{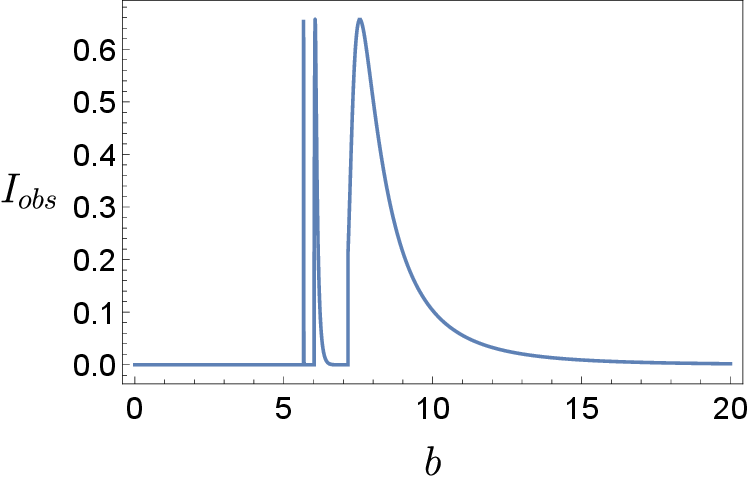}
 	     \end{subfigure}
      \begin{subfigure}[t]{.24\textwidth}
 	         \includegraphics[width = \textwidth]{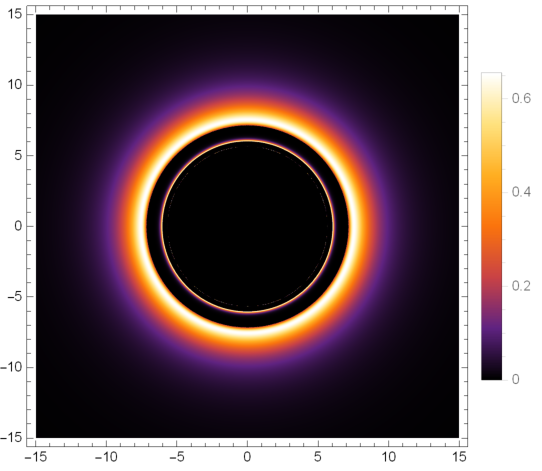}
 	     \end{subfigure}
      \begin{subfigure}[t]{.24\textwidth}
 	         \includegraphics[width = \textwidth]{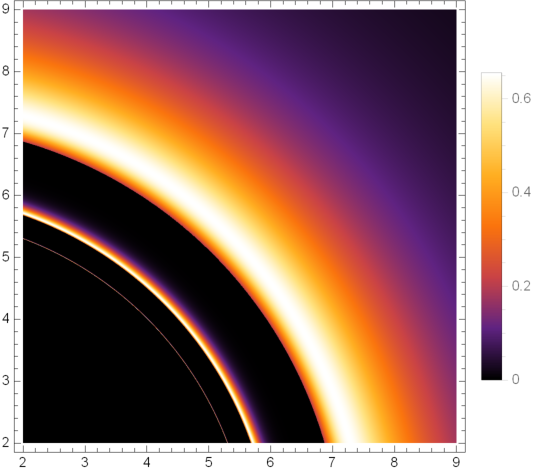}
 	     \end{subfigure}
 
      \begin{subfigure}[t]{.24\textwidth}
 	         \includegraphics[width = \textwidth]{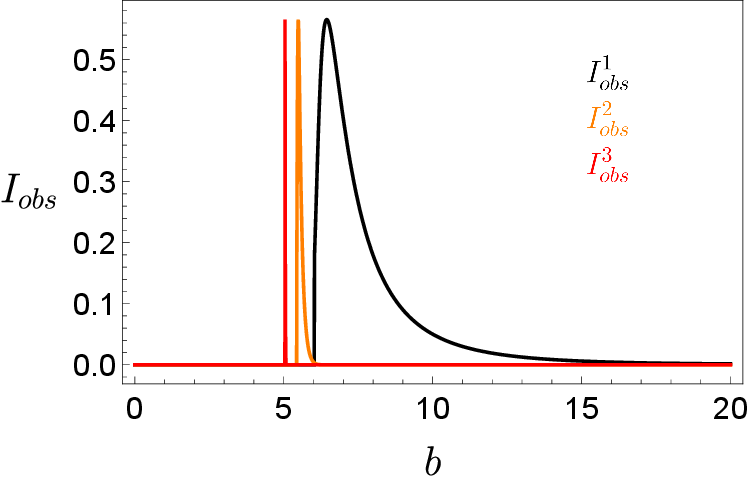}
 	     \end{subfigure}
      \begin{subfigure}[t]{.24\textwidth}
 	         \includegraphics[width = \textwidth]{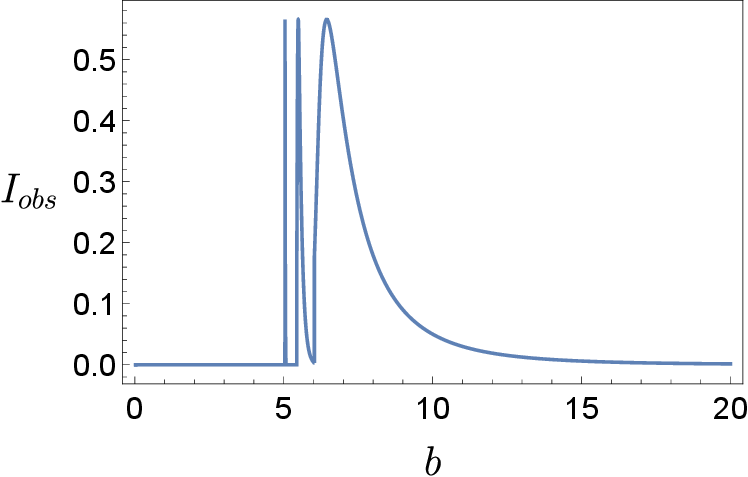}
 	     \end{subfigure}
      \begin{subfigure}[t]{.24\textwidth}
 	         \includegraphics[width = \textwidth]{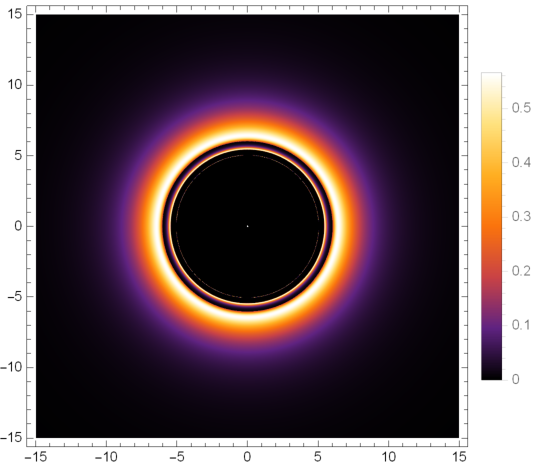}
 	     \end{subfigure}
      \begin{subfigure}[t]{.24\textwidth}
 	         \includegraphics[width = \textwidth]{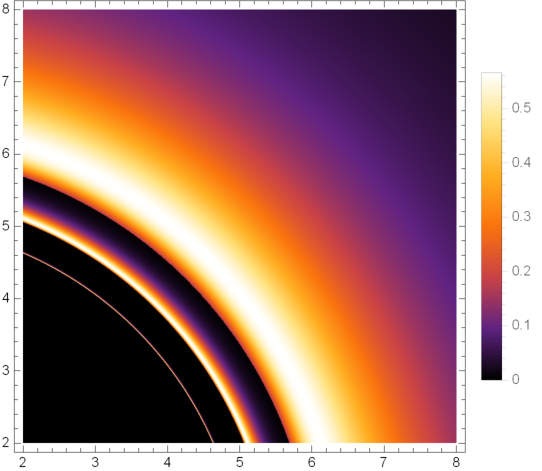}
 	     \end{subfigure}
      \caption{Optical appearance of $q$-deformed objects illuminated by thin accretion disk profile in \textbf{Case~I} of the GLM model. 
      From top to bottom row, the magnetic charge is varies as $\tilde{P}=0.5,1$ and $1.2522$ with fixed $\lambda=0.5$ and $\tilde{D}=-0.8$, respectively.
      The first column shows the individual observed intensity contributions from the direct (black), lensing ring (orange) and photon ring (red) regions. 
      The second column presents the total observed intensity. 
      The third column displays the resulting image of a light source near the $q$-deformed object and the close up images are shown in the last column.
      \justifying}
    \label{fig:shdwvaryPmodel1}
  \end{figure}

\begin{figure}[htbp]
      \begin{subfigure}[t]{.24\textwidth}
 	         \includegraphics[width = \textwidth]{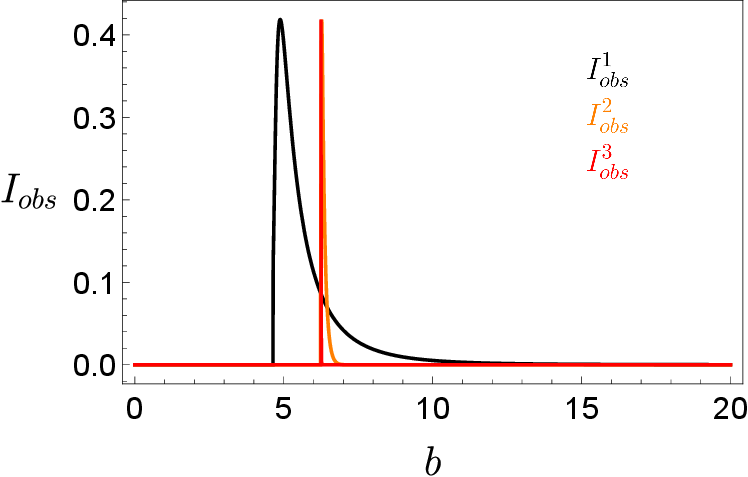}
 	     \end{subfigure}
      \begin{subfigure}[t]{.24\textwidth}
 	         \includegraphics[width = \textwidth]{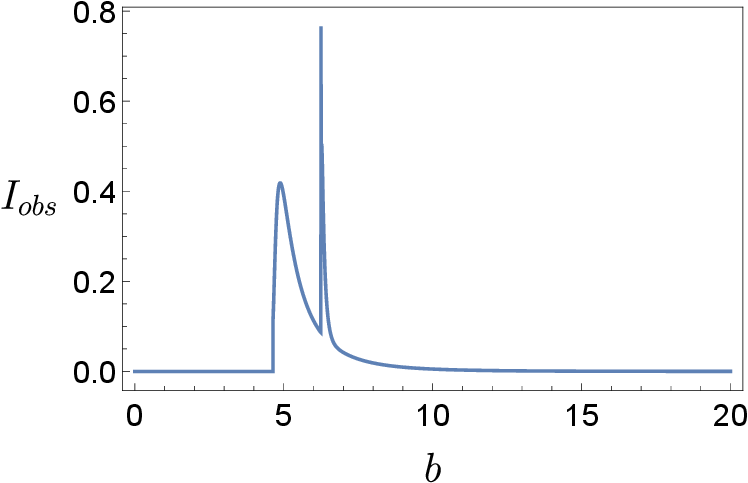}
 	     \end{subfigure}
      \begin{subfigure}[t]{.24\textwidth}
 	         \includegraphics[width = \textwidth]{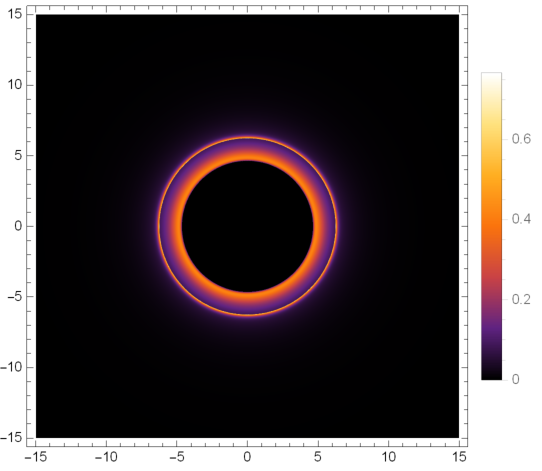}
 	     \end{subfigure}
      \begin{subfigure}[t]{.24\textwidth}
 	         \includegraphics[width = \textwidth]{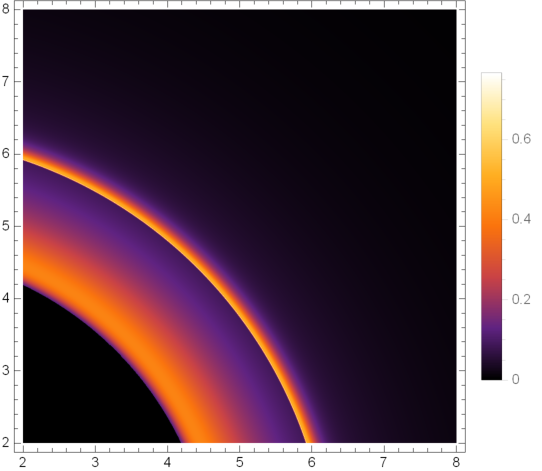}
 	     \end{subfigure}
 
      \begin{subfigure}[t]{.24\textwidth}
 	         \includegraphics[width = \textwidth]{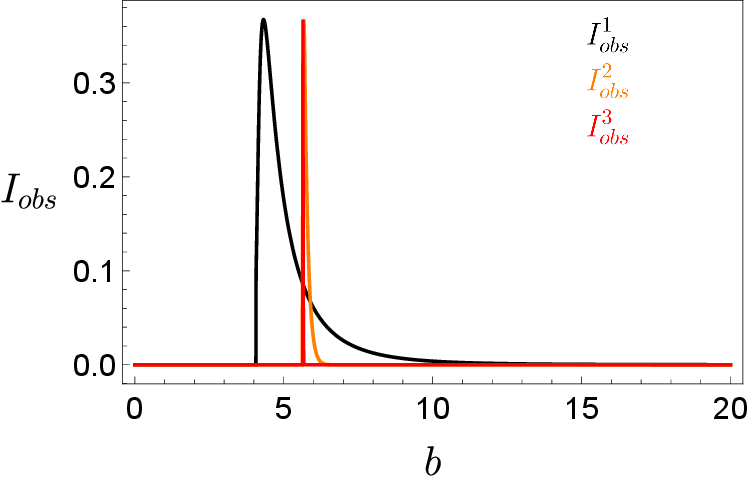}
 	     \end{subfigure}
      \begin{subfigure}[t]{.24\textwidth}
 	         \includegraphics[width = \textwidth]{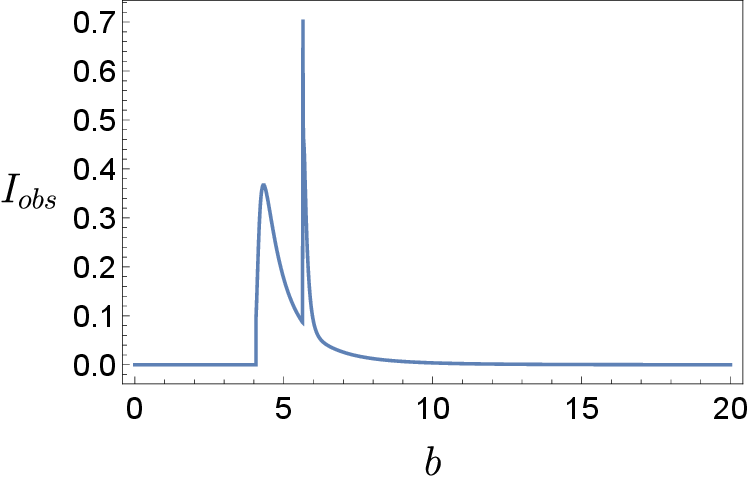}
 	     \end{subfigure}
      \begin{subfigure}[t]{.24\textwidth}
 	         \includegraphics[width = \textwidth]{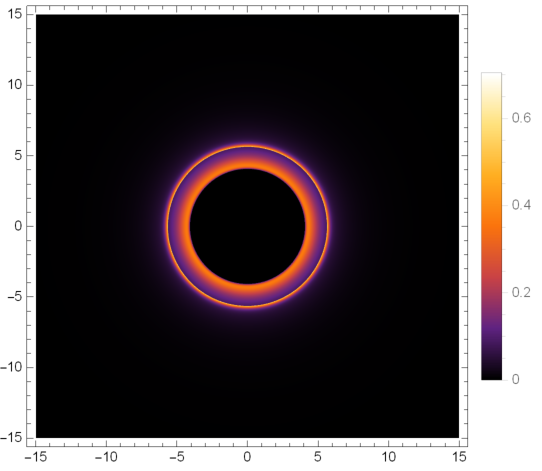}
 	     \end{subfigure}
      \begin{subfigure}[t]{.24\textwidth}
 	         \includegraphics[width = \textwidth]{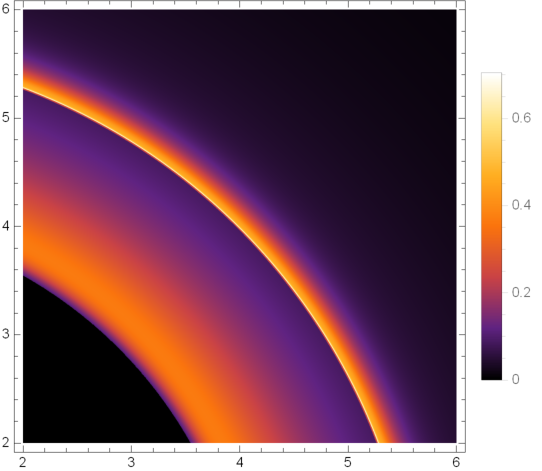}
 	     \end{subfigure}
 
      \begin{subfigure}[t]{.24\textwidth}
 	         \includegraphics[width = \textwidth]{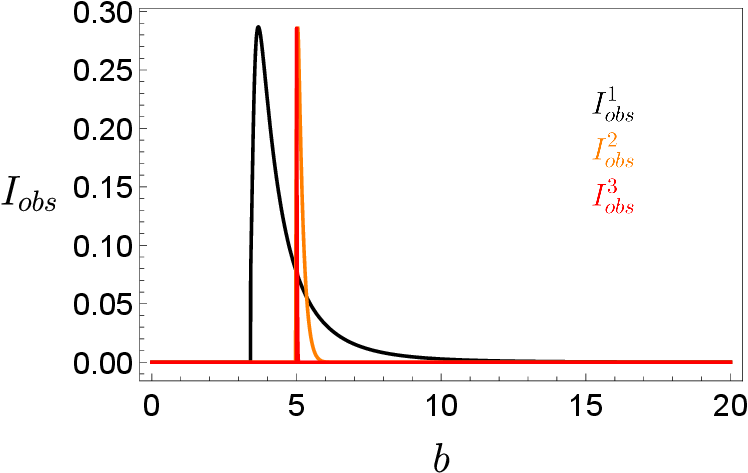}
 	     \end{subfigure}
      \begin{subfigure}[t]{.24\textwidth}
 	         \includegraphics[width = \textwidth]{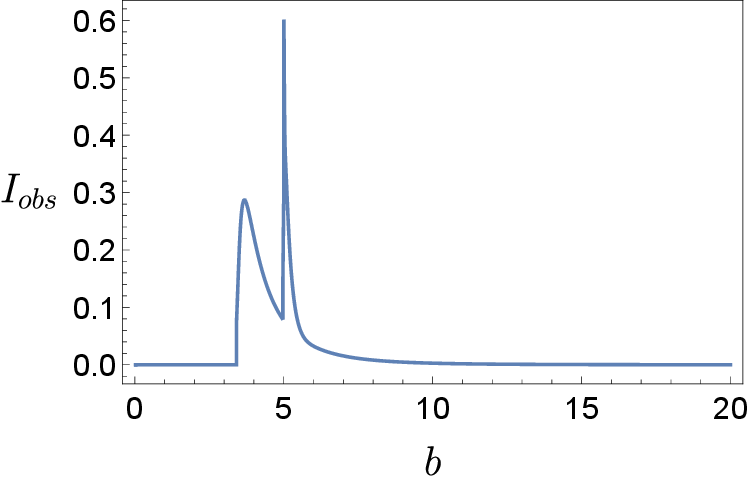}
 	     \end{subfigure}
      \begin{subfigure}[t]{.24\textwidth}
 	         \includegraphics[width = \textwidth]{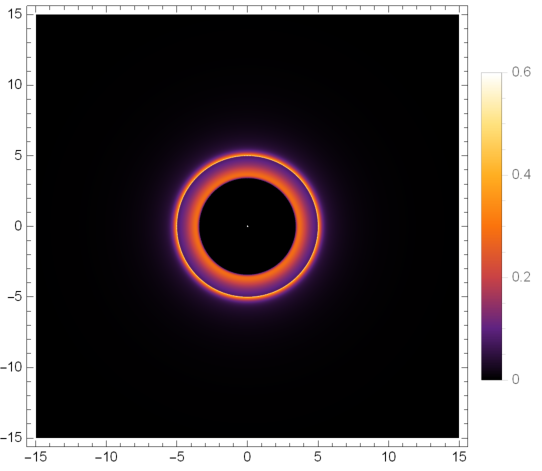}
 	     \end{subfigure}
      \begin{subfigure}[t]{.24\textwidth}
 	         \includegraphics[width = \textwidth]{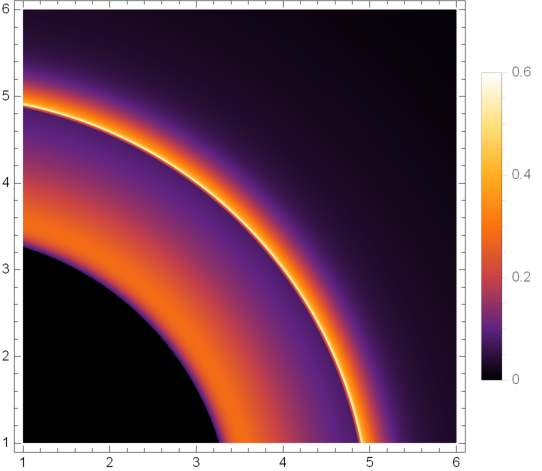}
 	     \end{subfigure}
      \caption{Optical appearance of $q$-deformed objects illuminated by thin accretion disk profile in \textbf{Case~II} of the GLM model. 
      From top to bottom row, the magnetic charge is varies as $\tilde{P}=0.5,1$ and $1.2522$ with fixed $\lambda=0.5$ and $\tilde{D}=-0.8$, respectively.
      The first column shows the individual observed intensity contributions from the direct (black), lensing ring (orange) and photon ring (red) regions. 
      The second column presents the total observed intensity. 
      The third column displays the resulting image of a light source near the $q$-deformed object and the close up images are shown in the last column. \justifying}
    \label{fig:shdwvaryPmodel2}
  \end{figure}

\begin{figure}[htbp]
      \begin{subfigure}[t]{.24\textwidth}
 	         \includegraphics[width = \textwidth]{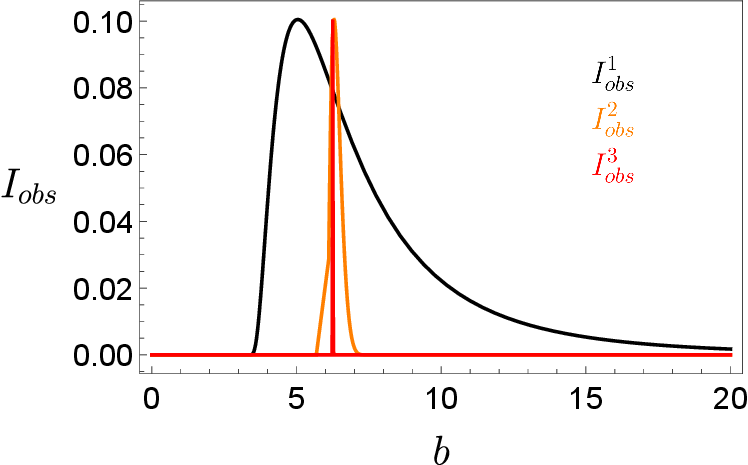}
 	     \end{subfigure}
      \begin{subfigure}[t]{.24\textwidth}
 	         \includegraphics[width = \textwidth]{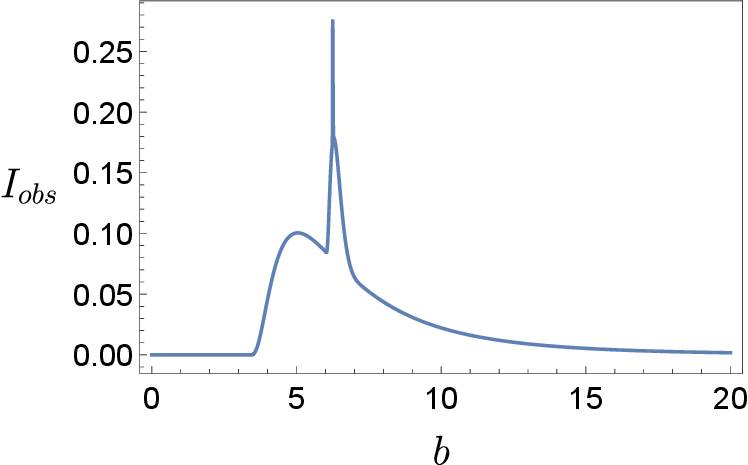}
 	     \end{subfigure}
      \begin{subfigure}[t]{.24\textwidth}
 	         \includegraphics[width = \textwidth]{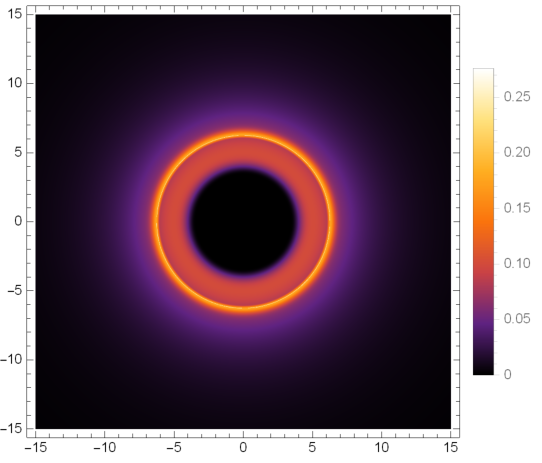}
 	     \end{subfigure}
      \begin{subfigure}[t]{.24\textwidth}
 	         \includegraphics[width = \textwidth]{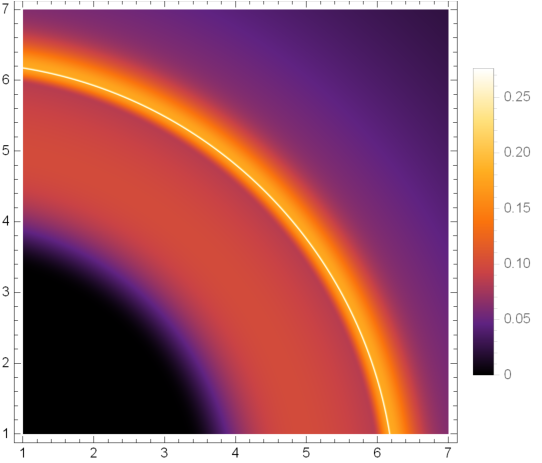}
 	     \end{subfigure}
 
      \begin{subfigure}[t]{.24\textwidth}
 	         \includegraphics[width = \textwidth]{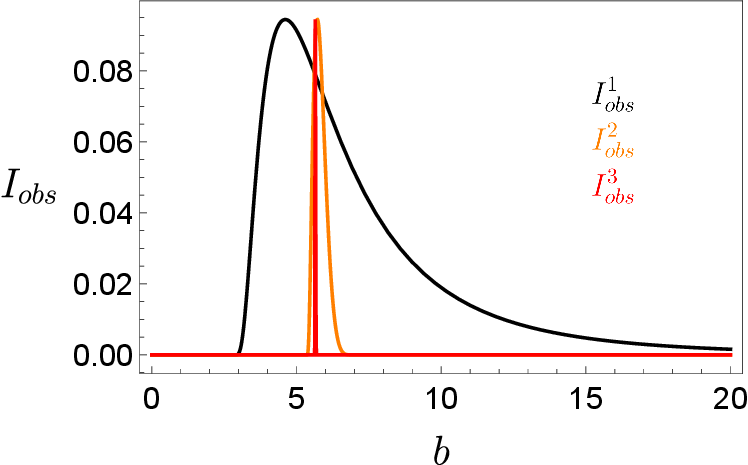}
 	     \end{subfigure}
      \begin{subfigure}[t]{.24\textwidth}
 	         \includegraphics[width = \textwidth]{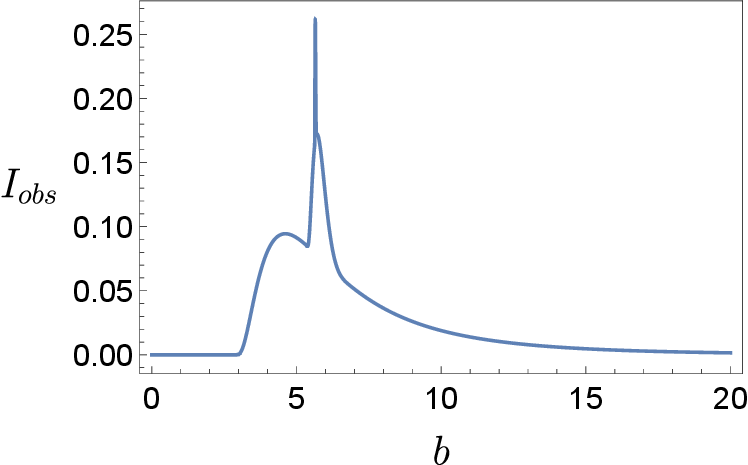}
 	     \end{subfigure}
      \begin{subfigure}[t]{.24\textwidth}
 	         \includegraphics[width = \textwidth]{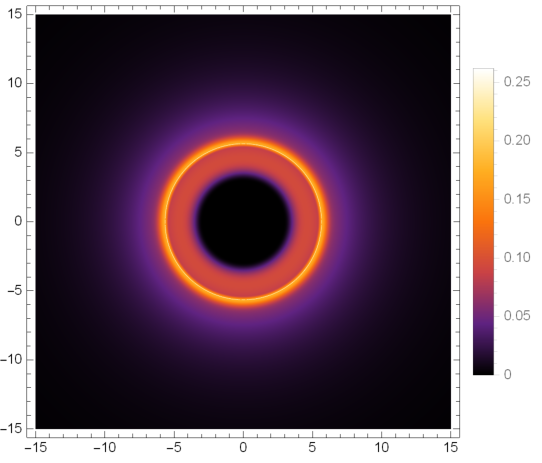}
 	     \end{subfigure}
      \begin{subfigure}[t]{.24\textwidth}
 	         \includegraphics[width = \textwidth]{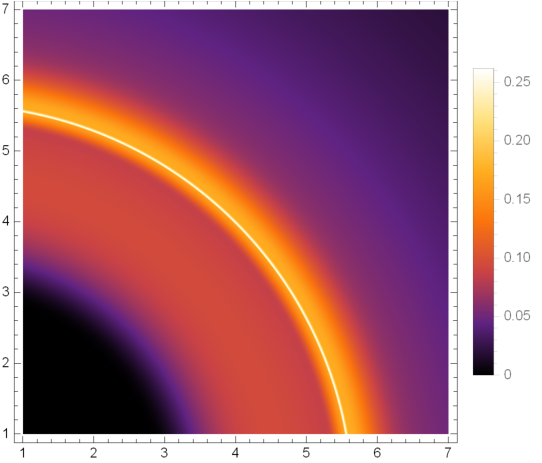}
 	     \end{subfigure}
 
      \begin{subfigure}[t]{.24\textwidth}
 	         \includegraphics[width = \textwidth]{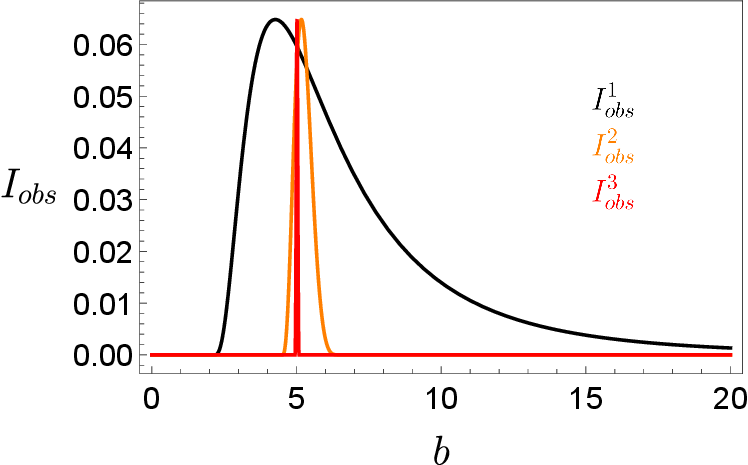}
 	     \end{subfigure}
      \begin{subfigure}[t]{.24\textwidth}
 	         \includegraphics[width = \textwidth]{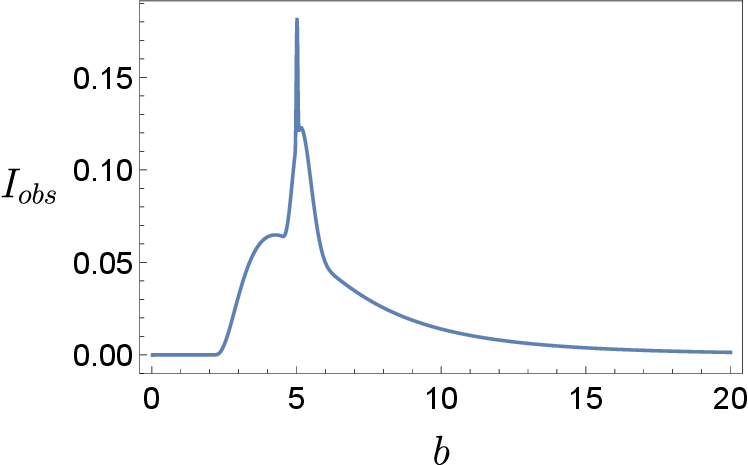}
 	     \end{subfigure}
      \begin{subfigure}[t]{.24\textwidth}
 	         \includegraphics[width = \textwidth]{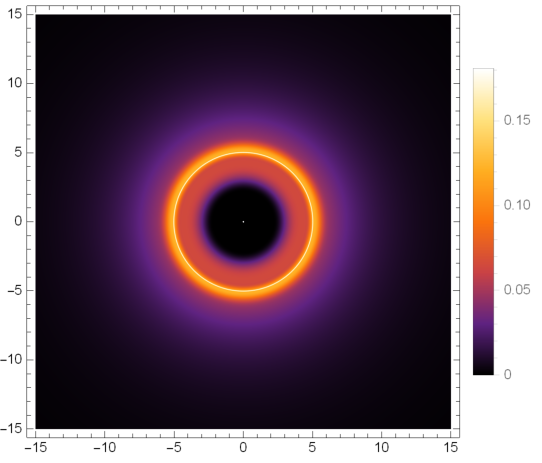}
 	     \end{subfigure}
      \begin{subfigure}[t]{.24\textwidth}
 	         \includegraphics[width = \textwidth]{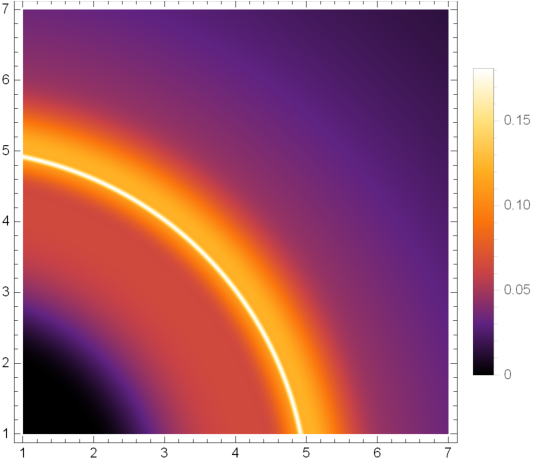}
 	     \end{subfigure}
      \caption{Optical appearance of $q$-deformed objects illuminated by thin accretion disk profile in \textbf{Case~III} of the GLM model. 
      From top to bottom row, the magnetic charge is varies as $\tilde{P}=0.5,1$ and $1.2522$ with fixed $\lambda=0.5$ and $\tilde{D}=-0.8$, respectively.
      The first column shows the individual observed intensity contributions from the direct (black), lensing ring (orange) and photon ring (red) regions. 
      The second column presents the total observed intensity. 
      The third column displays the resulting image of a light source near the $q$-deformed object and the close up images are shown in the last column. \justifying}
  \label{fig:shdwvaryPmodel3}
  \end{figure}

In this subsection, three cases of the GLM model are applied to investigate the intensity of emission near pointlike objects in EMD gravity. 
The results are shown in Figs.~\ref{fig:shdwvaryLmodel1}-\ref{fig:shdwvaryPmodel3}. 
The first column clearly illustrates the observed intensities, $I_{obs}^1(b)$, $I_{obs}^2(b)$ and $I_{obs}^3(b)$, corresponding to the first, second and third transfer functions and shown as black, orange and red curves, respectively.
The second column presents the total intensity $I_{obs}(b)$ as seen by a distant observer. 
The third column shows the density plot of $I_{obs}(b)$ in the plane, while the fourth column provides a zoomed-in view of a selected sector of this density plot.

First, we study the effect of varying $\lambda$ on the observed intensity in \textbf{Case~I}, \textbf{II} and \textbf{III} of the GLM model, as shown in Figs.~\ref{fig:shdwvaryLmodel1}, \ref{fig:shdwvaryLmodel2} and \ref{fig:shdwvaryLmodel3}, respectively. 
In these figures, we consider $\lambda = 0.5$ (first row), $\lambda = 1$ (second row) and $\lambda = \sqrt{3}$ (third row), while keeping $D=-1$ and $P=0.5$ fixed.
As illustrated in the left panel of Fig.~\ref{fig:ISCO_bc}, the shadow radius decreases as the value of $\lambda$ increases over the range investigated.
We demonstrate how $\lambda$ influences the images as follows:

\begin{itemize}
    \item For \textbf{Case~I}, the emission profile begins as a peak near $r_{ISCO}$ and decreases monotonically with increasing $r$. As shown in the first column of Fig.~\ref{fig:shdwvaryLmodel1}, the observed intensities $I_{obs}^1$, $I_{obs}^2$ and $I_{obs}^3$ are clearly separated, with no regions of overlap. The total intensity $I_{obs}$ in the second column exhibits three distinct narrow peaks, corresponding to the photon ring, lensing ring and direct emission, respectively. Since the emission region lies entirely outside the photon sphere ($r_{ph} < r_{ISCO}$), the two-dimensional shadow region of $I_{obs}$ for $b < b_{ph}$ appears completely dark, as displayed in the third column of Fig.~\ref{fig:shdwvaryLmodel1}. 
    Interestingly, within a finite range of $\lambda$, the $b_{ph}-r_{ISCO}$ relation becomes multi-valued, namely three distinct shadow radii correspond to the same $r_{ISCO}$, as shown in the top left panel of Fig.~\ref{fig:ISCO_bc}. 
    Consequently, for a thin-disk model whose emission peaks at $6.5\lesssim r_{ISCO}\lesssim 8$, there are three distinct images of the $q$-deformed object corresponding to three distinct value of $\lambda$. 
    Moreover, two distincts $\lambda$ can potentially yield the images with the same shadow region as can be seen in the top left panel of Fig.~\ref{fig:ISCO_bc}.


    \item    
    For \textbf{Case~II}, the intensity profile of the accretion disk begins with a peak near $r_{ph}$ and decreases monotonically with increasing $r$. Although increasing $\lambda$ generally reduces the thickness of the photon ring region and suppresses the flux from the photon ring, in this case the overlap among $I_{obs}^1$, $I_{obs}^2$ and $I_{obs}^3$ as depicted in the first column of Fig.~\ref{fig:shdwvaryLmodel2} enhances the photon and lensing rings contribution to $I_{obs}$, as shown in the second column of Fig.~\ref{fig:shdwvaryLmodel2}. 
    In contrast to the \textbf{Case~I}, the results suggest that larger values of $\lambda$ enhance the brightness of both the photon ring and the lensing ring regions.
    Since the peak of direct emission occurs at $b < b_{ph}$, the interior of the shadow image is not completely dark, as illustrated in the third column of Fig.~\ref{fig:shdwvaryLmodel2}.

    \item For \textbf{Case~III}, the intensity profile peaks near $r_+$ and then monotonically decreases with increasing $r$.
    In this case, the individual observed intensities $I_{obs}^1$, $I_{obs}^2$ and $I_{obs}^3$ overlap. This overlap enhances the resulting photon and lensing ring regions within the total observed intensity $I_{obs}$, as illustrated in the first and second columns of Fig.~\ref{fig:shdwvaryLmodel3}.
    Notably, as $\lambda$ increases, i.e., $\lambda =0.5, 1$ and $\sqrt{3}$, $b_{ph}$ decreases while $r_+$ increases.
    This shift implies that the photon and lensing rings occur closer to the peak of the accretion disk's emission profile. 
    Consequently, the results lead to a sharper and clearer ring structure on the observed image.
    
\end{itemize}

Next, we examine how varying the scalar charge $\tilde{D}$ affects the images of spacetime near the $q$-deformed object in \textbf{Case~I}, \textbf{II} and \textbf{III}, as shown in Figs.~\ref{fig:shdwvaryDmodel1}, \ref{fig:shdwvaryDmodel2} and \ref{fig:shdwvaryDmodel3}, respectively. 
In these figures, we consider $\tilde{D} = -1.5$ (first row), $\tilde{D} = -0.5$ (second row) and $\tilde{D} =0.5$ (third row), while keeping $\lambda =1.5$ and $\tilde{P}=0.1$ fixed.
As evidenced by the middle panel of the bottom row in Fig.~\ref{fig:Rs} and the middle panel in Fig.~\ref{fig:ISCO_bc}, the shadow radius $b_{ph}$ increases along with $r_{ISCO}, r_{ph}$ and $r_+$ in the range from $\tilde{D}=-1.5$ to $\tilde{D}=-0.5$ and subsequently decreases when $\tilde{D}=0.5$. 
We demonstrate how $\tilde{D}$ influences the images as follows:
\begin{itemize}
    \item In \textbf{Case~I}, the first column in Fig.~\ref{fig:shdwvaryDmodel1} shows that emission's peak from direct (black), lensing-ring (orange) and photon ring (red) regions remain clearly separated as $\tilde D$ increases.
    As illustrated in the third and fourth columns in Fig.~\ref{fig:shdwvaryDmodel1}, the shadow area increases from $\tilde{D}=-1.5$ to $\tilde{D}=-0.5$ and then slightly decreases when $\tilde{D}=0.5$, consistent with the middle column in Fig.~\ref{fig:ISCO_bc}.
    \item For \textbf{Case~II}, the emission peak, which is located near $r_{ph}$, follows the trend of $r_{ph}$ and decreases as $\tilde{D}$ increases.
    As depicted in the first column in Fig.~\ref{fig:shdwvaryDmodel2}, the overlap of the three individual observed intensities enhances the contribution of the photon and lensing rings to the total observed intensity $I_{obs}$, which is displayed in the second column in Fig.~\ref{fig:shdwvaryDmodel2}.
    Notably, as the scalar charge $\tilde{D}$ increases, the intensity contribution from the direct region separates from the photon ring and lensing ring regions. 
    Conversely, the emission from the photon ring and lensing ring themselves exhibit increasing overlap, as illustrated in the first column of Fig.~\ref{fig:shdwvaryDmodel2}. 
    Consequently, this behavior leads to reduced visibility of the photon ring sub-structure, as displayed in the third and fourth columns of Fig.~\ref{fig:shdwvaryDmodel2}.
    \item For \textbf{Case~III}, where the emissivity peaks near $r_+$, the peak location moves inward (to smaller $r$) as $\tilde{D}$ increases. 
    As shown in the first column of Fig.~\ref{fig:shdwvaryDmodel3}, the direct-emission peak shifts to smaller $b$, whereas the photon ring and lensing ring peaks move outward to larger $b$ with increasing $\tilde{D}$. 
    Consequently, at $\tilde{D}=-1.5$ the three peaks are closely spaced, making the sub-ring structure more pronounced than at larger $\tilde{D}$, see the third and fourth columns in Fig.~\ref{fig:shdwvaryDmodel3}.

\end{itemize}

Finally, we investigate the influence of the magnetic charge $\tilde{P}$ on the resulting images for GLM \textbf{Case I-III}, while $\lambda =0.5$ and $\tilde{D}=-0.8$ were fixed. 
The third column of the bottom row in Fig.~\ref{fig:Rs}, together with the right column of Fig.~\ref{fig:ISCO_bc}, shows that the critical impact parameter $b_{ph}$ and the characteristic radii, i.e., $r_{ISCO}, r_{ph}$ and $r_+$,  decrease monotonically with $\tilde{P}$. 
The effects of $\tilde{P}$ on the optical appearance for three cases of GLM were illustrated as follows:
\begin{itemize}
    \item For \textbf{Case I}, the direct image, the lensing ring and the photon ring remain clearly separated, as shown in the first column in Fig.~\ref{fig:shdwvaryPmodel1}.
    As $\tilde{P}$ increases, the intensity peaks corresponding to these three regions shift toward a smaller $b$. 
    The resulting $I_{obs}$ and images in $X-Y$ plane show a shrinking shadow region as seen in the second, third and fourth columns in Fig.~\ref{fig:shdwvaryPmodel1}.
    \item In \textbf{Case II}, the three characteristic peaks of emission intensities are overlap, as seen in the first column in Fig.~\ref{fig:shdwvaryPmodel2}.
    The resulting $I_{obs}$, displayed in the second column in Fig.~\ref{fig:shdwvaryPmodel2}, exhibits two peaks, one occur at $b<b_{ph}$ with lower intensity and $b\sim b_{ph}$ with higher intensity. 
    These two peaks are correspond to the direct image and the photon ring regions, respectively.
    Furthermore, as illustrated in the third and fourth columns in Fig.~\ref{fig:shdwvaryPmodel2}, the width of the emission intensity originating from the  photon ring region broadens as $\tilde{P}$ increases.
    \item Finally, we investigate the optical appearance of the $q$-deformed object in \textbf{Case III} of the GLM model as $\tilde{P}$ is varied.
    The first column in Fig.~\ref{fig:shdwvaryPmodel3} illustrates that $I_{obs}^1$, $I_{obs}^2$ and $I_{obs}^3$ are overlap in such a way that the total observed intensity $I_{obs}$ exhibiting two distinct peaks as seen in the second column in Fig.~\ref{fig:shdwvaryPmodel3}.
    The first peak located at $b<b_{ph}$ has a lower intensity, while the second peak situated near $b\sim b_{ph}$.
    These first and second peaks corresponding to the direct image and photon ring, respectively.
    As $\tilde{P}$ increases, these two peaks move toward each other, which makes the photon ring visibly sharper in the resulting images, as demonstrated in the third and fourth columns in Fig.~\ref{fig:shdwvaryPmodel3}.

\end{itemize}

  \clearpage

\section{Conclusions}\label{sec:conclude}





In this work, we have performed a comprehensive investigation on the optical signatures of $q$-deformed objects, which represent a novel class of solutions within the EMD gravity. 
The deformation parameter $q$ is intrinsically linked to three fundamental parameters of the solution: the dilaton coupling $\lambda$, the scalar charge $\tilde{D}$, and the magnetic charge $\tilde{P}$. 
Notably, in the limiting case where $q=1$, our model recovers the well-known results of the dilaton black hole.

To characterize the observational features, we first analyze the effective potential and the behavior of null geodesics in these $q$-deformed backgrounds. 
By employing the total number of orbits $n$ to classify light ray trajectories, we examine the contributions from direct emission, the lensing ring and the photon ring. 
Our analysis reveals that the width of the photon ring region decreases as $\lambda$ increases, whereas it expands with higher values of $\tilde{D}$ and $\tilde{P}$. 
This behavior is found to be in agreement with the variations of the angular Lyapunov exponent $\gamma$, which dictates the instability of the photon orbits.

In addition, we have studied the shadow cast by $q-$deformed objects. By varying $\lambda$ with fixed $\tilde{D}$ and $\tilde{P}$, we find that $b_{ph}$ exhibits non-monotonic behavior with respect to $\lambda$.
Specifically, $b_{ph}$ initially starts below $5.20$ which marks the shadow radius of the Schwarzschild black hole at $\lambda=0$, and slowly increases from $b_{ph}=4.75$ in $\lambda \geq1$ region as illustrated in Fig.~\ref{fig:Rs}. 
The shadow radius exhibits a non-monotonic dependence on $\tilde{D}$ when the parameters $\lambda$ and $\tilde{P}$ are held fixed. 
As $\tilde{D}$ varies from negative to maximum values of the extremal limit, $b_{ph}$ is found to be larger than the Schwarzschild case (at $\tilde{D}=-2,b_{ph}=7$) and evolves to $5.11$ at the extremal limit. Furthermore, $b_{ph}$ monotonically decreases from $6.42$ to $5.00$  as $\tilde{P}$ increases when $\lambda$ and $\tilde{D}$ are fixed.  
This behavior highlights the significant influence of three parameters on the shadow region for $q-$deformed objects.

Finally, we investigate the observational appearances of $q$-deformed objects under illumination of thin accretion disks with three distinct GLM emission profiles. 
We find that the model parameters $\lambda$, $\tilde{D}$, and $\tilde{P}$ play a crucial role in determining the peak positions and the distribution of the emission intensity profiles. Specifically, the shadow radius $b_{ph}$ exhibits a complex, non-monotonic dependence on the peak locations, i.e., $r_{ISCO}, r_{ph}$ and $r_+$, associated with each thin accretion disk profile, as illustrated in Fig.~\ref{fig:ISCO_bc}. 
Consequently, the size of the dark regions within the shadow, as modeled by the GLM accretion disk, demonstrates a non-monotonic evolution. 
This behavior distinguishes $q$-deformed objects from standard dilatonic black holes \cite{Promsiri:2023rez}, where such variations are typically monotonic.

Our results show that nontrivial dilatonic hair in $q-$ deformed EMD solutions can produce distinct optical signatures beyond the classical no-hair theorem.
This identifies $q-$ deformed objects as viable black hole mimickers.
Since EMD gravity naturally emerges as a low-energy limit of string theory, our findings open a potential observational window onto string-inspired physics in the strong-gravity regime.

\begin{acknowledgments}
This research has received funding support from the NSRF via the Program Management Unit for Human Resource and Institutional Development, Research and Innovation grant number $B39G680009$. 
\end{acknowledgments}

\appendix

\section{Timelike geodesics: $r_{ISCO}$}\label{app:isco}
\begin{figure}
    \centering
    \begin{subfigure}[t]{.3\linewidth}
        \includegraphics[width=5.4cm]{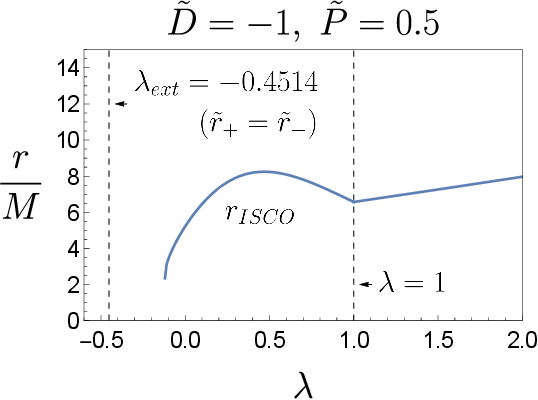}
    \end{subfigure}
    \hspace{6mm}
    \begin{subfigure}[t]{.3\linewidth}
        \includegraphics[width=5.4cm]{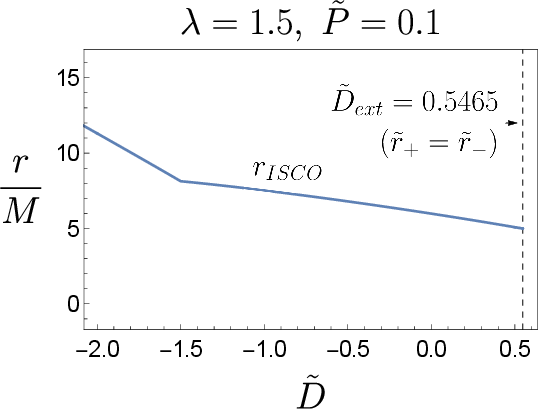}
    \end{subfigure}
    \hspace{6 mm}
    \begin{subfigure}[t]{.3\linewidth}
        \includegraphics[width=5.4cm]{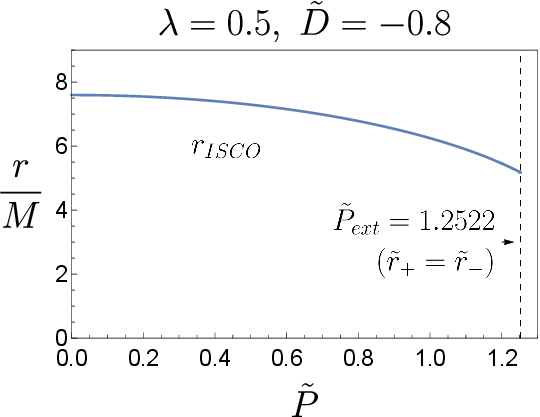}
    \end{subfigure}
    \caption{The innermost stable circular orbit radius $r_{ISCO}$ for different variation of parameters $\lambda$, $\tilde{D}$ and $\tilde{P}$. The black dashed line represents the extremal case $r_+=r_-$ for each variation.  \justifying} 
    \label{fig:risco}
\end{figure}

This appendix is devoted to a study of the radius of innermost stable circular orbit (ISCO). Let us consider, the equation describing massive particle \eqref{eq22} i.e., $\delta=0,\theta=\pi/2,\mathcal{Q}=0$
\begin{align}
    \left(\frac{dr}{d\phi}\right)^2 &=\frac{h(r)^2}{g(r)}\left(\frac{1}{b^2f(r)}-\frac{1}{h(r)}-\frac{1}{L^2}\right) \equiv v_{eff}.
\end{align}
Here $v_{eff}$ can be considered as an effective potential of massive particle. The ISCO radius is simultaneously satisfied by the following
\begin{align*}
    \nu_{eff}=0 ~,\frac{d \nu_{eff}}{dr}=0 ~,\frac{d^2 \nu_{eff}}{dr^2}=0.
\end{align*}
The first condition allow us to exactly determine the impact parameter
\begin{align}
    b^2&=\frac{L^2h(r)}{f(r)\left[L^2+h(r)\right]}\biggr\rvert_ {r=r_{ISCO}}.\label{eqa3}
\end{align}
Thus, we substitute it into the second condition to obtain the constant $L$
\begin{align}
    L^2&=\frac{f'(r)h(r)^2}{f(r)h'(r)-f'(r)h(r)}\biggr\rvert_{r=r_{ISCO}}.\label{eqa4}
\end{align}
Lastly, the third condition implies
\begin{align}
\left[h''(r)+h'(r)\left(\frac{2f'(r)}{f(r)}-\frac{2h'(r)}{h(r)}-\frac{f''(r)}{f'(r)}\right)\right]_{r=r_{ISCO}}&=0 .\label{eqa5}
\end{align}
The ISCO radius can be determined from this equation. The behaviour of the ISCO radius under influence of $\tilde{P}$, $\lambda$ and $\tilde{D}$ is demonstrated in Fig.~\ref{fig:risco}. 

In the left panel, we shows the ISCO radius that varies with $\lambda$ while $\tilde{D}=-1$ and $\tilde{P}=0.5$ are fixed. As shown, the radius cannot be plotted from the extremal case $\lambda_{ext}=-0.4514$ because in the absence region, the radii become complex value. We obtain the real-value radii at about $\lambda\sim-0.1$. Interestingly, the ISCO radius increases and reaches a local maximum peak at $\lambda\sim0.47$ and then decreases to $\lambda=1$. After that, the radius steadily rises when $\lambda$ increases. For the middle panel, the ISCO radius is plotted as a function of $\Tilde{D}$ with fixed $\lambda=1.5$ and $\tilde{P}=0.1$. In general, we see that the radii decrease with increasing $\Tilde{D}$. However, interestingly, the declining rate changes at $\Tilde{D}\approx -1.5$. Finally in the right panel, we fix $\lambda=0.5$ and $\tilde{D}=-0.8$ and vary $\Tilde{P}$. We can see that the ISCO radius monotonically decreases as $\tilde{P}$ increases and reaches its minimum value at at the extremal value of $\tilde{P}_{ext}=1.2522$. 

We remark that the ISCO radii of the Schwarzschild black hole and extremal Reissner-Nordstr\"om black hole are $6$ and $4$, respectively \cite{Chandra} . In our case, the ISCO radius can be either smaller and larger than those of the Schwarzschild and RN black holes. It is obvious from the right panel of Fig.~\ref{fig:risco}, that the ISCO radius decreases as one approaches the extremal limit $\tilde{P} \to \tilde{P}_{ext}$ but never reduces to the RN case ($4$). In contrast to our result, in the case of electrically charged black hole in EMD gravity, the ISCO radius reduces to $6$ as black hole's charge $q=0$ for generic value of $\lambda$ (see Fig. 19 of \cite{Promsiri:2023rez}). More detailed analysis on ISCO radius of q-deformed objects can be found in \cite{Sakkawattana_Ponglertsakul_2025}.



\section{Lyapunov exponent}\label{app:lypunov}
Here in this appendix,  we explain the Lyapunov's exponent, which relates to the width of the photon ring emission. We consider a region around photon ring by introducing $\delta r=r-r_{ph}$. The circular motion equation \eqref{radialEq} is given by
\begin{align}
    \frac{d\delta r}{d\tilde{\sigma}}&=\frac{1}{\sqrt{f(\delta r+r_{ph})g(\delta r + r_{ph})}}\sqrt{\frac{1}{b^2}-V_{eff}(\delta r+r_{ph})}, \nonumber \\
    &\approx \frac{1}{\sqrt{f(r_{ph})g(r_{ph})}}\delta r\sqrt{-\frac{1}{2}V''_{eff}(r_{ph})}.\label{eqb1}
\end{align}
With \eqref{eq9}, the equation of photon circular motion $\delta r$ in term of $\phi$ is expressed as 
\begin{align}
    \pi \frac{d\delta r}{d\phi}&= \delta r~\frac{\pi h(r_{ph})}{{\sqrt{f(r_{ph})g(r_{ph})}}}\sqrt{-\frac{1}{2}V''_{eff}(r_{ph})}.
\end{align}
This equation has the following solution
\begin{align}
    \delta r(\phi)&=\delta r_{0}e^{\frac{\gamma\phi}{\pi}},\label{eqb4}
\end{align}
where we have defined the \textit{angular Lyapunov exponent} $\gamma$ \cite{Broderick:2023jfl,Johnson:2019ljv}
\begin{align}
    \gamma& \equiv \frac{\pi h(r_{ph})}{\sqrt{f(r_{ph})g(r_{ph})}}\sqrt{-\frac{1}{2}V''_{eff}(r_{ph})}.\label{eqb3}
\end{align}
The $\delta r_0$ measures an initial deviation from the circular orbital motion. Therefore, from \eqref{radialEq}, when $dr/d\tilde{\sigma}=0$, we obtain
\begin{align}
    b^2&=\frac{1}{V_{eff}(r_{ph}+\delta r_0)}, \nonumber \\
    &\approx \frac{1}{V_{eff}(r_{ph})}+\frac{1}{2}\left[\frac{1}{V_{eff}(r)}\right]_{r_{ph}}''\delta r_0^2. \label{eqb5}
\end{align}
We introduce the number of half-orbit $m=\frac{2\phi}{\pi}$. Hence, \eqref{eqb4} can be written as
\begin{align}
    m\approx \frac{2}{\gamma}\ln\left(\frac{\delta r_{max}}{\delta r_0}\right),\label{eqb6}
\end{align}
where $\delta r_{max} \approx 1$ defines the maximum value of $\delta r$ which the Taylor series expansion remains valid. Remark that, the shadow radius is given as $R_s=b_{ph}$. Thus, one defines $R_m$ as the $m-$th order photon subring that associates with $\delta r_{max}$. Then, the width of photon ring can be derived as follows
\begin{align}
    R_m-R_s&\approx\frac{1}{4}V_{eff}(r_{ph})\left[\frac{1}{V_{eff}(r)}\right]''_{r_{ph}}\delta r_{max}^2 e^{-m\gamma}.
\end{align}
It is clear that the width of photon ring is exponentially suppressed by $e^{-m\gamma}$ factor.

\begin{figure}[htbp]
    \centering
    \begin{subfigure}[t]{.3\linewidth}
        \includegraphics[width=5.4cm]{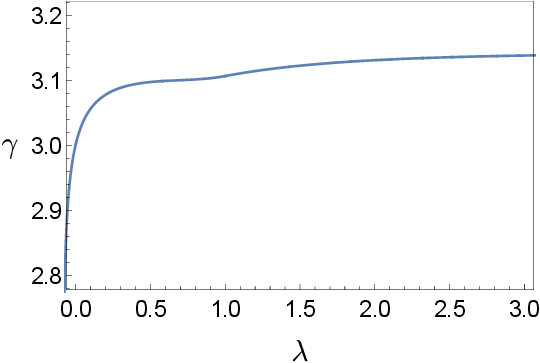}
        \caption{Vary $\lambda$ $,\tilde{P}=0.5~,\tilde{D}=-1$}
    \end{subfigure}
    \hspace{5mm}
    \begin{subfigure}[t]{.3\linewidth}
        \includegraphics[width=5.4cm]{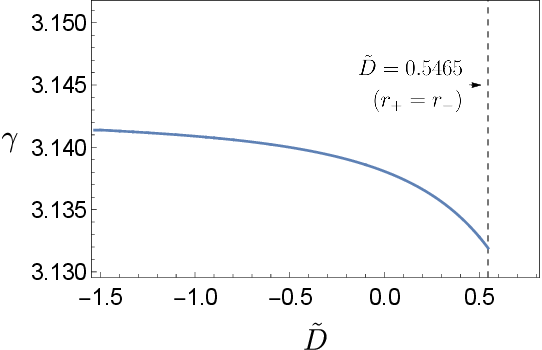}
        \caption{Vary $\tilde{D}$ $,\tilde{P}=0.1~,\lambda=1.5$}
    \end{subfigure}
    \hspace{5mm}
    \begin{subfigure}[t]{.3\linewidth}
        \includegraphics[width=5.4cm]{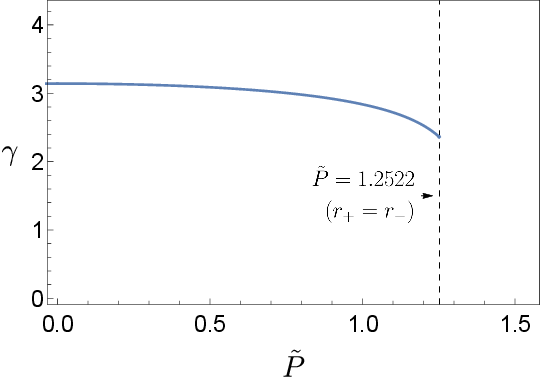}
        \caption{Vary $\tilde{P},\lambda=0.5~,\tilde{D}=-0.8$}
    \end{subfigure}
    \caption{The angular Lyapunov exponent for different variation of parameters $\tilde{D}$, $\tilde{P}$ and $\lambda$. The dashed line represents the extremal case $r_+=r_-$ for each variation.  \justifying } 
    \label{fig:newlypunov}
\end{figure}
In Fig.~\ref{fig:newlypunov}, we plot the angular Lyapunov exponent for the variation of parameters $\lambda$, $\tilde{D}$, and $\tilde{P}$ while the other two parameters are fixed. In the left panel, the Lyapunov exponent shows the variation of $\lambda$ with fixed $\tilde{P}=0.5$ and $\tilde{D}=-1$. The Lyapunov exponent becomes smaller when $\lambda$ is decreasing to zero. In the middle panel, the Lyapunov exponent is ploted as a function of parameter $\tilde{D}$ with fixed $\tilde{P}=0.1$ and $\lambda=1.5$. We observe that as $\tilde{D}$ increases, the Lyapunov exponent generally decreases and eventually reaches its minimum value $\gamma=3.1319$ at the extremal limit i.e., $\tilde{D}=0.5465$. In the right panel, the Lyapunov exponent is illustrated as a function of $\tilde{P}$ with fixed $\lambda=0.5$ and $\tilde{D}=-0.8$. The Lyapunov exponent is decreasing while $\tilde{P}$ increases and reaching to the finite value $\gamma=2.3588$ at the extremal $\tilde{P}=1.2522$. Interestingly, from further numerical exploration, we notice that the Lyapunov exponent asymptotically approaches $\pi$ as $\lambda$ gets larger and $\tilde{D},\tilde{P}$ get smaller.

\bibliography{EMD}

\end{document}